\documentclass[INR,biber]{nowfnt} 

\usepackage{hyperref}
\usepackage[totoc, font=footnotesize]{idxlayout}

\usepackage{paralist}

\AtBeginBibliography{\footnotesize}

\newcommand{\ie}{\emph{i.e.,~}}
\newcommand{\eg}{\emph{e.g.,~}}

\title{Mathematical Information Retrieval}
\subtitle{Search and Question Answering}

\maintitleauthorlist{
Richard Zanibbi \\
Department of Computer Science\\
Rochester Institute of Technology \\
rxzvcs@rit.edu
\and
Behrooz Mansouri \\
Department of Computer Science\\
University of Southern Maine\\
behrooz.mansouri@maine.edu
\and
Anurag Agarwal \\
School of Mathematics \&  Statistics\\
Rochester Institute of Technology\\
axasma@rit.edu
}

\issuesetup
{%
 copyrightowner={R. ~Zanibbi and and B. ~Mansouri and A. ~Agarwal },
 volume        = xx,
 issue         = xx,
 eisbn         = xxx-x-xxxxx-xxx-x,
 doi           = 10.1561/XXXXXXXXX,
 firstpage     = 1, 
 lastpage      = 18
 }

\author[1]{Zanibbi, Richard}
\author[2]{Mansouri, Behrooz}
\author[1]{Agarwal, Anurag}

\affil[1]{Rochester Institute of Technology; \{rxzvcs,axasma\}@rit.edu}
\affil[2]{University of Southern Maine; behrooz.mansouri@maine.edu}

\articledatabox{\nowfntstandardcitation}

\DefineBibliographyStrings{english}{%
  mathesis = {Master's thesis},
}

\usepackage{enumitem}

\usepackage[utf8]{inputenc}
\usepackage[dvipsnames]{xcolor}
\usepackage{booktabs}
\usepackage{verbatim}
\usepackage{amsfonts}
\usepackage{amsmath}
\usepackage{amssymb}

\usepackage{amsbsy,amsmath,amsthm,dsfont,latexsym,alltt,array}
\usepackage{mathtools}
\usepackage{pgfpages,tikz}
\usetikzlibrary{shapes,arrows,matrix,decorations.markings}
\usepackage{hf-tikz}
\usepackage[most]{tcolorbox}
\usepackage{geometry,caption,wrapfig}

\usepackage{tikz-cd}
\usepackage{listings}

\usepackage{fancyvrb}
\usepackage{float}

\makeatletter
\renewcommand{\boxed}[1]{\text{\fboxsep=.2em\fbox{\m@th$\displaystyle#1$}}}
\makeatother

\usepackage{subcaption}

\newtcbtheorem[number within=chapter]{myexample}{Example}{
                lower separated=false,
                colback=white,
colframe=white!50!black, fonttitle=\bfseries,
colbacktitle=white!50!gray,
float,
floatplacement=tb!,
coltitle=black,
enhanced,
boxed title style={colframe=black},
fonttitle=\small,
fontupper=\small,
}{myexa}

\newtcbtheorem[auto counter]{myneed}{Strategy}{
                lower separated=false,
                colback=white,
colframe=white!50!black, fonttitle=\bfseries,
colbacktitle=black,
float,
floatplacement=tbp,
coltitle=white,
enhanced,
boxed title style={colframe=black},
fonttitle=\small,
fontupper=\scriptsize,
attach boxed title to top center={yshift=-2mm}
}{myneed}

\newcommand{\refchapter}[1]{Chapter \ref{#1}}
\newcommand{\reffig}[1]{Figure \ref{#1}}
\newcommand{\reftab}[1]{Table \ref{#1}}

\newcommand{\refneed}[1]{Strategy \ref{myneed:#1}}
\newcommand{\refexample}[1]{Example \ref{myexa:#1}}

\newcommand{\imagefillwidth}[1]{
	\resizebox{\columnwidth}{!}{
		\includegraphics{ #1 }
		}
}

\newcommand{\ourfigure}[4]{
    \begin{figure}[#1]
    \begin{center}
        #2
    \end{center}
    \vspace{-0.15in}
    \caption{#3}
    \label{#4}
    \end{figure}
}
\newcommand{\ourfigscaled}[5]{
    \begin{figure}[#1]
    \begin{center}
    	\scalebox{#2}
	{\includegraphics{#3}}
    \end{center}
    \vspace{-0.15in}
    \caption{#4}
    \label{#5}
    \end{figure}
}
\newcommand{\ourtable}[4]{
    \begin{table}[!#1]
        \caption{#3}
        \label{#4}
        \begin{center}
        #2
        \end{center}
    \end{table}
}

\newcommand{\forceline}[0]{\mbox{}\\}
\newcommand{\fheader}[1]{\subsection*{\mbox{#1}}}
\newcommand{\ffheader}[1]{\forceline \forceline \forceline \fheader{#1}}

\begin{document}

\makeabstracttitle

\begin{abstract}
Mathematical information is essential for technical work, but its creation, interpretation, and search are challenging.
To help address these challenges, researchers have developed multimodal search engines and mathematical question answering systems. 
This book begins with a simple framework 
characterizing the information tasks that people and systems perform 
as we work to answer math-related questions.  
The framework is used to organize and relate the other core topics
of the book, including interactions between people and systems, 
representing math formulas in sources, and evaluation. 
We close by addressing some key questions and presenting directions for future work.  
This book is intended for 
students, instructors, and researchers
interested in systems that help us find and use mathematical information.
\end{abstract}


\begin{acknowledgements}

\textbf{From Richard.}
 Thanks to our colleagues from the MathSeer project (Douglas W. Oard, C. Lee Giles, Jian Wu, and Shaurya Rohatgi), and the MathSeer advisory board  (Marti Hearst, Michael Kohlhase, and Frank Tompa) for providing helpful insights over many years. We also wish to thank the Alfred P. Sloan Foundation and National Science Foundation (USA) for supporting the MathSeer project (2017-2022). 
 
Many students and staff have completed MIR-related research in the Document and Pattern Recognition Lab (dprl) at RIT. The approach taken in this book has been heavily influenced by their good work and good humour, and by their comments at over a decade of lab meetings. 
Before RIT, I had two excellent doctoral advisors at Queen's University in Canada, Dorothea Blostein and James R. Cordy. Their support and guidance were critical early in my career when I worked on recognizing  mathematical notation, and then again when I began doing math retrieval work after I had arrived at RIT. The SLT and OPT formula representations used in this book were originally developed in collaboration with Dorothea and Jim.

Also, the influence of George Nagy can be seen in both the subject matter and form of this book: his mentorship profoundly impacted my career and approach to research, and without his utterly unique style of intervention this book would not exist.\footnote{A research visit that opened with a promise to shop for a replacement house fan if the first author's ideas were uninteresting. We later went shopping, and then discussed other research directions, of which math-aware search became the focus.} George, I hope that something in this book interests you, and ideally, also makes you laugh. 

Another instructive experience was an invited talk that I gave at RMIT in Melbourne, at the invitation of Justin Zobel. Justin, thank you for the helpful discussions and support over the years, even when that talk turned out to be a sleep-deprived ramble. Thanks also for providing comments on the book draft despite being tremendously busy.
 
 In terms of bringing the topics of this book into better focus, we thank Jimmy Lin, who suggested adapting Broder's taxonomy to  characterize mathematical information needs more concretely, and Susan Dumais, who helpfully reminded me that variable names carry little information without additional context. Her comment made us ask ourselves, `exactly what information does a \emph{formula} carry?' which led to the approach taken in Chapter 2.
 
 The anonymous reviewers had numerous helpful comments, and their observations led to an improved organization of the book.
Other students and colleagues offered helpful comments on drafts including Harold Mouch\`ere, Bryan Amador, Patrick Philippy, Wei Zhong, and Nickvash Kani. Bryan in particular was instrumental in helping prioritize topics and improve the organization of the early chapters.
 
I am deeply grateful to Aikiko Aizawa and Yiqun Liu for providing the opportunity to write this book, and particularly to Yiqun for his tremendous patience as I struggled to finish it. 
Thanks also to the good people at Now Publishers, including Alet Heezemans, Mike Casey, Mark de Jongh, and Anahita Gupta, who were also helpful and patient despite repeated delays.

Behrooz and Anurag: thank you both for your many contributions to this book; the content is much more current, better illustrated, and better grounded mathematically as a result. Thanks also for hanging in there as what was supposed to be a 1 year project exceeded 3 years.

Most of all, I need to thank my wife and partner Katey, without whose reminders and support neither the time nor courage to finish this book would  have ever been found.  Thanks also to my father whose influence motivated me to write a monograph in the first place, and my mother and daughters Alix and Evalyn for their love and patience as I worked...a lot...on this book.

\textbf{From Behrooz.} I am grateful to Richard and Anurag for providing this wonderful opportunity and for their collaboration. I extend my thanks to my Ph.D. advisors, Richard Zanibbi and Douglas W. Oard, whose support and guidance have been instrumental in shaping my career. Their mentorship enabled me to contribute to this book. Lastly, I would like to thank my students at the Artificial Intelligence and Information Retrieval (AIIR) Lab at the University of Southern Maine for their valuable feedback.

\textbf{From Anurag.} The support and encouragement of many wonderful individuals has been instrumental in this journey.  My deepest gratitude is to God for giving me an opportunity to learn mathematics and to my parents for instilling a love of knowledge in me. Nothing would have been the same had it not been for their infinite love and support in everything I did. 

I would like to extend special thanks to Richard (the first author) who introduced me to the world of MIR. His support has been a constant source of strength. He offered invaluable advice, and was a constant source of inspiration throughout the writing process. 

To Behrooz (the second author) and all the students who worked with us over the years on related projects, I am deeply grateful for your time, honesty, and thoughtful suggestions. You helped me see this story from perspectives I hadn’t considered.

\end{acknowledgements}
\chapter*{Preface}
\addcontentsline{toc}{chapter}{Preface}

This book provides an introduction to the foundations and current developments in
\emph{mathematical information retrieval} research, or \emph{math IR}. 
In this area we focus on  
systems designed to assist with finding, collecting, and using mathematical information.
With the advent of Large Language Models (LLMs), mathematical question answering is of
particularly keen interest at the moment.
Systems that combine LLMs with logic-based systems have been making news by
solving problems from the International Mathematical Olympiad, for example.

The authors have spent more than a decade working on systems and interfaces for
math-aware search engines in web pages, PDF documents, and even videos.
We have more recently worked with interactive \emph{conversational} systems for math IR,
and looked into applications of LLMs.
In this book we try to summarize what we have learned about systems  
for searching existing sources, and briefly introduce emerging math question answering
systems that automatically generate responses based on usage patterns for 
text \emph{and} formulas.

Our intended audience includes students, instructors, and researchers of information science,
computer science, and mathematics.\footnote{With this said, we have often been `unintended' readers of books, and warmly welcome anyone with even the slightest interest in what we have to say here.}
Because the goal of this series is to introduce a topic from its fundamentals and then build up to the state-of-the-art,
we needed to prioritize making the presentation as clearly and concretely as we could manage.
This means that 
we also had to make difficult decisions about what material to include.
We have focused on covering what we understand to be core topics and concerns, rather than provide an exhaustive survey of techniques.  
For those interested in learning more than we could cover here, we refer you to other math IR surveys that
are available
\citep{zanibbi_recognition_2012,guidi_survey_2016,10.1145/3699953}.

Our approach in this book is searcher-centered, \ie focused on models and systems used directly by people.
 We acknowledge that there are \emph{vast} bodies of literature concerned with searching and discovering mathematical information  with little or no human interaction.
 Important examples include automated theorem proving, one of the oldest and most influential corners of artificial intelligence
 research, and work with mathematical knowledge databases in the Mathematical Knowledge Management community (\eg at the CICM\footnote{https://cicm-conference.org/cicm.php} conferences), from which much of the early influential work in math IR also originates from.
 In Appendix B we present related work from the theorem proving space, but this is
 only a brief overview, and just the tip of the iceberg.

Structured representations for formulas and computations precede the electronic computer. 
The tree-based representations for formula structure presented in Chapter 2 were designed to be simple and capture key
structural properties, \ie writing lines of symbols for formula appearance, and operation hierarchies for mathematical expressions represented by formulas. 
However, we make no assertion of their novelty; many analagous representations have been used in papers and systems.
We've found that Symbol Layout Trees (SLTs) and Operator Trees (OPTs) generalize the graph types used for recognition and retrieval of formulas well.

\begin{figure*}[t!]
\imagefillwidth{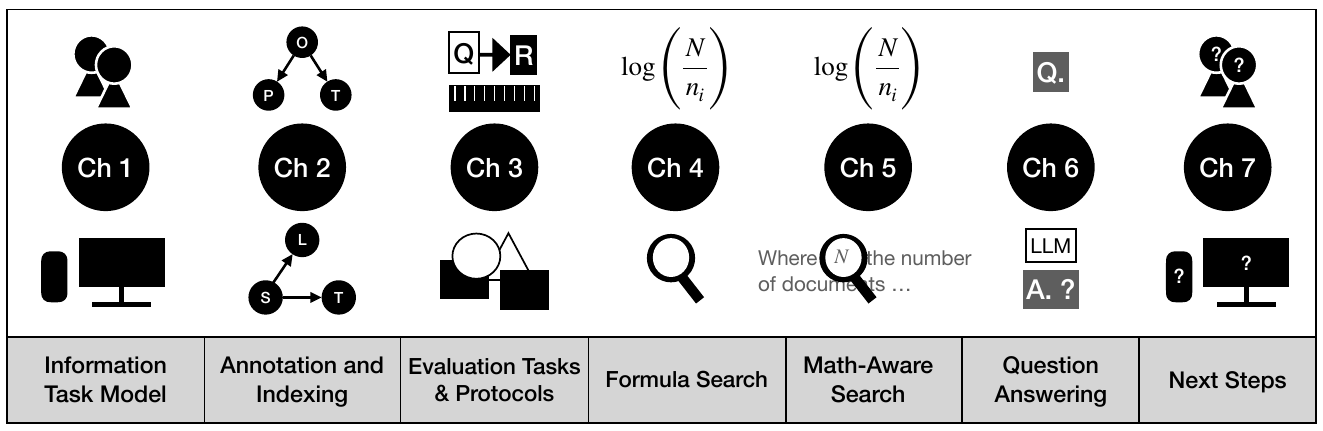}
\vspace{-0.2in}
\caption{Visual summary of this book's contents}
\label{fig:preface}
\end{figure*}

We'll close here with a brief summary of the contents of this book.
The chapters of the book are summarized visually in \reffig{fig:preface}, and in more detail below.
\forceline 
\begin{compactdesc}
\item[Chapter 1] introduces math IR, and presents a framework that unifies information seeking activities performed by people \emph{and} systems  in (1) the real world, with (2) `traditional' retrieval systems, and (3) question answering systems. 
The framework is organized around information needs, sources, and tasks, with an informal `source jar' model for human information seeking, and a structured task graph for systems. 
The systems-oriented model is used to organize material in Chs. 2-6.

\item[Chapter 2] considers the types of mathematical information present (and missing) in sources, and provides an overview of formula representations. The larger focus is annotating sources with additional information (\eg formula representations) and representing text and formulas in indexes for `sparse' (\ie discrete pattern-based lookup) and `dense' (\ie continuous vector space-based lookup) retrieval. 

\item[Chapter 3] presents the math IR tasks addressed in Chs. 4-6, along with procedures for creating \emph{test collections} for evaluation, and evaluation metrics used in benchmarks.  These are presented together to emphasize similarities and differences between retrieval and question answering tasks.

\item[Chapters 4-6] present past and current systems for formula search, text + formula search (`math-aware' search), and question answering. These systems depend upon indexing and evaluation techniques covered in Chs. 2 and 3. 
Each chapter begins with a summary of test collections for evaluation, followed by a presentation and comparison of methods.

\item[Chapter 7] provides a closing summary, a discussion of what math IR \emph{cannot} provide, and directions for future work organized around the information task framework from Ch. 1.

\end{compactdesc}

%

~\\
{
\flushright 

Richard Zanibbi, Rochester, NY, USA\\
Behrooz Mansouri, Portland, ME, USA\\
Anurag Agarwal, Rochester, NY, USA\\
December 2024\\
}

\chapter{Sources and Information Tasks}
\label{c-inf-needs} 

We often pause to search when something that we read, watch, or hear prompts questions that we want answers to. We then go about finding answers using additional sources of information: some already exist, some are created in response to requests (\eg emails or search results), and some are created to record and organize what we find.

In this way, information sources are the backbone of our information ecosystems.  The
sources available to us place a hard limit upon which questions we can answer. In addition to the information content in a source, its terminology, notation, writing style, and other factors determine the amount of information one can recover from a source, and how accurately and completely. This is a key reason why math instructors that communicate well are so highly regarded: they help us more easily understand topics by \emph{how} they speak, write, and present exercises. Through course materials, lectures, and conversations, these instructors provide multiple sources tailored to their students' level of understanding and communication style.

Outside the classroom, we still often find ourselves in need of mathematical information. It might be as simple as finding a formula to convert temperatures in Fahrenheit to Celsius, or the formula associated with a name (\eg  inverse document frequency). Or the goal may be more complex, such as understanding a proof of the \emph{sensitivity conjecture}.

As we look for answers, we will in some way annotate and organize the sources we find in order to identify and apply pertinent information, \eg  to find other sources, choose different search terms, execute suggested exercises, and make notes about partial answers to our questions. The effort needed for these tasks depends largely on the content and presentation in the sources that we access.
To save time, we often create additional sources of our own (\eg bookmarking a web page, placing notes in a file, or highlighting a PDF document).

In this book, when we speak about \textbf{sources}, we are usually referring to individual documents, recordings (\eg videos) or other artifacts that contain information. Libraries and other people are of course also information sources, in the sense that they can provide information, but here we use `sources' to refer to records of specific information.

To help organize our study of mathematical information retrieval, in this chapter we introduce a framework for information tasks based on sources.
The framework is built upon two main ideas:\\
\begin{compactenum}
\item Search begins, progresses, and ends with sources.
\item Tasks other than search are often needed to find information. \\
\end{compactenum}
The key components of the framework are:\\
\begin{compactitem}
\item \textbf{information needs} that individuals have,
\item \textbf{sources of information} that we search, consult, and create, 
\item information \textbf{tasks} performed to address information needs, and
\item their roles in \textbf{search algorithms and user-interfaces}. \\
\end{compactitem}
In the next section, we consider how these components interact when we have a mathematical question that we wish to answer.

\section{When and where do we search?}

Some short answers to this question are (1) when we have a question, and (2) wherever is easiest. While not very satisfying, these answers are basically correct.
Search is generally performed as part of some larger information task, and not for its own sake.\footnote{A fact that is both important and humbling for IR researchers.} This  motivates finding quick paths to answers. 

However, technical subjects such as mathematics can be complex. Finding and understanding information on math may require multiple activities, such as web search, reading sources (\eg Wikipedia pages and textbooks), taking notes, talking to instructors or colleagues, and doing exercises. 
As a result, when retrieving technical material on math and other specialized topics (\eg law, chemistry, music history), it is helpful to understand how search interacts with other information tasks. 

To illustrate, consider the more general problem of \emph{sensemaking}, which learning about detailed mathematical topics is closely related to.\footnote{See \citep{hearst2009search} for an overview of early research on sensemaking.}  In sensemaking, we construct a conceptual understanding of a topic with many sources, usually along with communicating this understanding. Common examples include writing a school term paper on an unfamiliar topic (\eg applications of category theory), or summarizing a complex historical event from multiple news reports.

Sensemaking tasks are challenging because information must be found in multiple sources, but also because this information must be analyzed, compared, and integrated. These thinking activities often require most of the effort for sensemaking. To manage these thinking tasks, we record plans, notes, and outlines to organize our work. These working documents may be checked repeatedly as we work, and as we write our final summary. They are themselves important information sources that provide the scaffolding needed to focus and ultimately complete work on a sensemaking task. 

To further illustrate information tasks that complement search, imagine taking handwritten notes on eigenvectors as described in a linear algebra textbook. The notes allow us to  annotate this source with our own observations, and record them for reference at a later time.  The analysis and insights in the notes come from  applying information that we know and find.  These notes communicate a new information source to a specialized audience: ourselves. 

For our notes to be useful, we organize them. Perhaps this is a purple sticky note that we attach to a monitor to check later in the evening. Or, perhaps we use a paper notebook with separate sections for different subjects, along with other organizational devices (\eg sticky notes acting as bookmarks). We might instead be using a tablet computer, which also provides handwriting recognition to convert the notes to computer-searchable data (\eg using {\tt Ctrl-f}).


The information tasks above are distinct from a basic search task where we submit a query, post a question, or send an email to obtain new information sources.
However, it turns out that search engines implement variations of the same information tasks described above: they need to \emph{index}, \emph{communicate}, \emph{annotate}, and \emph{apply}  information in sources to be effective. 
For example, we organize sources when we arrange sticky notes by topic and color on a wall, or construct an \emph{inverted index} mapping words or formulas to their document locations: these are both forms of \emph{indexing}. As another example, search engines produce Search Engine Result Pages (SERPs) summarizing documents matching a query, and question answering systems or AI `bots' produce answers. These retrieval system outputs and our notes are \emph{communications} creating new information sources.

Making notes on a passage requires us to \emph{apply} information to create an \emph{annotation}: additional information associated with the passage. In turn, if those notes were handwritten on a tablet computer, a system converting these to text and \LaTeX{} for math applies information captured in an algorithm, annotating the notes themselves. We end up with a hierarchy of annotations: the notes annotate a passage, while a recognition algorithm annotates the notes.  

In our framework we will distinguish different source types, based largely on what information tasks they are primarily used for. More specifically, we distinguish:\\
\begin{compactenum}
\item available sources on a topic including search queries and results,
\item information added to sources (\textit{annotation}), and
\item structures and organizations created for search (\textit{indexing})\\
\end{compactenum}



Getting back to our motivating question,
when we have identified a mathematical \emph{information need}, we generally start with questions, and hope to end with one or more information sources 
that we feel address or ideally answer those questions (\ie \textit{relevant} sources for the information need).  Where we search is motivated by the types of sources we expect to find from places online and/or the physical world (\eg conversations and post-its). Unless we are casually browsing resources on a topic, the places and order in which these sources are found will generally reflect attempts to reduce our time and effort.\footnote{Information foraging theory \citep{Pirolli1999} suggests we evolved to gather and consume information similar to food, governed by cost-benefit analyses.} Relevant sources are often of different types: perhaps a passage in a web page along with a SERP page, an answer from an online AI system, an email from a friend, and a green sticky note on your monitor.  

From this perspective, math-aware search engines and question answering systems are important tools, but only one among many resources for finding math information, and only a small part of what happens when we search for mathematical information.



\section{Information task framework}

While we focus in this book on information retrieval using computers, we wish to address sources in their broadest sense here. Not all sources are text documents, and not all sources are recorded in documents. Consider an informal conversation about Bayesian decision theory in the hallway, or observing that there are no clouds in the sky: often, your only record of important information is your own memory. 

In addition to textbooks, technical papers, and web pages, in recent years the types of resources used to locate mathematical information has grown to include substantial amounts of video \citep{DBLP:journals/access/DavilaXSG21} and audio, \eg~for course lectures, tutorials, and technical talks. Community Question Answering sites, and direct question answering is provided by resources such as Math Stack Exchange,\footnote{\url{https://math.stackexchange.com}} Wolfram Alpha, and large language models such as the Generative Pre-trained Transformer (GPT).

It is worth noting that when we have found or produced information we want to share or reuse, we usually produce a source of information ourselves. For example, when we have an answer to a math homework question, we create a physical or digital document, so that this can be checked by ourselves and graded by our instructor. If we found a helpful video while doing the homework, we might share it in a text message, which is itself a form of `micro-source.'

\paragraph{Differences in Information Sources.}
Especially when we include information obtained directly from our environment along with modern computing and communication devices, information sources may come in many forms. Sources vary in the dimensions listed below, among others.
\begin{compactitem}
\item immediacy, \eg having a conversation vs. reading a transcription
\item authorship, \eg human, machine generated, environment
\item interactivity, \eg a human/chatbot conversation vs. a document
\item audience, \eg grade school students vs. math professors
\item modality, \eg text, video, audio, or a web page combining these
\item purpose, \eg textbook, search results, or a search index
\item structure, \eg free text in a sticky note, vs. a book with chapters
\item length
\item formality, \eg proof vs. text message
\item style, \eg how concepts and examples are communicated
\item correctness, \eg correct vs. incorrect definition or proof
\item completeness, \eg partial vs. full search index
\item type of information, \eg technical, notes, communications\\
\end{compactitem} 
For the \emph{immediacy} of a source, we are referring to whether the source comes directly from observing a person's environment (\eg through conversation, experimentation, travel, etc.) or is recorded, as in a document or audiovisual recording.
Particularly with the advent of large language models, authorship and correctness are important concerns. In many cases LLMs and other sources may appear credible but are incorrect. Knowing how a source is created can help us determine how trusting we should be of it when we are uncertain about validity (\eg from the perceived expertise of an author or system). 

The intended audience and style of a source are also critical concerns. They determine the prerequisite knowledge needed to decide whether a source is relevant and to interpret and use information in the source.


\paragraph{Information task types.}
The primary task types we will use for people and systems come from common descriptions of sensemaking and simpler information tasks: retrieving, analyzing, and synthesizing. For this framework, these \textbf{tasks are considered at the source level}. For example, \emph{analyzing} a source refers to analysis that produces additional recorded information for a source (\eg in a note) rather than reading and interpreting a source without producing an observable artifact. 

\ourfigure
{t}
{
\hrule ~\\
\flushleft
{\bf Retrieve Information}
~\\
\begin{compactitem}
\item[R1] \textbf{Query} to request sources of information
\item[R2] \textbf{Consult} and interpret available sources, \emph{examining} and \emph{navigating} within and across sources 
\end{compactitem}

~\\
{\bf Analyze Information} 
~\\
\begin{compactitem}
\item[A1] \textbf{Annotate} sources with additional information, \eg notes, add formula locations  
\item[A2] \textbf{Index} sources by organizing them for retrieval 
\end{compactitem}

~\\
{\bf Synthesize Information}
~\\
\begin{compactitem}
\item[S1] \textbf{Apply} available information that we know, have in available sources, or is encoded in algorithms, etc.
\item[S2]  \textbf{Communicate} information by creating new sources\\
\end{compactitem} 
~\\
\hrule
~\\

}
{Information Task Taxonomy}
{fig:taxonomy}

We subdivide each of these into two subtasks based on how sources are created and used, producing six tasks in total, as shown
in \reffig{fig:taxonomy}.
The \emph{apply} task is critical, and used for all other tasks (\eg creating queries, consulting or creating information sources, or generating annotations and/or indexes). It is distinct because people often apply information without producing observable sources. For example, recognizing what a variable represents does not involve creating an annotation, index, or source outside of our own minds. The \emph{apply} task also identifies an important commonality between thought and computation (\eg algorithms): both apply information but using differing levels of formality, flexibility, and automation.


Not all information needs require queries. If we have  a helpful document describing the inverse document frequency on our laptop, we may simply consult it to review previously highlighted passages. This locating of items in an available source or across available sources through references and links is known as \emph{navigation}, which is distinct from submitting a \emph{query} to a system or person to find new sources.
As another example of using navigation to satisfy an information need, in some case we me may simply use the contents of available sources directly (\eg copying-and-pasting into an online form). 


\ourfigure
	{!t}
	{
		\scalebox{0.8}{
			\includegraphics{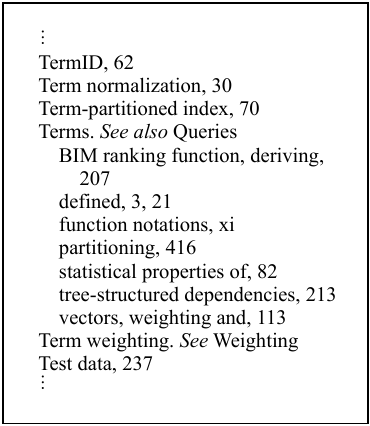}
		}
	}
	{Excerpt from the index to ``Introduction to Information Retrieval'' by Manning, Raghavan, and Sch\"utze. }
	{fig:BookIndex}

The process of analyzing a source and recording a map for use in retrieval is known as \emph{indexing}. 
Consider \reffig{fig:BookIndex}, where a book index provides a map for the book, so that a reader can quickly \emph{navigate} to parts of the book discussing `Terms,' for example. Contrast this \emph{subject index} with the index used in a traditional term-based search engine, which provide a much simpler map known as a \emph{concordance} recording where specific terms appear in documents  \citep{duncan}.
While these different indices are both used for retrieval, they differ in their scales (one document vs. a collection) and intended audiences (human reader vs. search algorithm). Other forms of indexing are less formal, such as collecting and organizing notes on different sticky notes for easier use.


As discussed earlier, we distinguish tasks for analyzing sources in terms of organizing them for use and retrieval (\emph{indexing}), and adding information to sources (\emph{annotation)}.  Annotations are often used in indexing sources, such as adding formula locations for PDF documents.

\ourfigure
	{!t}
	{
		\scalebox{0.5}{
			\includegraphics{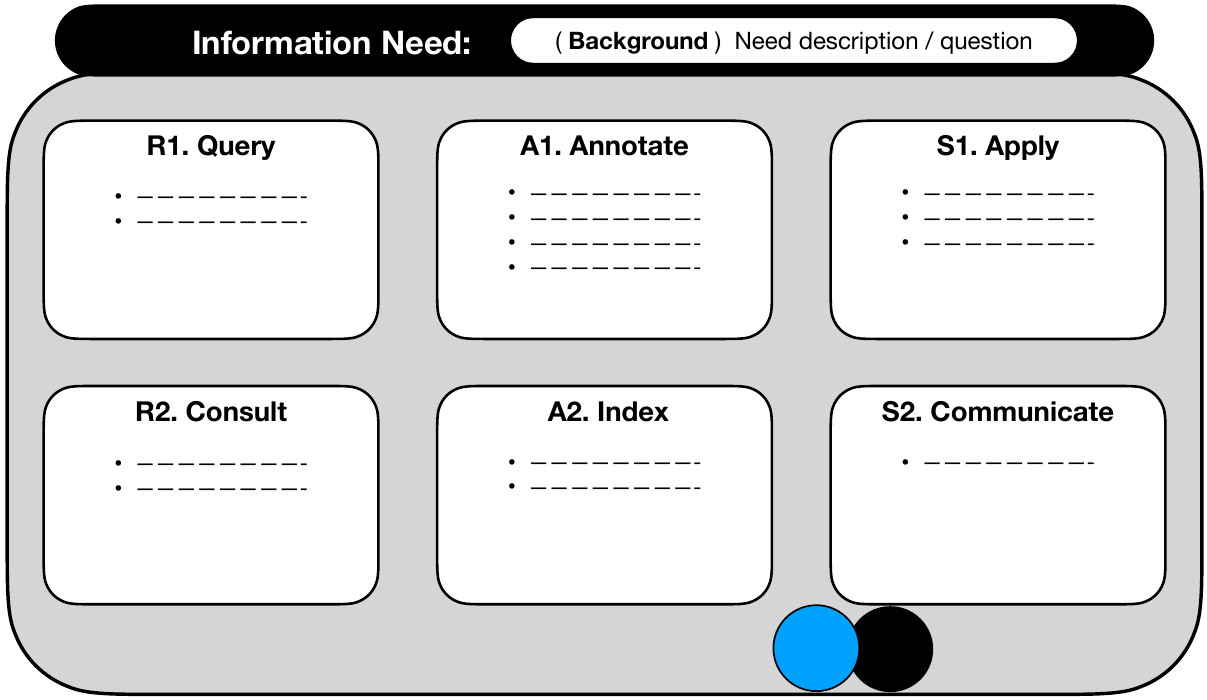}
		}
	}
	{Information Task Framework: The Source Jar. The jar contains source `marbles.' As we work we add, create, annotate and organize the sources in the jar, and record completed information tasks on the jar labels. }
	{fig:placeMatTasks}

\paragraph{Source Jar Framework.}
To put sources and the tasks used to create them in a more intuitive relationship, \reffig{fig:placeMatTasks} visualizes our task framework as a jar of sources with a lid.
 The jar contains immediately available sources as marbles in the jar. Each marble has an identifying color and shape. The source marbles contain information of different types, and may refer to other sources inside and outside of the jar. Sources that are directly available are either with us, or inside the jar. 
  
 The jar 
lid is labeled with the background of the searcher, and the need that information is being retrieved, analyzed, and synthesized for. 
Stickers on the outside of the jar record information tasks that we perform to address this need.
 When we find or create a new information source, we add a marble to the jar. 
 
 If a new source \emph{annotates} another source, we place it in a container with the source it describes inside the jar (\eg using a small plastic box). \emph{indexing} produces a marble containing a description of which sources it organizes, and how. We take source marbles and containers out of the jar to use them, and return them to the jar when they are no longer useful.  It is also possible to lose sources when the jar is accidentally left open and `spilled.'\footnote{\eg `the dog ate it,' `my internet is down,` or `I know it's here...somewhere.'}
When we stop working to find and create sources for our information need, we select any sources we might wish to use, and then close the lid.

We can imagine having a shelf of these jars for different information needs. 
For a new information need we create a jar, adding any potentially useful initial sources to the jar (possibly from other jars). 
To reuse or get additional information for a need we worked on previously, we open a jar from our shelf.\footnote{\eg `Wait; I forgot one of the types of category theory applications I wanted to discuss in my paper from my notes...'}  
Just as in the real world, not replacing sources in a jar runs the risk that we lose track of it, and have to pick up an old jar and work to \emph{refind} that source, or find a replacement for the information in the lost source.
 
This informal jar model is intended to roughly capture how people experience working with information in a simple way. It captures observable sources and observable task actions. We tend to move from source to source, performing tasks of specific types with a goal in mind. We are often unaware of why we performed tasks in a particular order, and so this is not represented explicitly, other than as marbles moving in and out of the jar and notes for which information tasks were executed being written on the outside of the jar.

\section{Information needs and information task strategies}


When searching for mathematical information, what we need to find will vary from finding definitions for terminology, math symbols, formulas, operational knowledge such as proof techniques, applications of mathematics (\eg information retrieval models), resources for instructors, and detailed information on mathematical spaces, theorems, etc.

\begin{myexample}{Differing Information Needs}{differ}

\textbf{Query:} What does 
$a^2+b^2=c^2$ represent and how is it useful?
\begin{description}
    \item[\it Students] might use this query to learn the Pythagoras theorem, and perhaps find an example demonstrating the theorem, and a possible proof.
    \item[\it Educators] may have similar interests to  students, but may seek additional resources on how to teach this result.
    \item[\it Researchers] can have very different interests than the other audiences. They may be interested in one of more of the following:
            \begin{compactitem}
                \item For a mathematician: Is this true in a general metric space and/or a Hilbert space?
                \item For a physicist: How is it linked to the probability assignments in quantum mechanics?
                \item For an IR expert: How is it related to probability assignments in a Hilbert space used in describing interaction for information retrieval?
            \end{compactitem}

\end{description}
\end{myexample}


\refexample{differ} illustrates information needs that  different audiences may be seeking to address using the same query, along with a list of sources that might be used to address their needs. These needs vary from finding definitions to exploring sophisticated relationships between the generalization of the theorem in different mathematical structures (Hilbert spaces) and applications in other fields (quantum mechanics).

As a result, the types of sources needed by each audience differ dramatically, but the initial (admittedly vague) query is identical: the \emph{query intent} differs for these audiences.  
For math information needs, we have found it important to consider information needs both in terms of the desired information, as well as who Some places where relevant information might be found for these information needs include web-pages (MSE, Brilliant.org etc.), YouTube videos, online lecture slides, text documents and digital books (\eg OpenStax, LibreTexts), articles, and online notes (\eg MIT OpenCourseWare).

For a broader sense of the types of mathematical information needs users have online,
\reftab{tab:broder} illustrates information needs for math organized by Broder's taxonomy of needs/intents behind web search queries \citep{DBLP:journals/sigir/Broder02}. 
While some question the usefulness of the transactional class in Broder's model, for math, the transactional class is a useful distinction. For example, a user may be looking to \emph{refind} a web page they used to enter formulas in \LaTeX~(i.e., a navigational intent). Or, they may instead be looking to find such a web page for the first time, thereby looking to interact/transact with as-yet unknown websites (i.e., a transactional intent).

\ourtable
    {!t}
    {
    \fbox{
    \small
    \begin{tabular}{l l}
         \bf Navigational: &  Find a specific source (`known item' retrieval) \\
         \hline
         & Web page (\eg for formula entry)\\
         & Document (\eg Book, Technical Paper) \\
         & YouTube or Khan Academy Video \\
         & Podcast \\
         &\\
         
        \bf Transactional: &  Find online resources for use/interaction\\
         \hline
         & Formula entry\\
         & Evaluating and plotting a formula\\
         & Simplification of a formula\\
         & Interactive theorem proving\\
              &\\
              
         \bf Informational: &  Find information for a topic or question \\
         & \textit{Sub-categories}: computation, concepts, and proofs\\
         \hline
         & How to compute an expression (\eg integral)\\
                 & Symbol and operation definitions (\eg $\zeta$, $n \choose k$)\\
         & Concept name(s) associated with a formula\\
         & When is a function not differentiable?\\
         & Who was Gauss?\\
         & Proof drafts for P = NP

    \end{tabular}
    }
    }
    {Examples of Mathematical Information Needs within Broder's Taxonomy \citep{DBLP:journals/sigir/Broder02}). A user's math background is another dimension. \vspace{-0.2in}}
    {tab:broder}

 Within the \emph{informational} needs class, a distinct subclass of \emph{computational} information needs exist. These include needs to evaluate or simplify a formula, or to produce a proof for a statement using logical operations. 
 It was useful to distinguish questions that were seeking \emph{concepts,} \emph{proofs,} and \emph{computation} for the ARQMath shared tasks \citep{arqmath2022} that we discuss in \refchapter{c-eval}.
 
 In our work we have found it useful to consider math information needs in two dimensions, based on the type of information need as shown in \reftab{tab:broder}, and the user's mathematical background. More formally, we have a space/set of mathematical information needs $N$ defined by a Cartesian product of possible information needs ($T$)ypes and user/audience ($B$)ackgrounds ($N \in T \times B$). How these types and backgrounds interact is illustrated in \refexample{differ}.

\paragraph{Information task strategies.}
For a given information need, it helps to think about \emph{strategies} that might be used to satisfy it. 
We can sketch these in strategy `jar' diagrams as seen in the panel labeled \refneed{square}. 
The diagrams identify an information need, initial queries, expected tasks, readily available sources, and a list of where other relevant sources might be found.
We can imagine beginning a new search by writing the information need on the lid, putting already available source `marbles' in the jar, and then writing planned tasks on the jar labels. 
For readability, we use informal descriptions for the three main task types along with initial queries.

\begin{myneed}{(Student) Completing the Square}{square}

    \begin{description}
        \item[Retrieve:]     
        	~\\\textbf{Query:} how to complete the square\\ \emph{\mbox{~~~~}OR} $ax^2+bx+c=(\star)^2+ \text{constant}$?

		Search using the text query or possibly the symbolic query; $(\star)$ is a wildcard for any subexpression.
		Identify where the general method can be found, and examine the proof of the result.

        \item[Analyze:] Mark-up/bookmark sources to identify useful information.  Use a notebook to summarize key details found in sources.  Save examples for different cases, \eg $a,b>0$ and $c \leq 0$.
              
        \item[Synthesize:] Solve an integration problem on paper, such as $\displaystyle \int\frac{1}{x^2+4x+3} \, dx$.\\
            \hrule
            
           \item[Initial Sources:] Textbook
           \item[Possible Sources:] ChatGPT, YouTube, Prof. X?
    \end{description}
\end{myneed}

Let us first consider search strategies that might be used by undergraduate students, for learning how to complete a square, and to change the base of a logarithm (\refneed{square} and \refneed{log}).  
In both examples, two queries that might be used are given, and the \emph{Synthesis} tasks clarify the specific information need: the source they want to produce. 
In the first example, this involves completing an exercise on paper, and in the second example, obtaining a value from a calculator. Note that for the queries containing formulas, students might find it difficult or be unable to express the formulas in queries using a standard text query box, particularly if they are unfamiliar with \LaTeX{}. 

\begin{myneed}{(Student) Log Base Change}{log}

    \begin{description}
        \item[Retrieve:] 
            ~\\ \textbf{Query:} log base change \emph{OR} how to convert $\log_bx$ to $\log_cx$?
        
            The student may use the text or symbolic query. Find sources giving the conversion rule with general bases.

        \item[Analyze:] Markup sources and note down where relevant sources are located in a list (\eg in a text file).
          Save some special cases like converting $\log_{10}x$ to $\ln x$.         
        
        \item[Synthesize:] They use this to compute $\log_{4}13$ on a calculator as the $\boxed{\log}$ button on most calculators only represents $\log_{10}(\cdot)$.\\
            \hrule
            \item[Initial Sources:] Web page on log conversion (hard to read)
            
            \item[Possible Sources:] Somewhere online?
     \end{description}
\end{myneed}

\begin{myneed}{(Researcher) Sensitivity Conjecture}{sensitivity}

    \begin{description}
        \item[Retrieve:] 
        ~\\    \textbf{Query:} What is \textbf{Sensitivity Conjecture}? Has it been proven?

        Find papers/books defining the conjecture and providing proofs.

        \item[Analyze:] Since the conjecture is very technical, retrieved material is annotated with sources where  terminology in the conjecture can be comprehended. An index (graph) is made capturing the chronological account of progress on the proof.
        \item[Synthesize:] Results and the methods for proving this conjecture are used for similar problems, and new articles/material are created to disseminate the findings.\\
            \hrule
        \item[Initial Sources:] Email from colleague suggesting this might be relevant for my work.
         \item[Possible Sources:] online encyclopedias (Wikipedia, Wolfram MathWorld), online Q\&A sites (MathOverflow.net, AoPS, sciencedirect.com), YouTube videos, online lecture notes, text documents (\eg digital books, research articles), online science \& math magazines (Quanta Magazine), online math databases (Cornell's mathematics library, zbMATH Open, \AmS: Math Reviews)
     \end{description}
\end{myneed}

Now let's consider more advanced information needs for researchers.
The researchers may be interested in following progress on an old conjecture (\eg Riemann Hypothesis). Or, they may be interested in learning about a new possible proof of the problem, or perhaps they were unfamiliar with the problem but are curious to know more about it. \refneed{sensitivity} seeks information and a proof for a problem that was posed in 1994. It became a major unresolved question in mathematical computer science until 2019, when Hao Huang solved it.
Notice that our researcher is aware of \emph{many} places where relevant sources for their information need may be found in comparison with the undergraduate examples.

As another example, imagine that another researcher encounters a technical statement for the sensitivity conjecture, but which does not name it.
They want to know the status of the statement, and if there are associated results they can use in their own work. 
Here the searcher only wants to learn the conjecture's name, properties, and proofs for later reference.
The strategy from \refneed{sensitivity} needs to be altered, as reflected in \refneed{sensitivityv2}.  
In this second case, the researcher has a document summarizing the key findings and where sources may be found.

\begin{myneed}{(Researcher) Unknown Conjecture}{sensitivityv2}

    \begin{description}
        \item[Retrieve:] 
        ~\\    \textbf{Query:} Any set $H$ of $2^{n-1}+1$ vertices of the $n-$cube contains a vertex with at least $\sqrt{n}$ neighbors in $H$.
                
        The search is done using a textual query with \LaTeX{} for the formulas. 
        Related papers/books are collected and consulted for theorem definitions and proofs.
        
        \item[Analyze:] Retrieved sources are annotated with links to other sources where terminology used can be comprehended. Highlight the name of the statement when it is found.
        
        \item[Synthesize:] Create document summarizing the theorem name and key details, with cites/links to key sources found. Include link to a file directory on a laptop where additional notes in text and \LaTeX{} files can be found, if any.\\
            \hrule

           \item[Initial Sources]: Research paper with technical statement of interest
        \item[Possible Sources:] online math databases (Cornell's mathematics library, zbMATH Open, \AmS: Math Reviews), \ldots 
     \end{description}
\end{myneed}

A related challenge is that the interpretation of most mathematical expressions is \textit{context-dependent}, i.e., the same formula may refer to different concepts in different contexts. For example a student looking to understand the formula $\pi(m+n)$  using search will likely end up with multiple interpretations, which might represent:
                \begin{compactitem}
                    \item the distributive law: $\pi(m+n)=\pi m + \pi n$, or 
                    \item the value of the \emph{prime-counting function} that counts the number of primes less than or equal to $m+n$.
                   \end{compactitem}
 This  property of a single object signifying multiple entities is known as \emph{polysemy}, such as the word `apple' being used to represent both a food and a company, and often poses challenges for both information retrieval and natural language processing. 

%
%
%
%
%

 \paragraph{User studies and use cases.}
There are a small number of papers examining math retrieval online. We know of just two studies looking at user behaviors in text-based search engines for math. The first was for the DLMF system \citep{10.1007/978-3-642-39320-4_19}, which supported text and formula search in a standard text box using queries in a \LaTeX{}-based formula syntax. Few users at the time visited the site intending to search using formulas, most likely because of its novelty, and because this capability wasn't prominently featured on the site. What math queries were used were often short, or even single symbols. There also tended to be fewer click-throughs to pages from search results, and more query reformulation for formula queries; whether users were browsing formula search results for interest, had challenges satisfying information needs, or some combination of these is unclear.

The second log study was for a standard text-based search engine  \citep{mansouri_characterizing_2019}. Query logs from a Persian general-purpose search engine were used. Compared to the general case, search sessions for math topics were typically longer with more query refinements (\ie changing queries to try and improve results) and were less successful. In contrast to the DLMF study, queries were also longer and more varied more than queries overall. This was partly because many math queries appeared to be questions copy-and-pasted from exercises or homework assignments. 

In another interesting study,  posts to threads in an online math Community Question Answering (CQA) site were studied (MathOverflow\footnote{\url{https://mathoverflow.net}}). The authors identified patterns in the collaborative actions they exhibit (\eg providing information, clarifying a question, revising an answer) and their impact on the final solution quality \citep{overflow}.

Earlier work considered use cases for math-aware search in a study of mathematics graduate students and faculty  \citep{zhao_math_2008}. Surprisingly the participants did not find formula search was useful overall, perhaps because they generally knew the names of entities they wanted to search on.  The study also points out that the type of a source is an important relevance factor (\eg exercises vs. code). Another analysis of expert use cases is also available  \citep{DBLP:conf/mkm/KohlhaseK07}, in which formula search was studied using the MathWebSearch tool.

\section{Retrieval systems}

\reffig{fig:system} provides an overview of retrieval system interactions with people, and the specific sub-tasks from the `jar' framework that they perform. 
Unlike the freely interacting tasks of the  `jar' model, retrieval systems generally perform information tasks in a fixed order, shown by arrows in \reffig{fig:system}. 
The figure has
two main information flows for the collection of sources that a retrieval system uses.\\

\begin{compactenum}
\item \textbf{Index construction (offline).} Information passes from the sources at top and flows to the bottom-right, as sources are annotated with additional information, and then used to compile a searchable index of patterns. The collection index is precomputed before the system is used for retrieval.
\item \textbf{Retrieval (online).} Submitted queries are annotated and then matched against patterns in the index, returning one or more matching sources.  The collection is generally consulted for passages, bibliographic data, and other contents when generating the result returned to the user.
\end{compactenum}

\ourfigure
	{!tp}
	{\imagefillwidth{figures/SystemTasks-mod} 
	\vspace{-0.2in}}
	{Information Tasks in Retrieval Systems (Backend). Arrows show the flow of information. All tasks in \reffig{fig:placeMatTasks} other than \emph{Apply} are shown.}
	{fig:system}

\paragraph{Consulting sources.}
Search engines that match queries to contents in sources are a type of filter. A standard search result is useful precisely because it contains sources with patterns of information shared with the query, omitting all other sources. 

The implementation of \emph{consult}  tasks that access sources is important for both index construction and retrieval, and is another way that sources are filtered in a retrieval system. Source contents shown in search results directly impact our impression of which returned sources are promising.  Source contents used for index construction define the available patterns for matching queries to sources. 

For example, omitting high frequency terms from queries and sources that do not signify a topic  (\ie skipping \emph{stop words} such as `the') can greatly reduce index sizes and increase retrieval speed, but at the risk of performing poorly on queries using these terms; a classic example is the phrase `to be or not to be' from Shakespeare's play Hamlet, which is instantly recognizable but composed entirely of stop words. 

For math-aware search, a similar decision would be omitting tokens and strings representing formulas (\eg in \LaTeX{} source files). Limitations on what can be consulted includes formulas in PDF documents, which are usually not represented explicitly \citep{DBLP:conf/icdar/ShahDZ21}. This and other missing information can be addressed by annotating sources.

\paragraph{Annotation and indexing.}
In addition to selecting source contents in the \emph{consult} step, we will also \emph{annotate} sources with additional information. This extra information can be used to add patterns for matching sources in the index, or to add information to retrieval results. 

For example, some neural net-based techniques such as SPLADE automatically add words that do not appear in a source to the inverted index \citep{splade}.\footnote{This augmentation is also applied to queries. Query annotations are called a \emph{query expansion} when they add tokens or other patterns for matching additional sources in the collection.} These additional terms are synonyms and other words appearing in similar contexts within a training collection. For math, a simple example is adding additional representations for formulas in sources, such as generating Content MathML for operator trees corresponding to formulas represented in \LaTeX{} or Presentation MathML, allowing formulas to be searched using both formula appearance and operation structure.

From the information obtained through \emph{consulting} and \emph{annotating} documents, an index of patterns for matching queries is produced. This can take different forms, but is generally one or a combination of:
\begin{compactenum}
\item \emph{inverted indexes} that map patterns to sources and source locations (\eg tokens or paths in graphs for math formulas), and 
\item \emph{embedding spaces} mapping patterns to points in a vector space, where entities with more similar contexts across a collection are closer (\eg words, sentences, and formulas).
\end{compactenum}
Embedding vectors have their own dictionary mapping vectors to specific sources or source locations (\eg when search is done on text \emph{passages} or individual math formulas). This allows sources matched in vectors to be consulted when communicating results to users. 


\paragraph{Retrieval: Querying sources and communicating system results.} We query a collection using the collection index and an annotated query containing additional terms and/or an embedding vector. 
Search using inverted indexes is referred to as \emph{sparse} retrieval, while search using embedding spaces is referred to as \emph{dense retrieval}, based on the underlying vector representations for each. In particular, term vectors representing the presence of words or formula structures in a document are mostly zeros. In general, sparse retrieval models such as BM25 that use tokens or other source contents directly for lookup are faster \citep{DBLP:journals/ftir/RobertsonZ09}, but dense retrieval models such as ColBERT \citep{10.1145/3397271.3401075} are more effective \citep{macdonald2023, giacalone24}. Some retrieval models use dense models to improve sparse models \eg SPLADE, mentioned earlier.

The improved effectiveness of dense retrieval models is partly from additional context used in defining patterns, \eg using the words referring to and surrounding a formula to represent a formula in a pattern vs. the formula alone. The use of a vector space also provides more holistic and flexible pattern matching, \eg finding source vectors with the most similar angles to a query vector, rather than matching query formula tokens individually to vocabulary entries in an inverted index. These help bridge the \emph{vocabulary problem} discussed in the next subsection.

How the final result of a query is \emph{communicated} (generated) can vary substantially, and often makes use of query and source annotations.
In a traditional search engine, specific sources are matched in the index for the \textit{query} task, with the index comprised of some combination of inverted indexes and embedding spaces. Source contents are then used to generate a result in the \textit{communicate} task, using sources and source locations matched in the index.

However, for a generative question answering or retrieval system, the result of the \emph{query} task may be a single vector capturing the similarity of patterns in the query to patterns within sources of a very large collection produced using a neural network. This vector is then used in the \emph{communicate} task as a starting point for generating the response, for example using a second recurrent neural network trained on the collection, possibly along with additional information from the original collection of sources  (\eg with references to specific sources).  Some are used to generate a list of ranked sources directly \citep{zamani2024}, ultimately producing an \emph{extractive} search result summary based on source contents. 

Other recent systems such as Google's AI search assistant produce \emph{abstractive} summaries of retrieved sources, which summarize matching sources but without limiting the summary to contents found in the matched sources or their annotations.

\paragraph{System design.}
System designers and IR researchers are interested in the efficiency and effectiveness of a retrieval system.  As seen in \reffig{fig:system}, these are observed in live systems through query and user interaction logs. For experiments, system results are computed using  simulated user interactions for a fixed set of queries, and relevance scores for sources, along with a description of the information needs associated with each query. Designers and researchers also make use of additional tools for evaluation, some of which we discuss in Chapter \ref{c-eval}.

\section{User interfaces and system interactions}


User interfaces play a very important role in mathematical information retrieval. 
In addition to executing queries and returning results, how queries are entered, how results are returned, and how other information tasks in \reffig{fig:placeMatTasks} are supported can help speed up or even limit a person's efforts to find and use information. 

We next present a user-centered view of retrieval systems in math information tasks. We then share some key challenges for retrieval system interaction, along with interface designs aiming to address them.

%
%
%


\paragraph{Interfaces-in-the-task-loop.}

\reffig{fig:ui} illustrates a student working to change the base of a logarithm (\ie \refneed{log}) using multiple retrieval systems.
At bottom-left of \reffig{fig:ui} is a jar holding sources the student had on hand when they started searching, new sources they find or create, along with other linked sources, \eg by web link, citation, or mention. 
In addition to these sources, their queries, results from queries, and handwritten notes (\eg from converting bases by hand) are also found in the jar.
Of these available sources, the ones currently being used are at bottom-right of \reffig{fig:ui}.  

\ourfigure
	{!tp}
	{\scalebox{0.935}{\imagefillwidth{figures/UI-user-mod}} }
	{Interacting with Multiple Retrieval Systems (Frontend). Each dotted arrow represents a retrieval system backend (see \reffig{fig:system}). 
	 Sources currently used to address the information need are shown in a separate container at bottom right.}
	{fig:ui}


Some selected sources being worked with partially or fully answer the student's needs, but others do not, such as sources later  deemed not relevant. 
Other selected sources might exercise knowledge such as shown for the \emph{Apply} task in \reffig{fig:ui}, or come from other  tasks such as annotating and indexing sources of interest.
Some selected sources may be added even after finding answers, perhaps because they provide a different perspective, or have a presentation that is easier to understand. 

For more complex tasks, 
we often see the focus of our selected sources drift.
In the \emph{berry picking} model  of retrieval \citep{bates1989design}, people see their queries and information needs change as they search and learn. Particularly for unfamiliar topics, our needs and queries may change dramatically as our understanding does \citep{belkin1980}. For example, this is likely to happen when a person explores unfamiliar concepts associated with  unfamiliar notation.  
In our jar model, information need changes involve changing the jar lid label, perhaps using an orange sticky note placed over the original description.\footnote{sticky notes: a versatile information tool in this chapter and in life.}

What we have in \reffig{fig:ui} is a person generating, selecting. and using sources for needs that may  change as they work. Retrieval systems are a part of this process, but not the focus. 

\paragraph{Interaction challenges.}
All systems embody design decisions and biases. Naturally, no one retrieval system will be ideal for all queries or subjects. However, users often have challenges in search that are more cognitive than system-related.  These are important considerations in creating usable systems, particularly for search interfaces \citep{hearst2009search, white2016interactions, holmes2019usability}. 

Norman identifies two `gulfs' that limit human task performance \citep{norman1988psychology}. Broadly speaking, for retrieval systems the two main categories of interaction challenges are with expressing queries (a \emph{gulf of execution}) and interpreting search results (a \emph{gulf of evaluation}). 
In both cases, for unfamiliar topics, the user may be unable to formulate an effective query or interpret results reliably precisely because of what they don't know, or because their understanding is incorrect (\ie their \emph{Anomalous State of Knowledge} \citep{belkin1980}). 

A common cause of a gulf of execution in query formulation is the \emph{vocabulary problem}, where the terms/patterns a person uses for search differ from those used to index sources. 
For example, in one study undergraduate students were challenged while trying to define the binomial coefficient `${n \choose k}$'  \citep{Wangari_2014}.
Because of this notation-based vocabulary problem, the students' were unable to find a definition using standard text search. When allowed to enter the expression by hand with automatic translation to \LaTeX{}, they found definitions using the same text search engines.

People sometimes also encounter a \emph{gulf of evaluation} when trying to identify relevant information in search results.  Aside from missing relevant items in results due to the vocabulary problem, 
an important factor here is how retrieval results are presented to a user. 
For example, \citep{reichenbach2014rendering} report statistically significant differences in the ability of participants to identify relevant sources in SERPs when excerpts present formulas as raw \LaTeX~vs. rendered formulas.  
Additional gulfs occur when selected excerpts are not relevant for a need, or a person lacks the math background to understand a result.\footnote{impatience, inattention, mental strain, and tiredness are also factors here.}




Learning new terminology and notation while searching allows a user to extend their patterns used to express queries and identify relevant results, bridging these gulfs of execution and evaluation. Some of these new patterns might be recorded explicitly in a source, \eg recording an unfamiliar notation for eigenvectors on a blue sticky note.

\paragraph{Query input: math-aware search bars.}


For the most part, math-aware search bars differ in how they include formulas.  
Perhaps the simplest design is for users to enter \emph{both} text terms and formulas as text. An early example is the Digital Library of Mathematical Functions\footnote{\url{https://dlmf.nist.gov}} which accepts \LaTeX{} commands for formulas along with text terms in queries \citep{miller_technical_nodate}. 
The more recent Approach Zero system\footnote{\url{https://approach0.xyz}} system uses
MathQuill\footnote{\url{http://mathquill.com}} to render \LaTeX{} formulas as they are typed in the search bar, and allows writing lines and argument positions to be reached with arrow keys rather than \LaTeX{} commands (\eg for superscripts and fraction denominators). 

To avoid remembering many names for operations and symbols, or to avoid unfamiliar \LaTeX{} or other syntax for creating formulas, usually a palette of buttons with images for symbols and operations accompanies the search bar.
Buttons add formula elements including operation structures (\eg fractions, integrals, and radicals) and symbols not found on a keyboard (\eg greek letters such as $\zeta$ (\emph{zeta})). 
Query bars with palettes often display formulas in a structured editor like those in document editors  (\eg Word).
Early examples of prototypes with symbol/operation palettes include  MathWebSearch  \citep{kohlhase_mathwebsearch_2013} and MIAS \citep{sojka_mias_2018}. 

As another way to reduce the effort and expertise required for formula entry, some search bars also support \emph{multimodal} formula entry. Multimodal query editors allow formulas to be uploaded from images or entered using handwriting in addition to standard keyboard and mouse-based entry. There are also multimodal tools such as Detexify\footnote{\url{http://detexify.kirelabs.org/classify.html}}, which looks up \LaTeX{} commands for symbols drawn using a tablet or mouse  \citep{detexify}. In addition to search, recognizing math in handwriting and images has been used for   interactive computer algebra systems and other applications, and is an active area of research dating back to the 1960's    \citep{zanibbi_recognition_2012,precsurvey2024}.

\ourfigure
    {!tp}
    {\imagefillwidth{figures/MathDeck}}
    {MathDeck query entry, formula chips, and cards \citep{diaz_mathdeck_2021}. Chips can be dragged,
    edited, and combined. Editing may be done using raw \LaTeX{}, or a combination of operations, chips, handwriting, and \LaTeX{} using the canvas at center.
    Formula cards (bottom left) contain chips, titles and descriptions. New cards can be created by users, and searched by formula \& title (video: \url{https://www.youtube.com/watch?v=XfXQhwIQlbc}). }
    {fig:mathdeck}

An example of a search bar with multimodal formula entry is the MathDeck system\footnote{\url{https://mathdeck.org}} \citep{diaz_mathdeck_2021}. As seen at top-left in \reffig{fig:mathdeck}, a text search box can be used to enter words and \LaTeX{} for formulas. Formulas can also be added from a visual formula editor shown at center-left in \reffig{fig:mathdeck}, and using formula `chips' with embedded \LaTeX{} (\eg blue oval at right of the query text box).
Like MathQuill and structured formula editors, MathDeck renders a formula as it is entered, but with more flexible subexpression selection and entry. MathDeck's query and formula entry interface is designed to:
\begin{compactenum}
\item support text entry; natural for text, and one can easily type `x + 2', or copy-and-paste \LaTeX{} with small changes (\eg $a \rightarrow x$)
\item provide  symbol palettes to help enter symbols and structures 
\item provide handwriting input for those who prefer it, and to avoid searching palettes for symbols \& structures
\item support formula reuse in chips; chips can be used in editing, and can be exported/shared as images with \LaTeX{} metadata  
\item construct formulas interactively using a structured editor, with larger formulas easily built up from smaller pieces.
\end{compactenum}
Other multimodal query entry interfaces have similar design goals, most commonly to support image and keyboard/mouse input.
 
 Other familiar ways to reduce query and formula entry effort are query suggestions and query autocompletion. 
 Their helpfulness is related to the principle of \emph{recognition over recall}: it is usually easier to recognize something we see than recreate the same thing from memory \citep{hearst2009search}. As a simple example, a query autocompletion might include a concept whose name but not formula we can remember, and allow us to quickly select a query containing both.


To illustrate the communication of retrieval results, we'll use a system for visual search that uses an inverted index. 
\reffig{fig:tangentv} shows handwriting in math lecture videos being queried with a \LaTeX{}-generated formula image. 
The inverted index uses pairs of symbols (\eg $(I,n)$, $(=,A)$) as the vocabulary for lookup.  
Before searching, the query image is annotated with a graph containing nodes for symbols and edges with angles between adjacent symbols.

The inverted index is queried by looking up \emph{all} adjacent query symbol pairs, to find their occurrences in the video collection.
Each entry in the \emph{posting list} for a queried symbol pair (\eg $(I,n)$) refers to an edge connecting the same symbols drawn in a video.  
Before indexing, videos are annotated with keyframes of drawn symbols that overlap in time along with an adjacency graph for each keyframe. 
Each edge in a keyframe graph is added to the posting list for its pair of symbols, as a \emph{posting} containing a unique identifier for the edge, its keyframe graph, and video.  
In \reffig{fig:tangentv}, zoomed-in video keyframe graphs are shown in results at left, and keyframe thumbnails are shown at far right.

\paragraph{Query response: communicating results.}
It is actually the drawn symbol keyframes annotated on videos that are searched.
Keyframes are scored by the similarity of matched adjacency subgraphs to the query graph, based on the similarity of matched symbols (nodes) and their angles (edges).
In the results shown in \reffig{fig:tangentv}, symbols matched in a query/keyframe have the same color, and matched graph edges are red.
To avoid missing symbols due to recognition errors, symbol pairs are indexed using all combinations of possible labels for each symbol. 
This is how $n$ matches $M$ in the second match shown.\footnote{A variation of adding tokens to queries and documents to increase possible matches in an inverted index. Symbol similarity is computed from all label probabilities assigned to each symbol. Tangent-v has also been used to search formula collections using unique symbol labels in PDF    \citep{davila_tangent-v_2019}.} 

\ourfigure
	{!t}
	{	\imagefillwidth{figures/Indexing/LOS-Video}
		\vspace{-0.1in}
	}
	{Tangent-V Formula Search Results (left) and Video Player Supporting Navigation (right)\citep{DBLP:conf/icfhr/DavilaZ18}. A rendered \LaTeX{} formula is used to search handwritten math symbols recognized in a video  \citep{DBLP:conf/icdar/DavilaZ17}. Here the user has clicked on the `=' of a matched formula on the whiteboard, and this advances the video to where it is first drawn (video: \url{https://www.youtube.com/watch?v=gn24qo1MLN0}).}
	{fig:tangentv} 

Let's consider the results in \reffig{fig:tangentv} more closely, with associated system tasks in \reffig{fig:system}. 
The query result shows the top-2 matching videos, and not keyframes.
To generate this view, the keyframe ranking from \emph{querying} the index is restructured as a video ranking, with videos ranked by best keyframe match. 
Also, the annotated query and videos have been \emph{consulted} to produce the graph matches shown for each video. 
The videos are consulted again for annotations not used for indexing: 
clicking the mouse on a symbol in a result keyframe makes the video player jump to where the symbol starts to be drawn.

How the search results filter and present the videos is motivated by tasks users carry out (see \reffig{fig:ui}).
For example, having symbols linked to frames can help people \emph{consult} videos by quickly navigating to where a formula is drawn and discussed.
Results rank videos rather than keyframes to make the search results more concise and easier to consult.
The \emph{communication}, \emph{annotation}, \emph{indexing}, and \emph{querying} tasks can also be supported from search results. 
In MathDeck  formulas in results can be used directly for search, selected for editing or export, or annotated in a card with a title and description.
Cards are also automatically indexed in a `deck' searchable by formula or title. 

Showing matching graphs in results is more helpful for designers than users; simple bolding or highlighting is more common.
In contrast, MathDeck's search results highlight matched query words and formulas located in PDF documents (\eg for papers from the ACL Anthology \citep{DBLP:conf/sigir/AmadorLDSZ23}).\footnote{ACL Anthology search demonstration: \url{https://drive.google.com/file/d/1fbiMyHtlfEYUJvmrbZsWzhfL0X_zo9-t/view}}

We've used just two systems here to illustrate search results that rank sources, and how they interact with human information tasks. However, results from other systems have different types. Some systems plot, simplify, and/or perform requested operations on formulas, or provide solutions for math problems posed in text and/or formulas directly (\eg using Wolfram Alpha or a math-aware chatbot). In these cases the response is an answer to a (possibly inferred) question, rather than a ranked list of sources. 
These are designated as \emph{question answering} systems, and interactions with chatbots addressing math queries are a type of \emph{conversational search} where clarifying questions and additional information may be provided or received in multiple rounds of query/result interactions.

Chatbots using Large Language Models (\emph{LLMs}) have proven intriguing and useful in some instances, but there is an increasing awareness of issues related to the validity of responses and other substantive concerns \citep{DBLP:conf/fat/BenderGMS21}. However, any retrieval result is only an information source -- understanding and verifying any source  requires additional work. Related to this, in Community Question Answering platforms (CQAs), many posts request clarification of a question, or clarify/correct posted answers and comments.\footnote{Including the classic, `my correction to @your correction.'}
This illustrates how human responses to math queries also often contain misunderstandings, ambiguities and errors.

Regardless of the result type, how information is chosen and presented in results is important. It has a real impact on the usefulness of the result as a source of information, and on how tasks other than  consulting the result itself are supported. In many cases usability testing can be used to check the effectiveness of result presentations and other interface design elements, and to discover refinements and alternatives.\footnote{\eg parts of the MathDeck were usability tested \citep{dmello_representing_nodate, nishizawa, diazthesis}, which led to substantial improvements.}

\paragraph{Supporting tasks for individual sources.} 
Programs used to view/consult sources can also help with the user tasks illustrated in \reffig{fig:ui}. 
\ourfigure
    {!tp}
    {\imagefillwidth{figures/ScholarPhi}}
    {ScholarPhi system showing definition for math symbols found within the same PDF paper \citep{head_augmenting_2021}. To assist skimming for details, text other than for definitions of a selected formula is greyed out (video: \url{https://www.youtube.com/watch?v=yYcQf-Yq8B0}).}
    {fig:scholarphi}
A nice example is the ScholarPhi system shown in \reffig{fig:scholarphi} \citep{head_augmenting_2021}. 
Reading formulas can be challenging, as symbols may be defined throughout a paper. ScholarPhi provides annotations decorating a selected formula/subexpression, providing symbol definitions in-place. Definitions are linked to where they appear, and text not associated with a selection is greyed out. 

To produce the definition views in ScholarPhi, sources need to be annotated with formulas, symbols, and definition locations, and then definitions need to be linked with associated entities where they appear in the paper (\ie symbols or subexpressions). Definition segmentation and linking entities are performed with natural language processing techniques. The original prototype identified math symbols with \LaTeX{} source files used to generate PDFs, simplifying formula detection.

A second example is the keyframe list at right in \reffig{fig:tangentv}. When viewing a video, all keyframes for handwritten content are available in a thumbnail list. Keyframes can be selected, and individual symbols clicked on to jump to where it is first drawn in the video (similar to the search results).  This requires annotating video sources with generated keyframes produced using computer vision techniques.

Both ScholarPhi and Tangent-v require generating additional information using automated inference (\ie AI), and their usefulness is limited by the accuracy and scalability of the methods employed. However, we believe that this is an important future direction for mathematical information retrieval, because the content and organization of mathematical sources can be complex. Particularly for non-expert users, mature versions of these techniques may be very helpful.


A number of well-known formats were devised or augmented to support detailed annotation with links and tags, including TIFF, PDF, and XML. Unfortunately detailed `semantic' annotation has proven difficult at scale despite significant efforts. Some possible reasons include the time required to create sources before annotations, the diversity of information needs (\eg which information do we annotate?), attaching large annotations makes files large and unwieldy, and overall progress in scalable AI has been slower than many anticipated.

As AI continues to improve, creating source annotations to support examining and navigating math sources and other user information tasks within UIs seems likely to be beneficial. Perhaps application-specific annotations such as used in ScholarPhi, Tangent-v, and MathDeck are a good starting point.

\chapter{Annotating and Indexing Sources}

\label{c-sources}



In this chapter we focus on indexing formulas and text in sources,
as illustrated in  \reffig{fig:system}. This requires consulting sources, annotating formulas and text with additional information, and then constructing `sparse' inverted indexes for looking up discrete patterns directly (\eg words or paths in formula trees), and/or `dense' indexes representing the same patterns in embedding vectors, which are searched based on the similarity of an embedded query pattern with items in the index (\eg using vector angles). With this data, a variety of sparse and dense retrieval frameworks can be used for a variety of search, question answering, machine learning, and evaluation tasks.

We will consider each of the three main tasks needed for indexing in turn, starting with the types of information and representations used in mathematical information sources.

\ourfigure
	{t}
	{\scalebox{0.50}{ \includegraphics{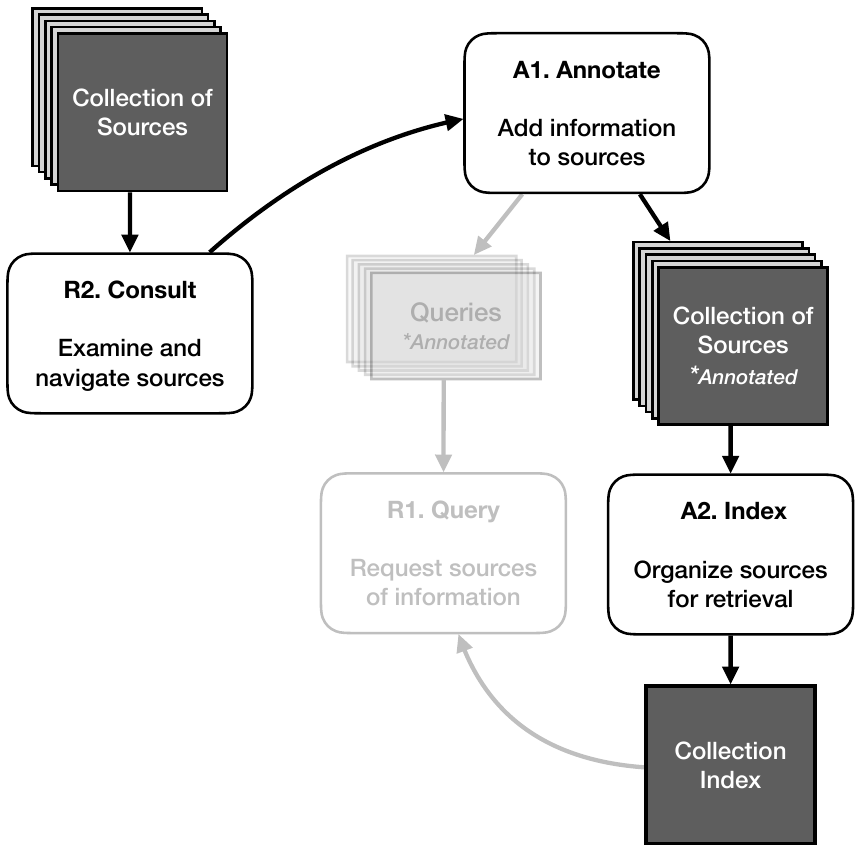}}
	}
	{Tasks from \reffig{fig:system} for Representing Formulas and Text in an Index. 
	For context, query annotation and querying the
	index are shown greyed-out. Content in sources is consulted, annotated to add information for text and formulas,
	and then indexed for later use in retrieval systems and system evaluation.}
	{fig:ch2overview}


\section{Consulting sources for mathematical information}


Let's first consider how we consult sources for mathematical information using some examples.
The examples come from technical documents and search queries, but much of what will be said applies equally well to videos, audio, conversations, text messages, and other source types.


We will start with a definition of the
 \emph{Inverse Document Frequency (IDF)} shown in \refexample{idfeg}. IDF is used in a number of influential sparse retrieval models including variants of TF-IDF (\emph{Term Frequency}-{Inverse Document Frequency}) and BM25 \citep{DBLP:conf/sigir/RobertsonW94,DBLP:journals/ftir/RobertsonZ09}.
Its utility comes from a simple but profound insight: 
query terms appearing in fewer documents are rarer and thus \emph{more specific}, and so should be given higher weight when ranking documents using matched query terms \citep{SprckJones2021ASI}.
 
For example, the term `BM25' is predominantly found in sources on information retrieval, while the term `weight' is used for many topics and in multiple senses, including the heaviness of an object and scaling numeric values.  
When scoring documents against the query `BM25 weight',  matches for `BM25' will have higher IDF scores than `weight,' reflecting the narrower usage of `BM25.'

\begin{myexample}{Inverse Document Frequency (\textit{IDF})  }{idfeg}
\textbf{Excerpt from \citep{Robertson:2004aa}}
\hrule
~\\
...Assume there are $N$ documents in the collection, and that term $t_i$ occurs in $n_i$ of them \ldots 
the measure proposed by Sparck Jones, as a weight to be applied to term $t_i$, is essentially

\begin{equation} 
\label{eq:idf}
\tag{1}
idf(t_i) = \log \frac{N}{n_i}
\end{equation}

\textbf{Variable and function definitions }
\vspace{0.025in}   
\hrule
\begin{center}
\includegraphics{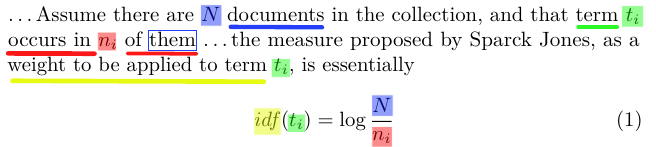}
\end{center}

\end{myexample}

Despite its brevity, the excerpt in \refexample{idfeg} contains a fair amount of directly represented and implied information. 
Recovering some of this information requires pattern matching and inference, \ie applying information known, found in the passage, or found in other sources.
To illustrate this, we annotated the definition at bottom of \refexample{idfeg}.
Colored highlighting is used to associate variables and function names with their definitions in the text.
`them' is placed in a box with a thin blue outline to represent the \emph{anaphoric} (backward) reference from `them' to `$N$ documents.' 
Knowing that `them' refers to documents, we can infer that $n_i$ is the number of documents containing term $t_i$.

In this way, gathering information from a source involves a combination of \emph{consulting} the source to 
identify stated information, and \emph{analyzing} the source to reveal additional information from explicit and implied linguistic patterns and relationships.
Both activities are informed by available knowledge, \ie readily available and actionable information that we have previously seen or inferred,
or find by \emph{navigating} to other sources or \emph{querying} for new sources (\eg following hyperlinks, or asking a person).

\begin{myexample}{Information extracted from definition in \refexample{idfeg} }{eg:info}
\forceline
{\bf \noindent Variables:} placeholders for a set of values, similar to common nouns~\\ \vspace{-1em}
\begin{compactitem}
\item The text identifies $N$ as the number of documents in a collection. $N$ is like a common noun, because the 
\underline{collection is not specified.} 

\item The text defines $t_i$ as \underline{any term} appearing $n_i$ times in a collection, with shared identifier $i$, \eg $(t_3, n_3)$ could be 
{\tt (`weight', 11)}. 
\end{compactitem}
~\\
{\bf \noindent Functions \& Operators:} create new from given values, like verbs~\\ \vspace{-1em}
\begin{compactitem}
\item 
$\log$: log function with \underline{unspecified base.} 
\item $idf(t_i)$: aside from the unspecified log base, a concrete function in Equation (1). The text says this gives a weight for term $t_i$. 
\item Division ($\frac{~~\cdot~~}{~~\cdot~~}$), application ($idf(\cdot)$, $\log \cdot$), and equivalence ($=$) appearing in Equation (1) that  are \underline{not defined in the excerpt.}
\end{compactitem}
~\\
{\bf \noindent Additional context:}~\\ \vspace{-1em}
\begin{compactitem}
\item  The text indicates Sp\"arck Jones introduced the $idf$ formula in a different, \underline{unspecified form} \citep{SprckJones2021ASI}.
\end{compactitem}
\end{myexample}






Following this process, we gathered the information shown in \refexample{eg:info}.
Note that the underlined missing details are \emph{deliberate} omissions:
\begin{compactitem}
\item By not specifying a collection, $N$ is defined for any collection. 
\item $t_i$ and $n_i$ give any term and a count for documents containing it.
\item Omitting the logarithm base emphasizes that logarithms increase with input size (\ie they are \emph{monotonic}) and so any base suffices. 
\item Function application and the operators used are common, and their definitions are assumed to save space and reader effort.
\end{compactitem}
These omissions are helpful, provided the reader can infer the missing details they need:
the definition would be longer and harder to read otherwise.
More generally, what an author chooses to omit 
is informed by (1) the context and focus of discussion, \ie items discussed earlier and the current topic, and (2) the assumed background knowledge of the audience.
These determine what can and \emph{should} be left out.

Note also that Equation (1) on its own conveys only partial information.
A formula presents a hierarchy of operations, but its symbol definitions and its \emph{purpose} generally come from surrounding text,
other formulas, and assumed knowledge.
In this example, the text defines all variables and the term weighting role of the $idf$ formula, but not the operations shown in the formula.


Let's next consider an alternative definition for the $idf$ function from Equation (1) that uses no mathematical symbols:
\begin{quote}
A term's \emph{inverse document frequency} is the percentage of documents containing the term, inverted and then log-scaled.
\end{quote}
This seems simple enough -- we invert the numerator and denominator in the percentage of documents containing the term (\ie `flip' the fraction) and convert this value to a logarithm.\footnote{The log scale reduces the rate at which $idf$ increases, which avoids having rare terms completely dominate rank scores.}  But we lose some useful patterns and information when we do not use math notation:
\forceline
\begin{compactdesc}
\item[\it Visibility:] Formulas are italicized and use distinct symbols, making them easier to find in sentences (\emph{inline}). Also, they may be offset from the main text and indented (\emph{displayed} like Equation (1)).
\item[\it Compact Reference:] Referring to symbol names is more efficient than reusing descriptions, \eg $N$ vs. `the number of documents.'
\item[\it Compact Structure:] Text describes relationships, while formulas \emph{show} relationships spatially, \eg Equation (1) vs. the textual definition above.  
A good example is the \emph{distributive property} from algebra, which is easily expressed using: $x(y+z) = xy + xz$.   
\item[\it Abstraction:]  
Formulas often define properties and patterns applicable in multiple contexts.
For example $idf$ can be applied to formulas if $t_i$ is redefined to refer to a unique formula, and $n_i$ the number of documents where the formula appears. We can also redefine $N_0$ and $\lambda$ in the decay function $N(t) = N_0 e^{-\lambda t}$ to estimate decreases in (1) a retirement fund balance, (2) the rate of a chemical reaction, or (3) the potential contained in a capacitor.
\end{compactdesc}
\forceline
Judicious use of formulas in technical writing makes mathematical information easier to find, analyze, and reuse/adapt. Which things are beneficial to formalize in notation again depends upon a document's focus and intended audience.

\paragraph{Another $idf$ definition and visual formula representations.}

The following is an alternative definition for IDF taken from the Wikipedia page for the \emph{tf-idf} retrieval model.

\begin{quote}
\it
The {\bf inverse document frequency} is a measure of how much information the word provides, i.e., 
how common or rate it is across all documents. It is the \underline{logarithmically scaled} inverse fraction
of the documents that contain the word (obtained by dividing the total number of documents by the number of 
documents containing the term, and then taking the logarithm of that quotient):

$$idf(t,D) = \log \frac{N}{ | \{d : d \in D ~\textrm{and}~t \in d \} |}$$
\end{quote}
This definition for the $idf$ formula is essentially equivalent to that in \refexample{idfeg}. 
However, the formula used to represent $idf$ has changed.
The document collection is given explicitly as a parameter $D$, and
the number of documents containing a term, previously $n_i$, is now the set size for  documents containing the term.
The term itself is noted as simply $t$, and not $t_i$.
Also, the wiki page includes a link to the \emph{Logarithmic scale} article that we can follow to review that concept. 

\begin{myexample}{Presentation MathML from Wikipedia \emph{tf-idf} article}{eg:wiki}

\begin{flushleft}

\begin{verbatim}
<math xmlns="http://www.w3.org/1998/Math/MathML"  
    alttext="{\displaystyle \mathrm {idf}...">
  <semantics>
    <mrow class="MJX-TeXAtom-ORD">
      <mstyle displaystyle="true" scriptlevel="0">
        <mrow class="MJX-TeXAtom-ORD">
          <mi>i</mi>
          <mi>d</mi>
          <mi>f</mi>
        </mrow>
        <mo>(</mo>
        <mi>t</mi>
        <mo>,</mo>
        <mi>D</mi>
        <mo>)</mo>
    ...
    </mrow>
    <annotation encoding="application/x-tex">
      {\displaystyle \mathrm {idf} (t,D)=
        \log {\frac {N}{|\{d:d\in D{\text{and}}t\in d\}|}}}
    </annotation>
  </semantics>
</math>
\end{verbatim}

~\\
 Source: \url{https://en.wikipedia.org/wiki/Tf-idf?oldid=1236851603}
 \end{flushleft}
\end{myexample}

\refexample{eg:wiki} provides an excerpt for the embedded formula in the \emph{tf-idf} article HTML page.
The formula appearance is represented using Presentation MathML, which is an XML encoding for the placement of symbols on writing lines along with their types (\eg {\tt <mrow>} for writing lines and token groupings of characters ($idf$), operators in {\tt <mo>}, variable and function identifiers in {\tt <mi>}).
The \LaTeX{} string used to generate the MathML using MathJax\footnote{\url{https://www.mathjax.org}} is included in the outermost {\tt <math>}
tag, and in the {\tt <annotation>} tag near the bottom of of the excerpt.\footnote{ The {\tt <semantics>} tag can be misleading: math operations are not represented.}
Many web browsers can render Presentation MathML directly, or can use javascript-based \LaTeX{} rendering tools, \eg applying MathJax to the {\tt alttext} attribute of the {\tt <math>} tag.

The vast majority of documents represent formulas by their appearance,
whether as raster images (\eg pixel-based PNGs), vector images (\eg SVGs with drawing instructions) images, 
\LaTeX{}, or Presentation MathML.
This is because it is easier to create the appearance of a formula than to formally define and represent its operations consistently and correctly.
We instead use a formula's appearance to \emph{suggest} operations and leave it to the reader to go through the process of consulting and analyzing the formula and its context to infer its meaning.

This preference for visual representations is equally true for text.
We write prose using a sequence of characters -- we almost never provide a fully annotated parse tree or other semantic representation for passages (\eg using first-order logic). 

In general, for human readers semantic annotations have complex structure and are often verbose.
For example, imagine reading an enhanced version of our list-based summary of the information in \refexample{idfeg} above, rather than the original passage.
How to select and \emph{ground} primitives and relationships in semantic representations is a slippery question, and using visual representations avoids this for authors.


\paragraph{Semantic representations for text and formulas.}
Despite their verbosity and complexity, semantic representations are useful when they capture information reliably enough for a task of interest (\eg search or question answering).
Creating semantic representations manually is difficult,  however we can use automated tools for this purpose.
\refexample{eg:amr}(c) illustrates such a representation for the query:
\begin{quote}
\emph{Find $x^n + y^n + z^n$ general solution}
\end{quote}
\begin{myexample}{Augmenting AMR Trees with Operator Trees}{eg:amr}

Abstract Meaning Representation (AMR) tree, with inserted operator tree for formula in
the query ``Find $x^n + y^n + z^n$ general solution."  
~\\

\centering 

\scalebox{1.05}
{
\hspace{-0.2in}
\includegraphics{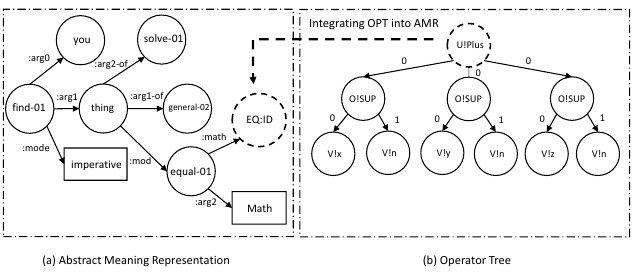}
}~\\~\\

\scalebox{1.375}
{
\includegraphics{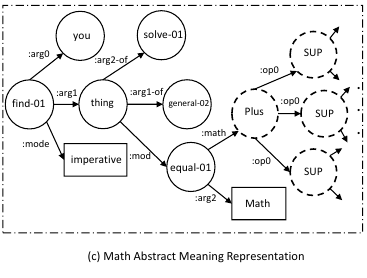}
}

\end{myexample}
As seen in \refexample{eg:amr}(a), an Abstract Meaning Representation (AMR) graph represents text semantics using a hierarchy of subjects, objects, actions, and attributes \citep{DBLP:conf/acl/LangkildeK98}.
\refexample{eg:amr}(b) shows an \emph{Operator Tree} (OPT) giving the hierarchy of operations in the query formula (\ie adding the exponentiated variables).
Both have node and edge annotations capturing types and argument orderings. 

Interestingly, the AMR graph and OPT have similar purposes and structure.
Both capture a hierarchy of events in a sentence or formula.
For example, the root of the AMR tree is a verb ({\tt find}) with a mode modifier ({\tt imperative}) used to indicate that the statement is a command rather than a request.
The verb has two arguments for who receives the command ({\tt you}) and what is requested: a {\tt thing} that is the general solution to the provided equation.
The root of the OPT is an unordered addition operator ({\tt U!Plus}) applied to subexpressions from exponentiation operators with ordered arguments ({\tt O!SUP}).

To produce the combined MathAMR tree in \refexample{eg:amr}(c), an AMR tree was produced using a neural network after replacing the formula with identifier {\tt EQ:ID} (see \refexample{eg:amr}(a)).
In \refexample{eg:amr}(c) the identifier node is replaced by the formula OPT with annotation changes to match the AMR syntax.
MathAMR was used to re-rank answers to Math Stack Exchange questions   
 that had been converted to this representation  \citep{amr2022}. 
MathAMR inserts formulas at leaf nodes, but for longer passages one can imagine adding additional information such as links between variables and their definitions.

\paragraph{Consulting sources in MathIR systems.}
As shown in \reffig{fig:ch2overview}, mathematical information retrieval systems require consulting sources for directly observable information, analyze and annotate their contents to generate additional information, and then organizing this information in an index for lookup, search, generating training data, and use in evaluation.

Unlike people, where we rapidly alternate between consulting and analyzing/annotating sources when trying to recover information, for large-scale systems we break processing up into steps to support batch processing. 
The design decisions made for each step are critically important.  
For example, word/sub-word and symbol vocabularies chosen to represent source contents impact the reliability of visual or semantic annotations and retrieval. 
We also cannot lookup or search using anything that we omit from these vocabularies, \eg if we remove frequent words and symbols such as `the' and $x$ to save space.

In the next section, we focus on annotating formulas and associated text with visual and semantic representations to enrich collections, and for later use in indexing.\footnote{
Diagrams and other graphics are frequently used in math, but outside of our discussion here;
\eg  commutative diagrams can be expressed as a matrix-like SLT container, but are really directed graphs with nodes/edges labeled by formulas: 
\[
 \begin{tikzcd}[ampersand replacement=\&]
    A \arrow{r}{f} \arrow[swap]{dr}{g\circ f} \& B \arrow{d}{g} \\
     \& C
  \end{tikzcd}
  \]
  }

\section{Annotating formulas: Representations and canonicalization}

In terms of visual structure, the excerpt in \refexample{idfeg}  can be represented by a sequential graph of word and formula tokens
shown in \refexample{eg:formulas}.
Each of the blue formula nodes/tokens contain a visual structure representable using a Symbol Layout Tree (SLT) as shown for the $idf$ formula in Equation (1).
The blue nodes for variable names in excerpt text each contain an SLT that can be found in the subtrees of the $idf$ SLT.

SLTs represent the placement of symbols on writing lines using spatial relationships between symbols and nested writing lines.
An SLT is a directed, rooted tree with a parent and child in every relationship -- the $idf$ node is the root node in our example.
The \emph{idf} and \emph{log} functions are single nodes with their characters grouped into tokens as
seen earlier in \refexample{eg:wiki}.
For easier reading, we have shown adjacent symbols on a writing line using undirected edges.
The eight spatial relationships used in STLs are: adjacent-at-right, sub/superscript, prefix sub/superscripts on the left side (\eg $^n_2C$ and $^{235}U$), inside (\eg $\sqrt{x}$), and above/below (\eg $\frac{N}{N_i}$). 


\begin{myexample}{Text tokens, formula tokens, and visual formula structure}{eg:formulas}
\textbf{Sequence of text (black) and formula (blue) tokens}
\hrule
~\\

\imagefillwidth{figures/TextAndMathRepresentation/TopLevelIDF}

\textbf{Symbol Layout Tree (SLT) for Equation (1)}
\hrule
\begin{center}
	\scalebox{0.425}{\includegraphics{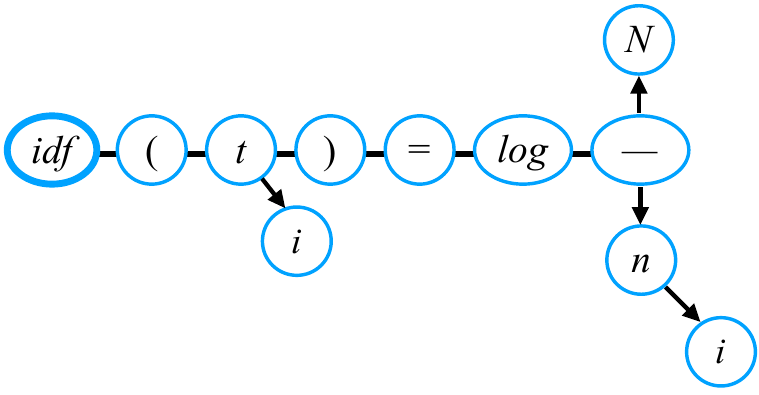}}
\end{center}
\end{myexample}

\begin{myexample}{Translating a Symbol Layout Tree to an Operator Tree}{eg:idfsltopt}

\begin{center}
	\scalebox{0.425}{\includegraphics{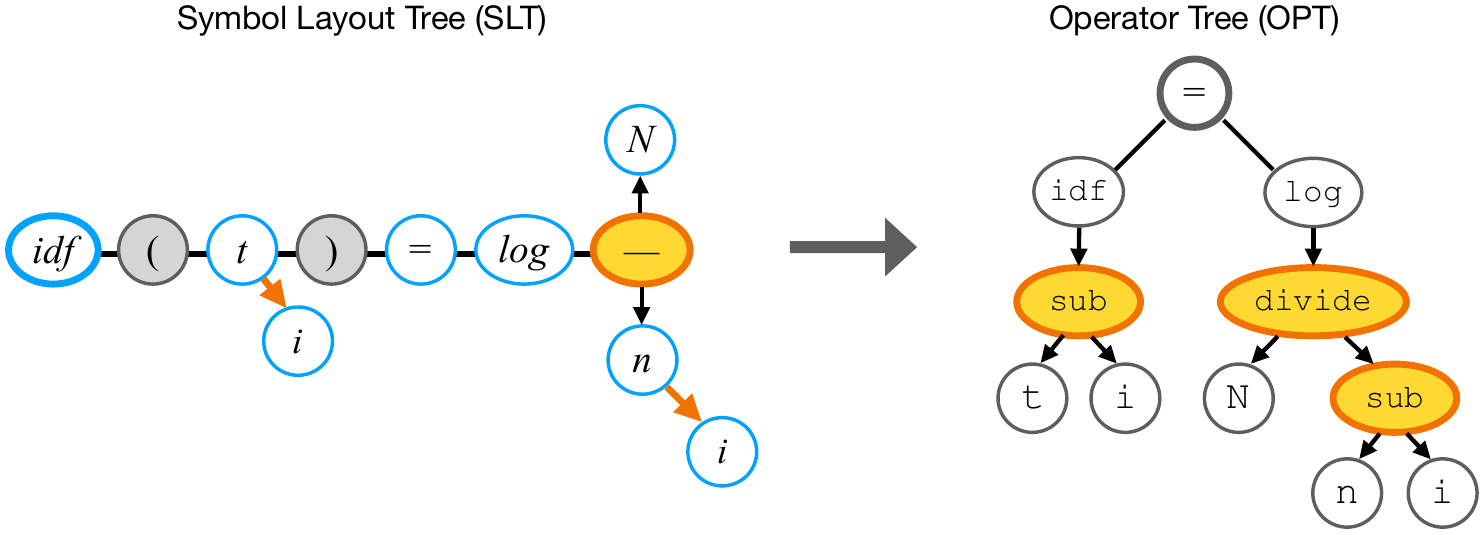}}
\end{center}

Grey nodes in the SLT indicate parentheses removed in the OPT, where they are redundant.
SLT subscript edges and fraction line are replaced by {\tt sub} and {\tt divide} nodes in the OPT.
\end{myexample}

In \refexample{idfeg}, one piece of missing information that we did not annotate is the hierarchy of operations in the $idf$ formula. 
While a reader is unlikely to think about this consciously, interpreting the formula essentially involves converting the SLT to an Operator Tree (OPT).
We saw previously that OPTs are (partial) semantic representation giving an operation hierarchy. 
\refexample{eg:idfsltopt} illustrates this conversion for the $idf$ formula.
In OPTs variables and other arguments appear at the leaves, with operations above the leaves in internal nodes. 
While SLTs are oriented left-right to reflect reading order, OPTs are oriented vertically to reflect operation order.
The order of operations is bottom-up in the OPT with precedence decreasing away from the leaves, \eg the `=' used to represent a definition is applied last for our $idf$ formula. 

Operations appear directly above their arguments in an operator tree.
If arguments have different roles (\eg $N$ and $n_i$ in $\frac{N}{n_i}$) this is captured by their left-right order below the operator.
When argument order is unimportant, arguments are normally arranged in reading order (\eg per the SLT). 
In our OPT directed edges indicate ordered arguments, and undirected edges indicate unordered arguments for `=.'

There are some subtleties with mapping formula appearance in an SLT to the operations in an OPT. 
For example, we use a {\tt sub} operator to represent subscripted variable names.
If we assume the intended semantic, these could be replaced by exponent nodes.
However, this is not true general, and so we map to a {\tt sub} operator instead (\eg $X^2$ may be a Cartesian set product).
Also, some operations without symbols in SLTs are explicit in OPTs, \eg $xy$ represents the operation $x \times y$.

In general, we work with fixed operation sets and SLT mappings when producing OPTs automatically using tools.
In some cases this leads to ambiguities or incorrect mappings.
This is unfortunately unavoidable, because the meanings of symbols are community and context-dependent, and symbols are frequently redefined by authors for their own purposes.
Despite these challenges with conversion,
we should note that some state-of-the-art formula search engines use OPTs rather than SLTs for search,
because SLTs do not fully capture the operation hierarchy.

There has been recent progress in this space. For example,
In one approach to translating SLT representations as shown earlier for Wikipedia \citep{DBLP:journals/pami/GreinerPetterSBSAG23}, noun phrases associated with SLTs are identified and then used in translating the SLTs to OPTs with additional information that allows computable functions to be recovered. An augmented \LaTeX{} syntax represents this annotated OPT, which is then translated to computable representations for Computer Algebra Systems (Maple and Mathematica).
Such semantically enriched formula encodings for use in CAS and theorem provers have long been a long-term goal in the Mathematical Knowledge Management community (MKM), and we anticipate that there will be strong renewed interest in this problem moving forward.

\paragraph{SLTs and OPTs in code.}
 \refexample{eg:code} provides code representing the $idf$ formula SLT in \LaTeX{}, and two versions of the OPT in Python.
The \LaTeX{} is shorter because it represents only formula appearance.
Commands such as {\tt $\backslash$frac} helpfully suggest operations, but define only the placement of symbols (\eg above and below a fraction line). 
We also see the \LaTeX{} subscript operator in this example ({\tt \_}).

For the OPT implementations, all implicitly defined or unspecified operators and values from \refexample{idfeg} must be provided 
to compute values.
For example, \eg {\tt math.log()} uses the base $e$, and the equals operator becomes Python's end-of-function-signature symbol ({\tt :}). 
We also add variables and data structures to hold input values for terms, term counts, and the number of documents in our collection.
The function signatures require additional arguments missing in the left-hand side of our OPT, because all values must be defined
(\eg for {\tt N} and term counts {\tt n}).

\begin{myexample}{$idf$ formula in \LaTeX{} and Python code}{eg:code}

\textbf{\LaTeX{}: Symbol Layout Tree representation}
\hrule
\begin{verbatim}
idf(t_i) = \log \frac{ N }{ n_i }
\end{verbatim}

\textbf{Python: Two Operator Tree representations}
\hrule
\begin{verbatim}
import math
t_all = [ "inverse", "document", "frequency" ]
n_all = [ 2, 100, 20 ]
D = 100
def idf(i, t, n, N):
   idf_weight = math.log( N / n[i] )
   return( t[i], idf_weight )

% Prefix form: ops before args to match OPT RHS
def divide(a, b): return a / b
def sub(a, b): return a[b]
def idf_prefix(i, t, n, N):
    idf_weight = math.log( divide( N, sub(n, i)) )
    return( sub(t, i), idf_weight )

for i in range(len(t_all)): 
   print(i, idf(i, t_all, n_all, D))

OUTPUT:  0 ('inverse', 3.912023005428146)
         1 ('document', 0.0)
         2 ('frequency', 1.6094379124341003)
\end{verbatim}
\end{myexample}

The function is defined twice, first using built-in infix operators for division ({\tt /}) and lookup (\eg {\tt n[i]}) and a second time using functions ({\tt idf\_prefix}).
While both produce the same output, note that the first version looks closer to the SLT, while the second matches the operation hierarchy in the right-hand side of the OPT.\footnote{Infix operators provide argument layouts more  similar to typeset formulas. This may partly  explain the popularity of languages with infix math operators vs. purely prefix-structured operations (\eg in {\tt idf\_prefix()} and Lisp).}  
idf values are assigned to the intermediate variable {\tt idf\_weight}, and annotated with their associated value of {\tt i} and text term in the output.

In the output {\tt "document"} has an idf value of $\log 100/100 = 0$.
This is valid, as the term appears in \emph{all} documents, and so doesn't provide any distinguishing information.
{\tt "frequency"} appears in 20/100 of the documents, and has an idf less than half the value for {\tt "inverse"},
which appears in only 2/100 documents. 
Note that without the log scaling, {\tt "inverse"} would have ten times the
idf score of {\tt "frequency"}.

\paragraph{MathML: Presentation (SLT) and Content (OPT).}
For MathIR systems and evaluation benchmarks, MathML is a file format commonly used to represent SLTs and OPTs.\footnote{\url{https://www.w3.org/Math}}
In MathML, OPTs are normally defined without the additional context needed to compute values that we saw for the 
Python in \refexample{eg:code}.\footnote{Some tools such as Maple and Mathematica provide MathML annotations and need to compute values from Content MathML.}
SLTs are given in \emph{Presentation MathML}, and OPTs in \emph{Content MathML}. 
Presentation and MathML generated from the \LaTeX{} for our $idf$ formula are seen in \refexample{eg:mml}

\begin{myexample}{MathML generated from \LaTeX{} using \LaTeX ML}{eg:mml}
	$idf$ is undefined in \LaTeX{}ML\footnote{\url{https://math.nist.gov/~BMiller/LaTeXML}} and so $i$, $d$, and $f$ are treated as variables.\\ 

	\imagefillwidth{figures/MathRepresentation/MML-top}
	\forceline
	\imagefillwidth{figures/MathRepresentation/MML-bottom}
\vspace{-0.15in}
\end{myexample}

Being XML-based, the syntax is similar to the prefix representation seen for {\tt idf\_prefix} in \refexample{eg:code}. 
Generally speaking, MathML commands begin with a start and matching end tag for the command, with a list of tags for arguments nested inside.
All tags may also contain a list of attributes, \eg {\tt xmlns} (XML namespace) or {\tt stretchy} (controlling the rendering of brackets). 
An example is the {\tt <apply>} command in Content MathML, where the first nested tag is an operator, and the remaining nested tags are the operator's
arguments.
MathML provides types for arguments, including {\tt <mi>} and {\tt <ci>} for variable identifiers, {\tt <mn>} and {\tt <cn>} for numbers.
In Content MathML, defined operations have predefined tags, and so {\tt log} appears as {\tt <log/>} in Content MathML but  as the identifier {\tt <mi>log</mi>} in Presentation MathML.

%
%
%
%
%
%
%
%
%
%
%

The \LaTeX ML tool used to produce \refexample{eg:mml} knows {\tt $\backslash$log} is an operator, and inserts an \emph{invisible} node in the Presentation MathML to capture its application to the fraction using hexadecimal Unicode value {\tt x2061}. 
This symbol does not appear when this formula is rendered.
In contrast,
$idf$ is not a defined operation or function, 
and is broken up into three variables in the SLT
and Content MathML.
In the Content MathML, these variables are multiplied with each other and $t_i$.

As discussed earlier, a fixed set of definitions muse be used to convert formulas in \LaTeX{} or Presentation MathML to Content MathML.
This means that in large collections inconsistencies such as those seen in \refexample{eg:code} are common along with the 
{\tt <cerror>} tag for unrecognized symbols and structures.
If these interpretations not intended by their authors are consistent in their representation, 
they still provide patterns useful for retrieval and other
information tasks.

\paragraph{Visual and spatial region-based formula representations.}
\label{sec:other-reps}

Let us now consider some other visual approaches to representing formulas. 
%
%
For raster (pixel-based) images symbol locations are unknown.
However, we can represent formulas directly as images, and compare formulas based on image similarity (we will return to this in the next section).

We can also capture symbol layout in raster images using an \emph{XY-cut tree}, as shown in \refexample{eg:appear}.
XY-cut trees partition touching pixel groups (connected components) by cutting at horizontal and vertical whitespace gaps.
The standard method strictly alternates the cutting direction, while the recursive version cuts the largest gap in either direction \citep{xycut-recursive,xycut}. 
Symbols can be recognized or features computed from sub-images at nodes for use in recognition and retrieval applications \citep{DBLP:conf/das/BakerSS10,DBLP:conf/icdar/ZanibbiY11}.
A variant of XY-trees was used in one of the earliest systems for parsing math formulas from images \citep{okamoto}.\footnote{Cutting thresholds and rules avoid splitting symbols with multiple components (\eg `i') and remove subexpressions from inside radicals (\eg $\sqrt{x}$).}
XY-trees can also be produced from known symbols, \eg by cutting around symbol bounding boxes.
They are also used to segment document pages into regions, which was the original purpose. 

\begin{myexample}{Region-based spatial representations for formulas}{eg:appear}

{\bf XY-Cut Trees (left: Recursive, right: Standard)} 
\hrule
\vspace{-0.1in}
\begin{center}
\scalebox{0.96}{\includegraphics{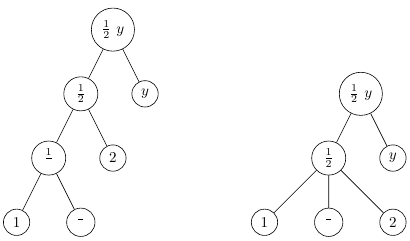}}
\end{center}

{\bf Pyramidal Histogram of Characters (XY-PHOC) }
\hrule
\begin{center}
\scalebox{0.125}{\includegraphics{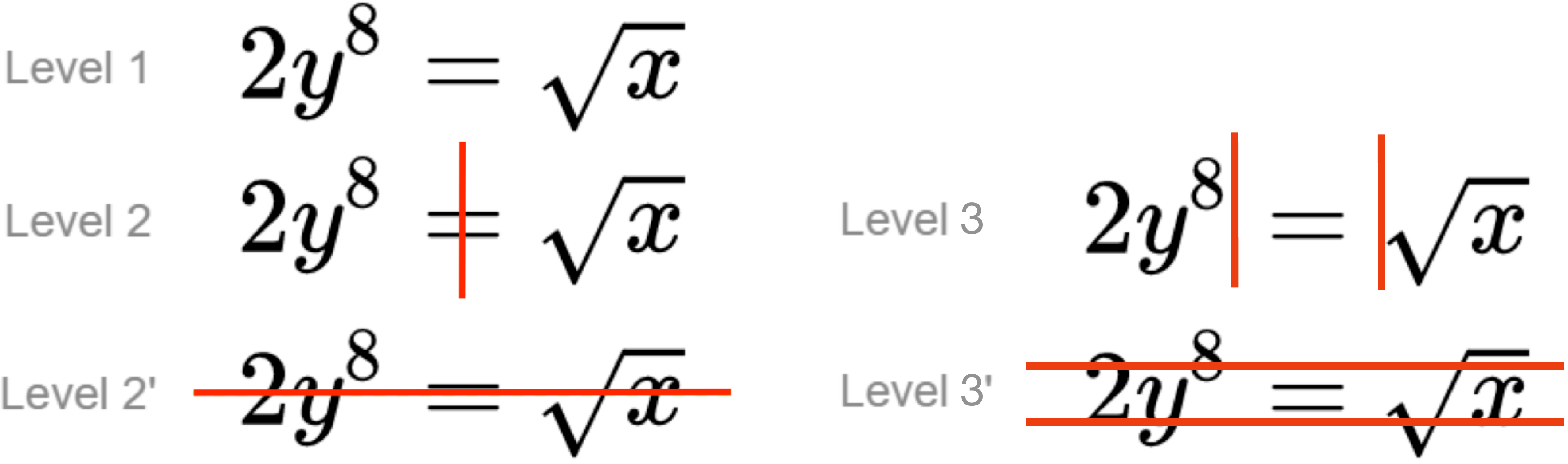}}

\end{center}
\end{myexample}

For images where symbol locations are known such as in SVG or PDF, or using symbol locations from OCR results,
we can produce additional spatial representations.
For example, in the previous chapter,
we saw an example of this where line-of-sight graphs over symbols were used to search handwritten and typeset math in video keyframes.
An alternative region-based spatial representation is the Pyramidal Histogram of Characters (PHOC), which identifies which symbols appear in a fixed set of recursively partitioned regions. 
PHOC was originally created for retrieving words in handwritten text \citep{DBLP:journals/pami/AlmazanGFV14} but can be generalized in a straight-forward way for representing two-dimensional structures like formulas \citep{DBLP:conf/clef/LangsenkampMZ22,DBLP:conf/clef/AvenosoMZ21,DBLP:conf/sigir/AmadorLDSZ23}.

As shown in \refexample{eg:appear}, for the XY-PHOC representation the first region contains all symbols in a formula.
Other regions split the formula into 2, 3, or more equal-size regions horizontally or vertically.
Later versions also used concentric ellipses or rectangles to better capture symmetry (\eg for $x+y$ and $y+x$).
The encoding is compact, using one bit vectors per unique symbol to record occupied regions.

\paragraph{Canonicalizing formulas and formula representation tables.}
\label{sec:canon}
\refexample{eg:enum} visualizes common ways to generalize variables and operations in an OPT,
which can also be applied to SLTs.
Applying these transformations reduces the number of unique formulas in a collection,
making additional formulas identical, and others more similar after these transformations are applied.
The process of reducing variation in representations is known as \emph{canonicalization}.

At middle in \refexample{eg:enum} we number each unique variable from left-to-right in the OPT, starting from 1.
This type of numbering for entities is known as an \emph{enumeration}. 
`2' for $i$ is repeated because it appears twice in the expression. 
We can use enumerations to capture formula structure while ignoring variable names. 
For example, using variable enumeration the \emph{pythagorean theorem} expressed as either
\begin{equation}
x^2 + y^2 = z^2
\end{equation}
or
\begin{equation}
a^2 + b^2 = c^2
\end{equation} 
has the same form: 
\begin{equation}
\boxed{1}^2 + \boxed{2}^2 = \boxed{3}^2.
\end{equation} 
We can apply this change for both the SLT and OPT representations.

For formulas, canonicalization often involves \emph{normalizing} symbols and structures as well, \eg replacing different names for the
same operation with a single name, or always ordering subscripts before superscripts in SLTs (\eg in \LaTeX{} or Presentation MathML).
Another normalization re-orders variable and constant names lexicographically for operators with unordered arguments, \eg to have both $x + y$ and $y + x$ represented by $x + y$.  

For MathML we often `flatten' nested tags,
such as repeated {\tt <mrow>} tags belonging to a single writing line and sequences of unordered operations. Both of these transformations for {\tt <mrow>} and {\tt <times>} are applied in \refexample{eg:mml}, flattening a chain of {\tt <times>} nodes into one node in the Content MathML and removing {\tt <mrow>} tags entirely from the Presentation MathML. 

\begin{myexample}{OPT Variable Enumeration and Symbol Types}{eg:enum}
\hspace{-0.25in}
\scalebox{1.05}{
\imagefillwidth{figures/MathRepresentation/OPT-Types-Eval-mod}
}
\end{myexample}

At right in \refexample{eg:enum}  we have an OPT with nodes replaced by an assigned \emph{type} for each symbol. 
\emph{I} indicates identifiers (\ie names) for variables and operations/functions like the {\tt idf} function identifier.
Note that identifier {\tt idf} must represent an operation, because it is an internal node in the OPT.
Other predefined mathematical operations are labeled to indicate whether their arguments are (O)rdered or (U)nordered.
Among other uses, types can be used to permit or constrain matches between constants, variables, and operations in two formulas.
More sophisticated typing schemes have been used, \eg to distinguish numbers from alphanumeric identifiers and greek letters, and relational operations from 
set operations and arithmetic operations.

Canonicalization can remove unhelpful distinctions,
however, too much canonicalization can also remove meaningful differences.  
A common compromise is to annotate all formulas with \emph{multiple} representations stored in tables, including one table for the original encodings (usually \LaTeX{} and/or Presentation MathML).

To annotate sources with different formula representations, each formula encountered in a source is assigned a unique integer identifier.
Tables used to hold each formula representation are sorted by these formula instance ids, so that one integer can be used to retrieve 
any representation that we have produced.
To save space, often the representation tables define only the unique formulas in each representation,
and we add a second lookup table. 
The lookup table is used to map formula instance ids to unique ids, and the 
representation tables map unique ids to detailed representations (\eg canonicalized OPTs).
This prevents detailed representations for $x$ or an isolated canonicalized enumerated variable (~\fbox{1}~) from being stored millions of times. 




\paragraph{Textual formula annotations: Math entity linking.} 
We can also use markup available in sources along with analysis tools to capture formula-text relationships such as shown in \refexample{idfeg} and \refexample{eg:amr}.
Textual formula annotations can be used for a variety of information tasks.
For example,  using formula symbol identifier descriptions as features for automatically generating Mathematics Subject Classification (MSC) subject codes \citep{10.1145/2911451.2911503}. MSC is a collaboratively-produced hierarchical classification scheme used
to identify subject codes for papers in math journals. 
Recent math-aware search engines have also explored using annotated formulas as their collection, including the math entity cards in MathDeck \citep{dmello_representing_nodate} that connect formulas to titles and descriptions from Wikipedia. 
Another retrieval system, MathMex \citep{DBLP:conf/ecir/DurginGM24} indexes formulas that appear with their textual descriptions in a document. 

When text annotations for formulas are not provided directly in sources, Math Entity Linking (MEL) systems can be used to connect formulas with their surrounding context.
Context may include descriptions for formulas and their symbols, other formulas defining symbols in a formula, and external sources (\eg linking formulas to Wikipedia pages).
We note that symbols like $x$ and $\lambda$ are frequently re-defined within a single paper, leading to multiple definitions  \citep{asakura-etal-2022-building}. This complicates the task of \emph{coreference resolution}, where multiple references to the same mathematical symbol or entity need to be identified and disambiguated when symbols are redefined \citep{ito-etal-2017-coreference}.\footnote{The Math Identifier-oriented Grounding Annotation Tool (MioGatto) \citep{asakura2021miogatto} provides a tool for annotating different roles for formulas and symbol, linking identifiers to pre-defined math concepts extracted from the document.}

Most early methods for MEL were rule-based due to the limited data available for training machine learning models. 
One of the earliest textual formula annotations linked math expressions to their corresponding Wikipedia page \citep{10.1007/978-3-319-49304-6_18}; unfortunately not all math expressions have Wikipedia pages, and  context provided in the document where a formula appears is likely more relevant and/or accurate for the formula. 
Another early system annotated formulas with descriptions and relationships to other formulas in dependency graphs \citep{DBLP:journals/ir/KristiantoTA17}. Textual descriptions are extracted using an SVM-based model to link description nodes to formulas and symbols \citep{kristianto2014extracting}.
References between formula nodes are captured through structural matching of formula sub-expressions.

Later systems including MathAlign \citep{alexeeva-etal-2020-mathalign} focused on textual annotations within the documents where formulas appear. 
There has also been work on automated variable typing, where pre-defined mathematical types (\eg integer, real) are assigned to variables in mathematical formulas  using sentences containing descriptions where a symbol appears \citep{stathopoulos-etal-2018-variable}. 

The \emph{SymLink} shared task at SemEval 2022 \citep{lai-etal-2022-semeval} required extracting math symbols with their textual descriptions from \LaTeX~source files collected from the arXiv. 
The main task requires this to be performed within a \LaTeX~paragraph. 
First, all text spans (contiguous excerpts) containing math symbols and descriptions are identified, and then symbols are matched with their descriptions.
The SymLink dataset provides more than 31,000 entities and 20,000 relation pairs, which allowed modern machine learning models (\eg BERT-based) to be proposed.

Math entity linking and other forms of annotating text-formula relationships are important directions for future research.
They are challenging because incorrect detections can corrupt intended meanings, and because
mis-detections can lead to cascading errors.
This fragility and the computational cost of constructing explicit semantic annotations are partly responsible for the popularity
of dense embedding models, which use language statistics to capture associations and usage patterns for tokens/formulas/passages etc.
However, information-wise embeddings capture associations rather than discrete entities and relationships, and
we expect that combining embeddings with constructing graph-based representations will prove beneficial in the future.\footnote{There is related research in knowledge graph construction \citep{DBLP:journals/csur/ZhongWLPW24}.}

\section{Indexing formulas and text}

Indexing is a critical component for both the implementation and evaluation of mathematical information retrieval systems discussed in
the later chapters of this book.
One might consider indexing to be mostly a brute-force compilation and reorganization of source data in a collection.
In fact, there are quite a number of
important encoding details (\eg character encodings, file formats, and their myriad variations), organizational and retrieval unit design decisions, and resource constraints such as speed and storage requirements that must be carefully addressed if downstream model and evaluation implementation efforts are to be reasonable and effective.
This is especially true for multimodal indexes used for math IR systems, where we may have multiple data representations for text, formulas, and their combination.\footnote{In our experience, implementing new indexing tools is a substantially larger effort than implementing retrieval and machine learning models using indexed data.}

\reffig{fig:ch2overview} illustrates the main tasks for indexing sources.
When we talk about indexing, we're actually referring to a process that consults and annotates sources with additional information, and then creates a collection index.
We discussed annotating formulas in the previous section, but annotations are needed for text as well.
For example, if we plan to use dense retrieval with sentences, we first need to record where sentences are found in sources, \eg in a table containing pairs of start and end character positions.
We often also create tables to hold metadata such as authors and logical regions such as titles, so that these may be quickly accessed or searched separately.

The collection index contains data structures for organizing both sources and their contents, along with search indexes that organize source contents by \emph{patterns} generated from sources and their annotations. 
More concretely, indexing involves:
\begin{compactenum}
\item consulting source text and formulas to generate dictionaries for fast lookup and analysis of their contents,
\item adding information to sources through additional dictionaries (\eg formula locations and representations), and
\item generating inverted index and/or dense vector index files from source contents and annotations.
\end{compactenum}

\paragraph{Locating and extracting formulas from sources.}
We often start by identifying formulas in our sources.
Videos and PDF documents generally do not identify formula locations, and so we create a table mapping integer formula identifiers to their locations. 
An  example is the Page-Region-Object tables used for ACL anthology PDFs in the MathDeck system \citep{DBLP:conf/sigir/AmadorLDSZ23}.
Each detected formula has an entry with  integer source and page ids, and two x-y coordinates for the top-left and bottom-right corners of a \emph{bounding box} containing the formula.
Formula representation tables are then created from the detected formulas.

For text documents with demarcated formulas (\eg \LaTeX{} and HTML with MathML), we extract encodings for formulas and
construct the formulas representation tables needed.
For some applications such as generating MathAMR trees or training transformer models such as BERT, it can be
helpful to replace formulas by an identifier, \eg {\tt <math [...] </math>} becomes {\tt EQ::42}. 

\paragraph{Vocabularies for text and formulas.} A simple but critical annotation are \emph{vocabularies}.
A vocabulary defines a set of unique objects/symbols seen in our collection
including words, XML tag types, \LaTeX{} commands, and other math symbols (\eg in unicode).  
Vocabularies for text and formulas may be stored separately, together, or both.
Different formula representations will generally have their own separate vocabularies.
In general we compile all unique `words' for each vocabulary, prune some of them
(\eg removing rare `words'), and then enumerate vocabulary items for use in table lookups.
\begin{myexample}{SLT and OPT paths for the $idf$ formula}{eg:paths}

\textbf{Symbol Layout Tree Paths (directed)}
\hrule
\begin{center}
\scalebox{0.475}{\includegraphics{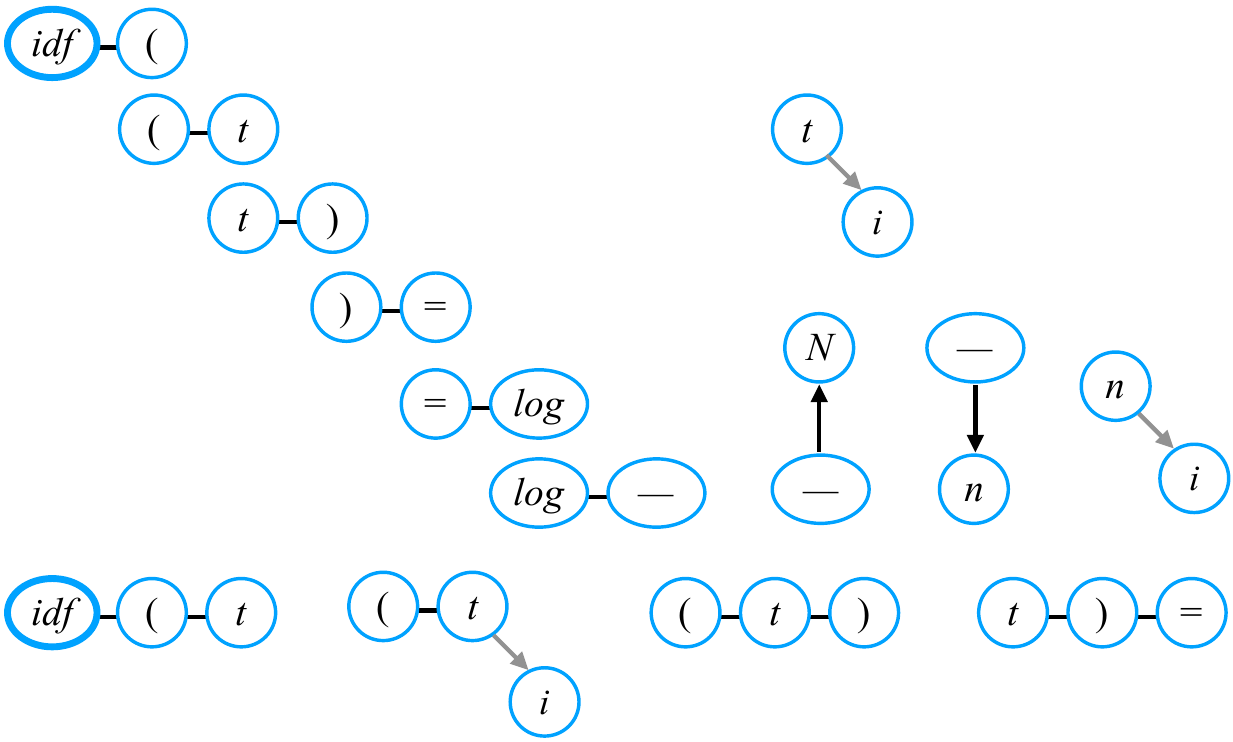}}
\end{center}

\textbf{Operator Tree Paths (leaf-root)}
\hrule
\begin{center}
\scalebox{0.41}{\includegraphics{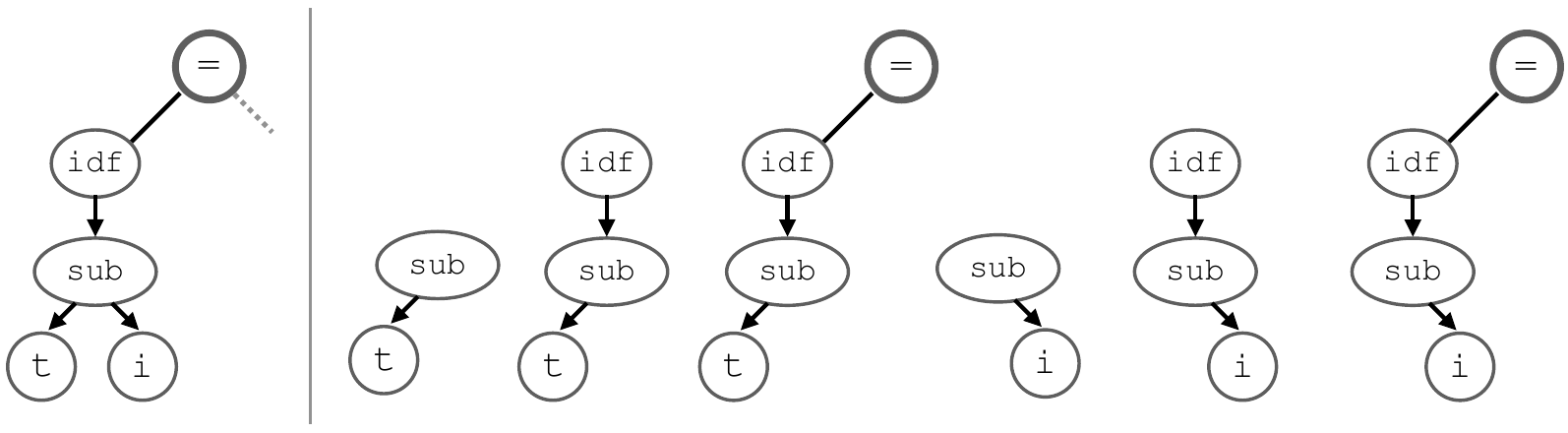}}
\end{center}


\end{myexample}

For formulas, in addition to individual symbols, we often need subexpressions or substructures in our formula vocabulary, \eg
for formula search.
\refexample{eg:paths} shows two common techniques for this, using unrooted and rooted paths.
The example at top shows directed paths in an SLT.
Shown are all unrooted paths of length 1, and four paths of length 2.
In the example at bottom, \emph{leaf-root} paths that start from leaves of an OPT are shown,
and we show all such paths for the left-hand side of the $idf$ formula.
Note that each OPT leaf-root path is a valid subexpression with one argument and one or more missing arguments and operations, while a number of the SLT
paths capture visual patterns, but are not valid subexpressions (\eg `( $\rightarrow$ t').
Both types of paths can be extracted from SLTs and OPTs.\footnote{Anchoring paths only at the root of an SLT or OPT is generally ineffective; many helpful patterns are missed. Leaf-root paths on SLTs are an interesting opportunity.}
One constraint is that indexing too many paths for a formula representation increases index size, and can lead to slow retrieval times.
So our choice of substructure vocabularies such as path types requires careful consideration.

For example, we can construct a text token vocabulary for use with BM25, and
a variety of path vocabularies to search OPTs, SLTs, and additional canonicalized versions described in the previous section.


\paragraph{Inverted indexes for sparse retrieval.}
Among other things, vocabularies define the patterns that can be used to retrieve information directly from index tables.
This type of pattern lookup in an index is often called `sparse' retrieval, because a standard query can be represented by a largely empty vector with bits or counts representing the vocabulary terms in the query \citep{10.1145/1132956.1132959}. BM25 is a sparse retrieval model, as are formula search models that retrieve formulas from tables using paths.
A table that maps vocabulary terms to lists of documents, formulas or other object identifiers is called an \emph{inverted index.}

As an alternative to inverted indexes, people have also used substitution indexing trees, originally developed for unification and matching of predicates in automated theorem proving \citep{10.1007/3-540-59200-8_52}. 
These trees group OPTs or SLTs with shared structure using a hierarchy of symbol and operation replacements and enumerated variables \citep{kohlhase_search_2006,SchellenbergYZ12}.
Substitution trees represent the complete set of possible operation sequences that produce concrete formulas in a collection at their leaves. 
Retrieval finds the most similar formulas in the substitution tree through transformations of the query.


\paragraph{Vector embeddings for dense retrieval.}
For dense retrieval patterns used for search are transformed into embedding vectors,
and retrieval involves finding some $k$ most similar items based on the geometry of the embedding space
(\eg using the cosine of the angles between the vectors). 
This may be for tokens, sentences, formulas, paths, or other objects.
A common approach for generating embedding vectors is having a neural network play an imitation game where we hide tokens (\ie mask them) in a text sequence or node/edge labels in an input graph (\eg an OPT or SLT), and have the network produce likelihoods for every alternative in a vocabulary.
During training, this game is played repeatedly using a training data set, with network weights updated to improve estimates.

Masking and other \emph{self-supervised} learning tasks capture a \emph{language model} reflecting the likelihood of objects appearing in similar contexts. 
The often large amounts of computation required is somewhat confusingly referred to as \emph{pre-training} because network weights are not optimized for retrieval or other tasks aside from the learned probability estimates.\footnote{`Pre-trained' language models often produce surprisingly strong retrieval baselines.
For example, symbols or sub-expressions that look quite different may have similar vectors if they are often used in similar contexts, because when masking these, one or the other will be more likely than other vocabulary items.}
To further improve dense retrieval performance, additional \emph{learning to rank} tasks are run, which require generating ranks for individual sources (\emph{point-wise}), comparing two items at a time (\emph{pair-wise}) or revising entire rankings (\emph{list-wise}). 
This training directly for retrieval or other `downstream' tasks after learning a language model is know as \emph{fine-tuning} of network weights for a specific task.

Learning to rank requires test collections where sources \emph{relevant} to specific queries have been identified, as described in the next chapter; normally relevance labeling is at most partially automated \citep{DBLP:journals/cacm/FaggioliDCDHHKKPSW24}.
As a result, data available for learning-to-rank is often much smaller than for language model learning.
In contrast, large amounts of pre-training data can be created by randomly hiding labels or other random manipulations without human involvement.

After the embedding network has been trained, items to be indexed are converted to embedding vectors and stored in one or more tensors.
Normally we also produce a companion dictionary mapping the tensor rows to integer identifiers for the objects that have been embedded, so
that we can recover the passage, word, formula or other object they come from.

\chapter{Math Retrieval Tasks and Evaluation Metrics}
\label{c-eval} 

In this chapter we introduce the retrieval tasks considered in this book.
There are two task types that we consider: \emph{search tasks}  requiring a ranked list of sources from a collection, and \emph{question answering tasks} (QA) requiring a single response that may or may not reference sources.
We categorize these tasks by expected responses below.

\begin{compactdesc}
\item  {\bf Search Tasks:} Return ranked list of sources
\begin{compactenum}
\item \textit{Formula search:}  ranked formulas 
\item \textit{Math-aware search:} ranked sources or excerpts with formulas and text (\eg passages)
\end{compactenum} 

\item
 {\bf QA Tasks:} Provide question answer possibly with justifications and/or step-by-step solutions
\begin{compactenum}
\item \textit{Multiple choice:}  selection from provided alternatives  
\item \textit{Value:}  numbers or strings, possibly in data structures (\eg lists)
\item \textit{Formula:}  formulas or short programs for  computing answers (often paired with a value response)
\item \textit{Open response:}  free response with formulas and text
\end{compactenum} 

\end{compactdesc}
\forceline
While we have selected these tasks to focus our discussion, this is by no means the exhaustive set of information tasks 
required or studied for Math IR.
For example,
other pertinent tasks include math entity linking \citep{10.1007/978-3-319-49304-6_18}, tasks related to theorem proving such as natural language premise selection \citep{ferreira2020naturallanguagepremiseselection} (see Appendix \ref{chap:proving}), and extraction and annotation of formulas that are not explicitly
demarcated in collections (\eg for handwriting  \citep{precsurvey2024} and PDF documents \citep{DBLP:conf/icdar/ShahDZ21}).
Additional tasks are described in other surveys of mathematical information retrieval
\citep{zanibbi_recognition_2012,guidi_survey_2016,10.1145/3699953}.

\section{Evaluation overview}

\reffig{fig:ch3overview} visualizes the people, data, and processes used to create retrieval task data and evaluate search and QA tasks.
The figure illustrates an important but easy-to-miss fact: it is  \emph{people} and not formal definitions, systems, or algorithms that define retrieval effectiveness, and the target responses that retrieval systems are designed to produce.
More specifically, these roles are:
\begin{compactitem}
\item \textbf{Users:} Realistic search queries and questions come from human \emph{users} either directly or from available data (\eg in query logs). 
\item \textbf{Assessors:} Identify relevant sources for search queries, and define correct question answers. 
\item \textbf{Designers/Researchers:} Collect queries, results, assessments, and set the measures and procedures that quantify performance.
\end{compactitem}
As a result, 
a chosen evaluation design and the human actors within it influence the data produced, along with any observations and conclusions made from this data.
It is important to consider this when reviewing evaluation results in the work of others, and when designing  our own evaluation frameworks and experiments.

\ourfigscaled
	{!t}
	{0.55}
	{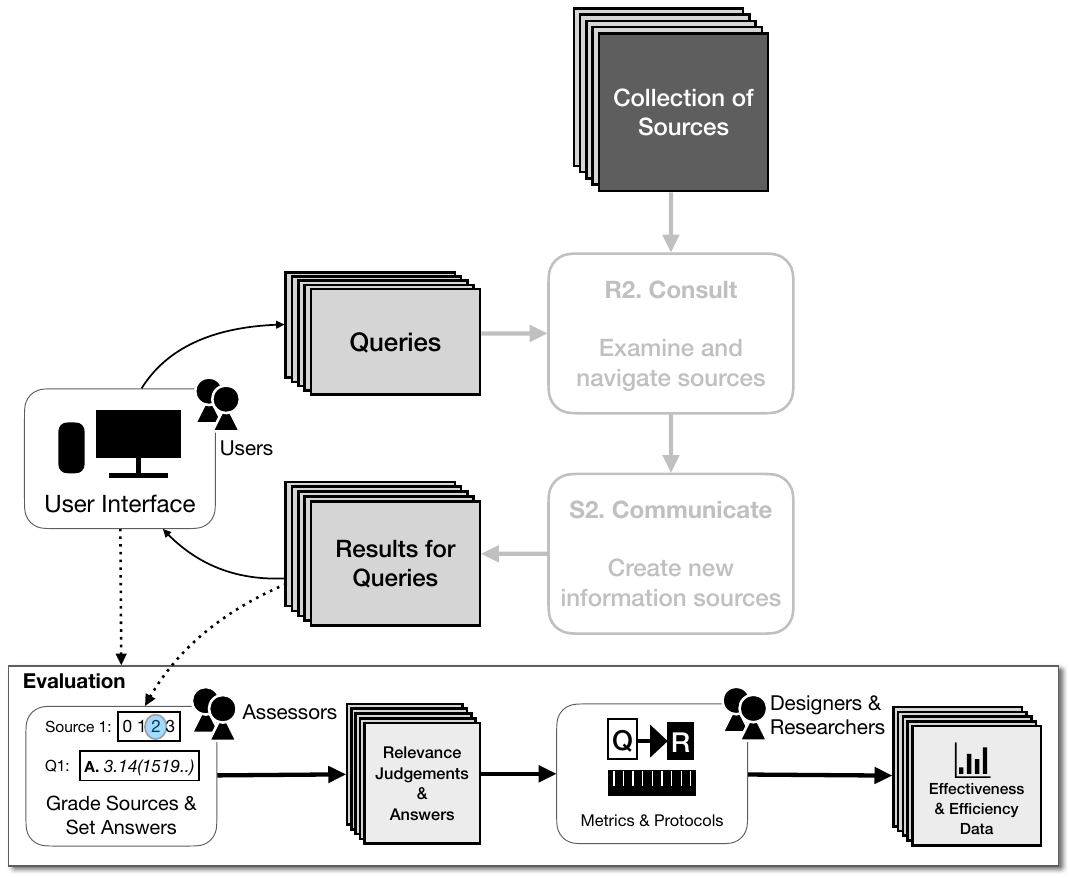}
	{People, Data, and Methods in Evaluation (expands  \reffig{fig:system}).
	Search tasks require rankings sources in a collection, while QA tasks require a single response.
	System information tasks are in gray/hidden: 
	systems are used to sample sources for grading search relevance, but are generally unused for QA evaluations. } 
	{fig:ch3overview}

This dependency of system evaluation on people is unavoidable.
The tasks retrieval systems perform are motivated by human information tasks, which system designers approximate through observation and then formalization within a system design. 
Also, large data sets recorded from users and sources authored by people are \emph{human} data. 
Effectiveness is measured by how well system outputs imitate human responses collected by researchers.
For systems utilizing machine learning, desired outputs are obtained by repeatedly playing imitation games scored by the distance between model outputs and human responses.
For these reasons, people determine or constrain nearly every aspect of system design and evaluation.\footnote{This is a feature, not a bug.}


For search tasks, systems also have a direct role in evaluation aside from producing results, sometimes even for their own evaluation.
This is because we normally \emph{pool} sources returned from multiple systems before assessors make relevance judgements.
This does introduce bias in evaluation, because only items returned by systems used in pooling are assessed.
However, most collections are too large to proceed any other way, and we can choose metrics to mitigate the effects of this bias on evaluation outcomes.
 
Evaluation-wise, we focus here on the effectiveness of query results as measured offline using \emph{test collections}. 
A test collection consists of:
\begin{compactenum}
\item \textbf{Collection of Sources:} sources to be searched for search tasks; optional, usually missing for question answering tasks.
\item \textbf{Topics:} queries to run. For search this also includes query information need descriptions and criteria for relevant sources. For QA questions may include context and/or explanations for answer requirements.
\item \textbf{Responses:}  includes pooled sources with relevance judgements for search tasks, and question answers for QA tasks.
\item \textbf{Protocol:} metrics and methods for producing evaluation data. 
\end{compactenum}
In addition to being aware of the roles people play in retrieval evaluation, it  is important to remember test collections provide a \emph{sample}, and not all possible queries and results for a task.
As such, the data that we collect provides \emph{evidence} for hypotheses (\eg system $A$ performs better than $B$ for metric $M$), and not proof \citep{fuhr17}.
With that said, many valuable things can be learned about system behaviors, evaluation data and frameworks, and even retrieval tasks themselves using test collections.
They allow us to address questions using direct observation in addition to observations reported by others.\footnote{Galileo expressed concern about getting information primarily by following entries in book indexes rather than experimentation (see \citep{duncan} pp. 9-10).}

Briefly returning to our `jar' model for information tasks in \refchapter{c-inf-needs} (\reffig{fig:placeMatTasks}) using test collections for evaluation and related experiments is an important information \emph{synthesis} task that involves \emph{applying} information  to \emph{communicate} new information sources such as research papers.
Reporting informative evaluation data requires significant effort: careful checking is needed at every step of data collection, measurement, analysis, and reporting. 
Test collections help ease this burden by providing standard data sets and methods for system comparison.

\textbf{Efficiency metrics.}
While discovering new retrieval models and understanding their information use and effectiveness is generally the focus of IR research,
 for large collections and real-world systems efficiency is also very important.\footnote{Quoting James Cordy: ``Get it right, then make it fast." We add a proviso: ``...but have a fast enough version for debugging and study, \eg using a small collection."  }
Metrics such as mean query response time ($MRT$, \ie average seconds/query), query throughput (\ie average queries/second), and index size on disk and in memory are used to evaluate system speed and resource utilization.
Efficiency metrics are also needed to check tradeoffs between time, space, and effectiveness.


\section{Retrieval Tasks: Search and Question Answering}

%
%
%


\begin{myexample}{Formula Search Tasks}{eg:ftasks}
\begin{scriptsize}
\fheader{Formula similarity search}
\centering 

\scalebox{0.9}{
  \small
	\hspace{-0.175in}
    \begin{tabular}{c | c  | c | c | c | c }
             & \multicolumn{5}{c}{\textbf{Result}}\\
             \textbf{Query} & 1 & 2 & 3 & 4 & 5\\
            \hline
            & & & & & \\
            $y=\frac{a+bx}{b-x}$  & 
            	$y=\frac{a+bx}{c+x}$ 	& 
            	$y=a+bx$ 			& 
            	$y=\frac{a-bx}{c-x}$ 	& 
            	$y=\frac{a+bx}{x+c}$ 	& 
            	$g(x)=\frac{x}{x-a}$ 		\\   
    \end{tabular} 
}

\ffheader{Formula similarity with wildcards (?w) \citep{aizawa_ntcir-10_2013} }


\scalebox{0.925}
{
	\small
    \begin{tabular}{c | l l }
         
        \textbf{Query} & \multicolumn{2}{c}{\textbf{Result}}\\ 
         
         \hline
         ~\\

        $
        \displaystyle
        \frac{\mathbf{?f}(\mathbf{?v}+\mathbf{?d})-\mathbf{?f}(\mathbf{?v})}{\mathbf{?d}}
        $ 
        
        &
        
        1
        
        &
        
        $
        \displaystyle
        g'(cx)=\lim_{h\to 0}\frac{\mathbf{g}(\mathbf{cx}+\mathbf{h})-\mathbf{g}(\mathbf{cx})}{\mathbf{h}}
        $\\
        & ... & ... \\
    \end{tabular}
    
}

\ffheader{Contextualized formula search} 


\begin{tabular}{p{1.75in} | c p{1.75in} }
       
            \textbf{\small Query} & \multicolumn{2}{l}{\textbf{\small Result}}\\ 
            \hline
            & \\


I have the sum
$$
\boxed{
\displaystyle
\sum_{k=0}^{n} {n\choose k} k 
}
$$
know the result is $n^2-1$ but I don't know how you get there. How does one even begin to simplify a sum like this that has binomial coefficients.

&

1

& 
\ldots which can be obtained by manipulating the second derivative of 
$$
\boxed{
\displaystyle
 \sum_{k=0}^n {n\choose k} z^k
}
$$
and let $z = p/(1-p)$ \ldots
\\

&

2

&

Yes, it is in fact possible to sum this. The answer is 
$$
\boxed{
\displaystyle
\sum_{k=0}^n{n\choose k}{m\choose k}={m+n\choose n}
}
$$
assuming that $n\leq m$. This comes from the fact that  \ldots

\\

\end{tabular}

\end{scriptsize}
\end{myexample}

Some example formula search queries and results are shown in \refexample{eg:ftasks}.
Top-to-bottom, the examples include a formula used directly as a query,\footnote{A `concrete' query}
 a formula query with wildcard symbols that can be replaced by subexpressions, and a \emph{contextualized} formula search query where the context the formula appears is included in the query, and returned formulas include their contexts. 

The concrete and wildcard formula queries are symbolic similarity searches, with relevance determined by just formula appearance (SLTs) and/or operations (OPTs).
This type of query is motivated by information needs including refinding a previously seen formula in a document collection, or browsing for similar formulas. 
Wildcards add boolean constraints to queries, as non-wildcard symbols and structures are ideally the same or as similar as possible in the formula returned. 
Wildcard names can also indicate repetition, \eg for wildcards \textbf{?f}, \textbf{?v}, and \textbf{?d} in the example.

Contextualized formula queries include the context where a formula appears, incorporating the types of formula-text interactions described in the previous chapter.
Here relevance is determined by both the formula and the text within which  it appears.
For example, two instances of the formula $X^2$ are distinct if in one context $X$ is defined as a number, and in the other it is defined as a set, whereas using symbolic search over SLTs, both formulas have the same representation.\footnote{For OPTs, the nodes for the variable and squaring operation may or may not differ, depending upon how the collection is created (see previous chapter).}

Examples of math-aware search and mathematical question answering tasks are shown in \refexample{eg:ft_tasks}.
We first show an \emph{ad-hoc} math-aware search task with queries that include formulas and text.
\emph{Ad-hoc} refers to the fact that queries can vary greatly, and may include patterns that are not proper phrases or sentences (\eg keyword queries, or the query {\tt `\underline{$x^2 + 5 = 30$} x value'}).
In the example shown, a full question post from the Community Question Answering (CQA) platform Math Stack Exchange (MSE) is used as a query, and a collection of MSE answer posts is searched.

The bottom portion of  \refexample{eg:ft_tasks} shows two question answering tasks.
The first is an \emph{open response} to an MSE question post, that was generated using GPT-3 \citep{10.1007/978-3-031-13643-6_20}.
The second shows \emph{math word problems} taken from two test collections. 
In both cases, the result should be an answer with two parts: an equation that can be used to compute the solution, and the solution value, here a number and a number list.
\emph{Multiple choice} questions are also common.
These require choosing from a small number of provided alternatives (\eg {\tt 4} for a fourth alternative \emph{(d) None of the above}).
Research-wise, multiple choice questions 
allow varying the complexity of information associated with questions and alternative responses, while constraining system outputs to an alternative from a small set. 
In some collections questions and multiple choice answers include visual elements such as tables or diagrams.
Using multiple choice questions makes it possible to study this type of \emph{multimodal} question answering without needing to change the response format.


\begin{myexample}{Math-Aware Search and Question Answering Tasks}{eg:ft_tasks}

\scriptsize

\fheader{Math-aware search (ad-hoc retrieval)}
\centering

\begin{tabular}{p{1.75in} | c p{1.75in} }
            \textbf{\small Query} & \multicolumn{2}{l}{\textbf{\small Result}}\\ 
            \hline
    & \\


I have the sum 
$$\sum_{k=0}^{n} {n\choose k} k$$
know the result is $n^2-1$ but I don't know how you get there. How does one even begin to simplify a sum like this that has binomial coefficients.

&

1

&

\ldots which can be obtained by manipulating the second derivative of 
$$
 \sum_{k=0}^n {n\choose k} z^k
$$
and let $z = p/(1-p)$
\ldots
\\

&

2

&
Yes, it is in fact possible to sum this. The answer is 
$$
\sum_{k=0}^n{n\choose k}{m\choose k}={m+n\choose n}
$$
assuming that $n\leq m$. This comes from the fact that \ldots
\\

\end{tabular}

\ffheader{Math Question Answering \citep{10.1007/978-3-031-13643-6_20}}
\centering

{\small
\begin{tabular}{p{1.5in} | p{2.25in} }
            \textbf{Query}  &  \textbf{Result}  \\  
            \hline
            & \\
 \scriptsize What does it mean for a matrix to be Hermitian? 
 & 
 \scriptsize A matrix is Hermitian if it is equal to its transpose conjugate.
\\
\end{tabular}
}

\ffheader{Math word problems} 
\centering
{\small
    \begin{tabular}{p{2.4in}  |   l  l }
              & \multicolumn{2}{c}{\textbf{Result}}\\
             \textbf{Query} & \textit{Equation} & \textit{Answer}\\
             \hline
             & & \\
\scriptsize Sarah has 5 pens, David has 3 pens. How many pens do they have? & \scriptsize $x=5+3$ & \scriptsize 8\\ 

\scriptsize 
Find two consecutive integers whose sum is 7. & \scriptsize $x+(x+1)=7$ & \scriptsize 3, 4\\
\hline
\end{tabular}
}
~\\~\\
From MathQA \cite{amini-etal-2019-mathqa} \& Dolphin18K \citep{huang-etal-2016-well}

\end{myexample}

\section{Creating test collections}




An important first consideration is where to collect queries from, and how to select which queries to include.
As our goal is evaluating performance on real-world tasks, it is usually best if search queries and questions come from real-world users and use cases.
For example, for search tasks topics may come from query logs or community question-answering websites. 
Questions on standardized tests such as the Math SAT are commonly used for question answering.
In some cases, topics generated by the test collection creators are designed to explore specific scenarios for new features (\eg wildcards in formula search queries).

The final topics sets ideally provide a \emph{representative} sample for the task being evaluated, while including some \emph{diversity} in topics so that different system capabilities are tested. 
Diversity is sometimes addressed using separate sub-tasks for a test collection.
For math retrieval, criteria to consider include mathematical subjects covered, modalities in queries and responses (\eg formulas, text, diagrams), and the complexity or mathematical difficulty (\eg target grade levels).

\paragraph{Train topics, test topics, and cross-validation.} Normally a test collection divides topics into \emph{training} and \emph{test} topics, so that systems can be compared using the same \emph{test} topics, while being tuned using a separate group of \emph{training} topics.
This way, all systems take the same `test,' without having seen the test search queries/questions previously (\ie not `cheating').
This allows us to observe and compare the information and task \emph{generalization} captured in system data structures (\eg network weights) and algorithms for the same unseen topics.
Systems should \textbf{never} be tuned on test topics when reporting test results. 
Published benchmarks for test collections are results for test topics by default.

Training topics are provided for tuning system parameters.
To obtain a more detailed characterization of system behavior using multiple train/test splits,  \emph{cross-validation} can be used, and  average metric values across splits reported, ideally along with a standard deviation to characterize variance (roughly, consistency across splits).
All train and test topics may be combined before generating splits for cross-validation,
but it must be made clear which topics are used and how train/test splits are produced (\eg leave-one-out treats every topic as the test sample separately; 5-fold cross validation randomizes the topic order, and then makes 5 equal splits, with each split being rotated as the test split, etc.). 
While cross-validation provides more robust evaluation measures, it is important to again note that official benchmarks for test collections are computed for \emph{unseen} test topics, and in this case test topics \emph{cannot} be used in training.\footnote{This can be easy to miss amongst multiple data sets versions. Care is needed.}
Unless noted otherwise, for benchmarking test topics are scored just once, without cross-validation.

Test collections sometimes include smaller train/test topic sets used primarily for development, and/or to make use of the collection easier for those new to a task. These can also be safely produced from small subsets of training topics (\emph{never} test topics), and are very helpful for fast testing and debugging.

\paragraph{Responses: Pooled relevance judgments for search tasks.}
For question answering tasks, answer data is often compiled from available sources by those creating the test collection.
In contrast, for search tasks relevance judgements are needed before a test collection can be released, as they are required to measure effectiveness.
The most common approach is \emph{shared tasks} in which multiple participants run their systems on provided search topics, and then share the outputs of their runs for pooling as illustrated in \refexample{eg:qrels}. 
After these assessments have been collected, relevance judgements are used to score participants' systems, and the relevance judgements produced are included when the final test collection is released.\footnote{Ideally, the system \emph{runs} (ranked responses for every topic by each participating system) are also included in the test collection for later study and comparison.}

In a shared task, all participants have their system(s) search the same collection of sources, and use the same topic queries.
Assessors assign relevance scores to pooled sources, and these assessments are stored in \emph{qrels} files (quantitative relevance assessments).\footnote{The standard qrels format is from TREC (\url{https://trec.nist.gov/data/qrels_eng})} 
Shared tasks are run frequently at conferences including TREC (Text REtrieval Conference), CLEF (Conference and Labs of the Evaluation Forum), NTCIR (NII Testbeds and Community for Information access Research), and FIRE (Forum for Information Retrieval Evaluation).

\begin{myexample}{Creating query relevance judgements (QRels)}{eg:qrels}
\centering
	\scalebox{0.6}{\includegraphics{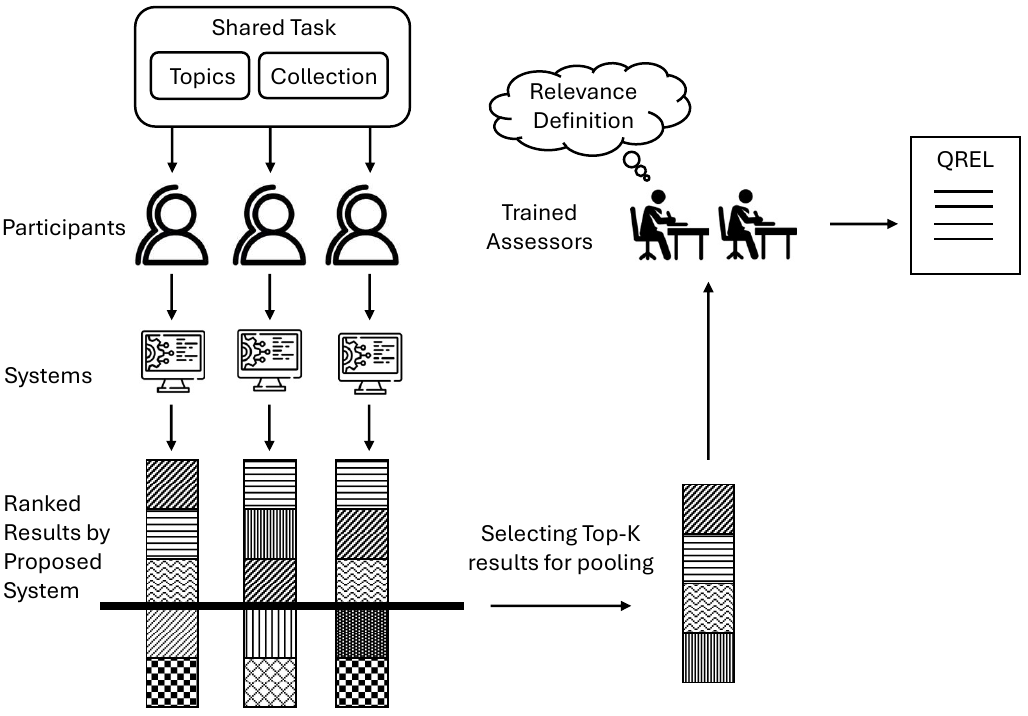}}

\end{myexample}


\refexample{eg:ratings} illustrates relevance assessments using a binary scale (\ie {\tt 1} is relevant, {\tt 0} non-relevant), and \emph{graded} relevance where an ordinal scale of three or more values is used, \eg {\tt Non-relevant} and {\tt Low, Medium, High} relevance. Graded relevances can be easily converted to binary relevances by thresholding.
For our graded relevance example, we might map {\tt Non-Relevant, Low} (0,1) $\rightarrow$ 0 (non-relevant), and {\tt Medium, High} (2,3) $\rightarrow$ 1 (relevant).
We also see an example of \emph{unknown} relevance for a formula search in the search results from system A at top (rank 3). 
Model  A retrieved a formula that was not included in the pool created during the shared task, and so it is missing in the published qrels file. We will come back to this later in the chapter.

Normally measures of agreement between assessors are reported for search test collections. 
This is done by providing the same set of topics among assessors and comparing their assessments with agreement measures such as Cohen's Kappa coefficient. Properly training assessors can help increase agreement among assessors. For instance, in ARQMath-3 \citep{10.1007/978-3-031-13643-6_20}, for the formula search task, the Cohen's Kappa value increased from 0.21 to 0.52 from the first training to the last (third) training session. 

\begin{myexample}{Relevance Assessments}{eg:ratings}
\scriptsize

\fheader{Binary and unknown relevance (?) }

\centering
{
	\small
    \begin{tabular}{ c| l  c |  l c}
        \toprule
            & \multicolumn{2}{c |}{\textbf{Model A}}  & \multicolumn{2}{c}{\textbf{Model B}}\\
            \textbf{Rank} & \textit{Formula} & \textit{Relevance} & \textit{Formula} & \textit{Relevance}\\
            \hline
            & & & &\\
            1 & $y=\frac{a+bx}{c+x}$ 	& 1			& $g(x)=\frac{x}{x-a}$ 	& 0\\
            2 & $y=a+bx$ 			& 0			& $y=\frac{a+bx}{c+x}$ 	& 1\\
            3 & $y=\frac{a-bx}{c-x}$ 	& \textbf{\fbox{?}} 	& $y=\frac{a+bx}{x+c}$ 	& 1\\
            4 & $y=\frac{a+bx}{x+c}$ 	& 1		 	& $y=\frac{a+x}{b+cx}$ 	& 1\\
            5 & $g(x)=\frac{x}{x-a}$ 		& 0			& $y=a+bx$ 			& 0 \\   
    \end{tabular}
}

\ffheader{Graded relevance (0-3: Non-, Low, Medium, High)}
\centering

\begin{tabular}{p{1.25in} | c p{1.25in} | l }
            \textbf{\small Query} & \multicolumn{2}{l |}{\textbf{\small Result}} & \textbf{\small Relevance}\\ 
            \hline


& & \\
I have the sum 
$$\sum_{k=0}^{n} {n\choose k} k$$
know the result is $n^2-1$ but I don't know how you get there. How does one even begin to simplify a sum like this that has binomial coefficients.
&

1

& 

\ldots
which can be obtained by manipulating the second derivative of 
$$
 \sum_{k=0}^n {n\choose k} z^k
$$
and let $z = p/(1-p)$
\ldots

&

\textbf{(3) High relevance}
\\

&

2

&
Yes, it is in fact possible to sum this. The answer is 
$$
\sum_{k=0}^n{n\choose k}{m\choose k}={m+n\choose n}
$$
assuming that $n\leq m$. This comes from the fact that 
\ldots

&
\textbf{(0) Non-relevant}
\\

\end{tabular}

\end{myexample}

\paragraph{Relevance assessment and tools.}
Assessing relevance for math search is inherently challenging. 
As discussed in earlier chapters, a  person's mathematical expertise influences their perception of relevance:
a highly technical document relevant to an expert might be irrelevant to someone with a basic understanding of math. 
It is necessary for assessors to have an appropriate mathematical background and to be trained for each search task that they will assess.
They should be provided with well-defined relevance definitions, including instructions on how to distinguish between different relevance degrees.
It is also a good idea to allow assessors to decline assigning a score when they are very uncertain, or to consult an expert.\footnote{In our own work, a math professor acted in this role.}

\refexample{eg:turkle} shows the Turkle interface used for contextualized formulas search in the ARQMath shared tasks.  
Each assessor has an account, and is assigned topics to evaluate. 
Relevance data is compiled automatically and converted to qrels files by the system.
In the example we see a query formula in its MSE question post on the left, and two formulas in their MSE answer posts taken  the assessment pool. 
Assessors were allowed to view the question threads that queries and results appeared using provided links.
At right we see buttons for the 4-level graded relevance scores, and two additional buttons for system failure (\eg when a result is unreadable), and when an assessor was uncertain how to rate the result.
A box for comments was also included, and was primarily used to explain why assessors selected "System failure" or  "Do not know."

\begin{myexample}{Relevance assessment for formula search}{eg:turkle}
	\hspace{-0.175in}
	\scalebox{0.66}{\includegraphics{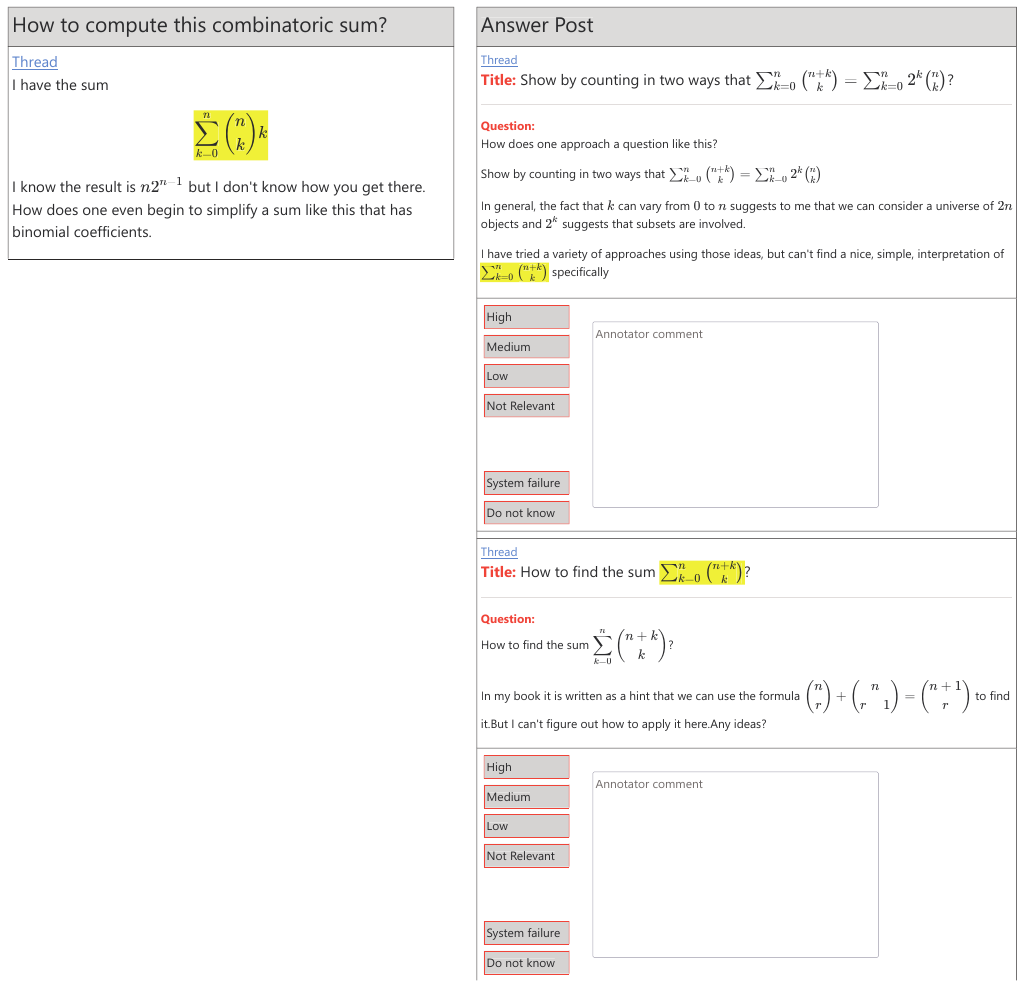}}
	\small
	ARQMath Turkle assessment interface (Formula Search). \textbf{Left:} formula query highlighted in a question post. \textbf{Right:} two question posts containing the same formula. Assessors consider posts when deciding relevance for each pooled  formula. They could also check question threads associated with posts using the `Thread' links.
\end{myexample}

An important practical consideration is the time and effort assessors require to generate answers or judge relevance for search results.
For example, in ARQMath the contextualized formula search task had an average assessment time of roughly 35 seconds for \emph{each} formula query and candidate formula. 
Using the same MSE question/answer posts there was a second answer retrieval task, where assessors had to decide how well an answer addressed a given question.
Assessors found this task much more difficult, and average assessment time was nearly twice as long as for formula search: 64 seconds per answer.

\paragraph{Responses: Question answers for QA tasks.}
Unlike search tasks, QA test collections may be created without the participation of QA systems (\ie shared tasks are not needed to create QA test collections). A list of QA test collections are provided in \refchapter{c-qa}.
Answers are created, annotated, and checked in a variety of ways, with differing levels of effort for assessors and test collection creators. 
For example, \LaTeX{} {\tt $\backslash$boxed\{\} } commands already identify final answers for in the AMC and AIME QA data sets, while the Amazon Figure Eight annotation platform was used to manually select operations and arguments from provided lists for MathQA.
Other collections have assessors generate answers for questions directly (GSM8k), or use
machine learning techniques to automatically segment final answer values from available CQA answers (\eg from answer posted in YahooAnswers using SVMs for Dolphin18K).
For AQuA-RAT, crowdsourced workers were used to modify questions (\eg changing variable values) and provided rationales as needed from available GMAT and GRE test questions.

In addition to final answers, some QA test collections include step-by-step answers or textual annotations.
For machine learning models, some of this textual data associated with \emph{training} topics can be used to improve system answers by providing additional contextual data/information, and used to improve generated explanations for answers or improve responses to queries and comments in interactive discussion (\eg for large language models).

In all QA datasets, there are also post-processing steps to normalize the format of answers, to avoid missing correct answers.
Where people are used to generate and annotate answers, the same concerns regarding assessor training, consistency, and effort as mentioned for search tasks apply here as well.

\section{Evaluation metrics}



To be focused in their purpose, in addition to collecting queries, responses, and assessments (\eg in qrels, or target answer files) test collections also need to define the measures of success and how they are computed.
This way people can use test collections independently, and compare their results in a consistent, meaningful way.
We want our measures to be automated for consistency and to avoid error as well. 

Selecting appropriate evaluation metrics is subtler than it may seem at first glance. 
For example, consider the top-5 retrieval results from two formula search models in \refexample{eg:ratings}. 
Which system would you prefer?
We might prefer model `A', because the first result is relevant.
However, for some information needs we may prefer model `B', because more relevant formulae are retrieved. 
This is a simple example of how information needs influence which effectiveness metrics are better suited to an individual topic or task.

As another simple example, suppose that we ask a QA system to {\tt `provide the value of $\pi$'}, and that we have stored the answer as $3.14159.$
If a single value is expected, we need to define the required number of digits for a response to be scored as correct.
We might accept $3.14$, but probably not $3$; if more digits of $\pi$ are returned than in our stored answer, we probably don't want to penalize this either.
The differences in answer formats mean that  measures of \emph{Accuracy} for math QA systems include normalizations of target responses and provided answers, which include tolerances and constraints to avoid penalizing correct answers.\footnote{This often loses detail, \eg $\pi$ is irrational, with infinite decimal places).} 

Some protocols for evaluation are more complex, such as the use of \emph{visually distinct} formulas for evaluating formula search in ARQMath (see \refchapter{c-formula-search}), which impacts pooling for assessment and the scoring of search results. 
To help people using test collections, normally provided evaluation scripts are used to run evaluation protocols automatically.

\paragraph{Search metrics.}
\reftab{tab:smetrics} presents metrics used to evaluate search effectiveness using relevance assessments (\ie qrels data).
Note that while most metrics are defined for a single query, they are often reported using their average value for a set of test queries.
It is helpful when standard deviations from the mean are also reported, to characterize how much the metric values vary across queries.
Most common metrics use binary relevance values for their computation.
As described earlier, we can binarize graded relevance values to compute these metrics.

\ourtable
	{!t}
	{
		\vspace{-0.2in}
		\scalebox{0.625}{
		\begin{tabular}{p{0.75in}  p{1.5in}  l  p{2.25in}}
		\toprule
		 & \bf Name & \bf Formula & \bf Description \\
		
		~\\
		\multicolumn{4}{l}{\bf Binary Relevance}\\
		\midrule
		RR 	& \textbf{Reciprocal Rank}			 
			& $1/k_f$ 
			& \mbox{Inverse of rank  $k_f$ for the} \linebreak first  relevant source 	 \\
		mRR & Mean Reciprocal Rank 		 
			& $\displaystyle \frac{1}{|Q|}\sum_{q \in Q} RR(q) $ 
			&  Avg. RR for query set $Q$ 			\\
		
		R \emph{or} R@hit & \textbf{Recall}
			& $|S \cap R| / |R|$
			& Percentage relevant in returned \\
			
		R@k & Recall at (rank) $k$
			& $|S_k \cap R| / |R|$
			& \mbox{Percentage relevant in first $k$} \linebreak sources returned \\
			
		P \emph{or} P@hit & \textbf{Precision}  
			& $|S \cap R| / |S|$		
			& Percentage returned in relevant\\
			
		P@k & Precision at (rank) $k$
			& $|S_k \cap R| / k$	
			& Percentage first k sources returned in relevant\\

		AP	& Average precision			 
			& $\displaystyle \frac{1}{|R|}\sum_{k \in K_r} P@k(S_k)$  
			& Avg. $P@k$ for relevant documents at ranks $K_r$ in sources returned\\

		mAP & \mbox{Mean Average Precision} 	
			& $\displaystyle \frac{1}{|Q|} \sum_{q \in Q} AP(q)$ 
			& Avg. AP for query set $Q$\\

		~\\
		\multicolumn{4}{l}{\bf Graded Relevance}\\
		\midrule	
		DCG@k & \textbf{Discounted \linebreak \mbox{Cumulative Gain}}
			& $\displaystyle r_1 + \sum_{i=2}^{k} \frac{r_i}{\log_2 i}$ 
			& Sums relevance scores $r_1$ through $r_k$ for first k sources returned using $\log$ discount from rank 3 on\\
		iDCG@k & \emph{Ideal} DCG   
			& 
			& DCG@k for first k pooled assessment scores after reverse sorting, \eg ($a_1, \ldots, a_5) = (3, 3, 2, 1, 0)$\\
		nDCG@k & \mbox{Normalized DCG} \linebreak \mbox{at (rank)} $k$ 
			& $DCG@k / iDCG@k$
			& Percentage of ideal DCG obtained for first k sources returned\\
		\mbox{nDCG or} nDCG@hit& Normalized DCG
			& $nDCG@k ~\textrm{for}~k = |S|$
			& nDCG for all returned sources $S$\\

		~\\
		~\\
		\multicolumn{4}{l}{\bf Scored Sources Only ($N$ : Non-relevant sources)}\\
		\midrule

		Bpref & \mbox{\textbf{Binary preference}} \linebreak \emph{(w. binary relevance)} 		
			& $\displaystyle \frac{1}{|R|} \sum_{k \in K_r} \left(1 - \frac{ \min( |S_k \cap N|, |R| )}{|R|} \right) $ 
			&  Avg. percentage relevant before non-relevant for relevant sources; treats $|N| = |R|$ to balance classes. \\
			
		M$'$ & \mbox{\textbf{Prime metric}} (\eg P$'$@5, nDCG$'$)  	
			& \emph{(as defined for metric $M$)} 
			& Compute $M$ with $S' = S \cap(R \cup N)$ rather than full ranking $S$ \\
		
		\bottomrule
		\end{tabular}
		
		}
	
	}
	{Search Metrics. By default, sources not in relevant set $R$ are in non-relevant set $N$. 
	$S_k$: first $k$ sources returned. \textbf{Note:} reported metrics are typically \emph{averaged} over queries but only mAP and mRR explicitly. 		Ch. 8 of Croft et. al's textbook provides an overview of these metrics \citep{DBLP:books/daglib/0022709}. 
	}
	{tab:smetrics}

There are  trade-offs that occur between some of  these metrics. 
A classic example is the tradeoff between \emph{recall} and \emph{precision}: the more items we return in search results, the more likely that relevant items will be included, which tends to increase recall metrics.
However, returning additional items tends to produce more non-relevant than relevant items, which decreases precision metrics.
Intuitively, this is because to retrieve more items, patterns used for matching need to be less constrained, making them more likely to match 
non-relevant sources.
In theory, we can return the entire collection for all queries to obtain an average recall ($R@hit$) of 100\% because all relevant items are returned.
However, the average query precision ($P@hit$) will be close to 0\%, because few/none of the returned sources are relevant for an individual query.

Another important trade-off occurs when systems are designed to maximize \emph{mean reciprocal rank} (mRR), the average inverse rank of the first relevant hit ($\frac{1}{r}$), or $precision@1$, the percentage of queries with a relevant source at rank 1 (\ie at the top of the ranking).
Doing this often reduces metrics for effectiveness over a full ranking such as mean average precision ($mAP$) or normalized discounted cumulative gain ($nDCG$).
This occurs because using retrieval patterns and scoring narrowly focused on matching high relevance sources can bias the model away from capturing rarer and/or partial relevance signals.

Not all of the retrieval metrics in \reftab{tab:smetrics} behave as expected, or are always applied or interpreted appropriately in the research literature.\footnote{see \citep{fuhr17} for a critique of mRR and a proposed replacement, as well as common uses of mAP.}
For example both $mAP$ and $nDCG$ ($nDCG@hit$) characterize relevance in complete rankings.
They are helpful for understanding differences between full rankings and the retrieval models that produce them.
However -- most users consider just a small number of results returned, and so other measures are more appropriate for user-oriented evaluations (\eg $P@5$ or $nDCG@5$). 
The tendency for users to examine few items also motivates the logarithmic discount used for $nDCG$.
$nDCG@5$ gives decreasing credit for relevant items starting at rank 3, in contrast to $P@5$ where all relevant items in the top 5 hits have the same weight.
$P@5$ uses binary relevance: items are identified as \emph{either} relevant or non-relevant. $nDCG@5$ instead uses graded relevance, where relevance scores may have different `levels' (\eg 1 for `Low,' 2 for `Medium,' and 3 for `High' relevance).

In the end, no metric is \emph{better} on its own -- it depends upon what we want to measure. 
If say we want to know how often relevant items are in the top-5 hits, $P@5$ is simpler to interpret than $nDCG@5$.
But if we instead want to characterize how often \emph{highly} relevant sources appear in the top-5 hits, then $nDCG@5$ is more helpful.

To avoid bias in model comparisons, even when there are disputed or incorrect relevance assessments in a qrels file, we report performance measures using the \emph{original} qrels file. 
`Cleaning up' a qrels file by adding or revising entries is to be avoided, as it prevents direct comparison with other published work using the collection: this changes the ideal response, and the specific corrections are likely motivated by improving performance for a specific system. 
It is acceptable to note in a paper where issues in qrels are found, and in some cases to do additional experiments using modified qrels.
However, this is only acceptable if results using the `official' qrels file from a test collection are reported, and the qrels changes are unbiased.\footnote{Automated qrel changes are preferred, \eg using topic and result data. Manual `cherry-picking' of qrel topics for a model family is weak science. }

\textbf{Qrels and items with unknown relevance.}
Qrels provide judgements for pooled sources used in assessment.
Unless the collection is extremely small, we do not have the relevance judgements for all collection sources per  topic: only pooled sources are evaluated.
This means that people using an existing test collection often retrieve sources that are unevaluated. 
This is illustrated for Model A in \refexample{eg:ratings}, where the 3rd formula retrieved was not included in the assessment pool for the query.
This unrated formula will be treated differently, depending upon the metrics that we use.
By default, unrated hits are considered non-relevant, \eg the \emph{Precision@5 (P@5)} for Model A giving the percentage of relevant items within the first 5 returned would be 2/5.

Some metrics ignore sources with unknown relevances, such as the Bpref and \emph{prime} ($'$) metrics shown in \reftab{tab:smetrics}. 
For example, if a 6th formula returned by Model A was graded relevant, then the $P'@5$ would be 3/5; if the 6th formula is non-relevant or no 6th  formula was returned, $P'@5$ is still 2/5. 
Ignoring unevaluated items allows evaluation using only graded items in qrels files, and avoids assuming unpooled sources are non-relevant.\footnote{These metrics are also helpful in early design, as models can simply rank scored sources. However, metrics obtained this way \emph{cannot} be compared with published systems, because the retrieval step that filters the collection is skipped.}

Bpref measures the number of consistent rank \emph{preferences} of relevant vs. non-relevant sources for a query.
A relevant source's preference is \emph{consistent} will all other sources at lower ranks; a non-relevant source is \emph{inconsistent} with relevant sources at lower ranks. 
Bpref gives no credit for a relevant source with $\geq |R|$ non-relevant sources above it in the ranking, where $R$ is the number of relevant sources for a topic/query. 
Relevant sources missing in a ranking are also given no credit.

%

\paragraph{Question answering metrics.}
The main metrics for question answering are simpler than for retrieval, as shown in \reftab{tab:ametrics}.
Most evaluations report the percentage of correct responses (with normalizations/tolerances as described earlier) and/or the number of responses that match the target answers in the test collection exactly.
Some test collections report \emph{perplexity} to characterize system uncertainty in making multiple choice question answer selections.
Perplexity for the correct answer $a_c$ in the answer probability distribution $A_k$ for question $q_k$ is converted to a random choice between $n$ options, using $n = 1/P(a_c | A_k, q_k)$ .
For example, if a model estimates the target answer is 25\% likely to be correct, then the perplexity is  $n = 1 / 0.25 = 4$.

For text responses, open response questions, and comparing explanations or step-by-step solutions against target answers in a collection, text similarity measures are used at the token, n-gram, and string/sequence levels.\footnote{Note: All string-based measures, including token F1 and BERTScore are affected by the method used to split words into tokens.} 
The token F1 measure is the harmonic mean for the percentages of target answer tokens in the response (\ie recall), and response tokens in the target answer (\ie precision).

For embedded tokens, a variation called the \emph{BERTScore} \citep{zhang2020bertscoreevaluatingtextgeneration} has been used.
This is similar to the token F1 score, but makes use of a trained embedding model (\eg BERT).
All target answer and response tokens are embedded using this model, and the highest cosine similarity between each target answer and response token is first computed.
From these maximum similarities, the average cosine similarity for a target answer to a response token is used for recall, and the average cosine similarity from a response token to a target answer token is used for precision. F1 is then computed as before.  
Embeddings capture token context missing in the token F1 measure, \eg this can avoid penalizing synonyms, but the token F1 measure is more easily computed and interpreted.

\ourtable
	{!t}
	{	
		\vspace{-0.2in}
		\scalebox{0.7}{
		\begin{tabular}{l      p{1.1in}  p{1.8in}  p{2.25in}}
			\toprule
			\bf & \bf Name  & \bf Formula / Reference & \bf Description \\
		~\\
		\multicolumn{4}{l}{\bf Correctness and Uncertainty}\\
		\midrule		
		EM & Exact \linebreak match rate
			& $\displaystyle \frac{1}{|Q|} \sum_{q_k \in Q} \delta(a_k, q_k)$
			& Percentage answers identical to target answers ($\delta$ returns 0/1)\\
			
		Accuracy & Correct \linebreak answer rate 
			& $\displaystyle \frac{1}{|Q|} \sum_{q_k \in Q} e(a_k, q_k)$
			& Percentage answers within tolerance of function $e$ (returns 0/1)\\

		Perplexity & Avg. correct answer perplexity 
			& $\displaystyle \frac{1}{|Q|} \sum_{q_k \in Q} \frac{1}{P(a_c | A_k, q_k)}$
			& Correct answer prob. uncertainty as \#items chosen from randomly \\
			
		~\\
		\multicolumn{4}{l}{\bf Similarity to Text Answers (including Rationales, Step-by-Step) }\\
		\midrule
		
		\textbf{Token Level}\\
		Token F1 & Token F1 score
			& $\displaystyle \frac{2RP}{R+P}$
			&  Harmonic mean: \% target answer tokens in response ($R$ecall), \% response tokens in target answer ($P$recision)\\
		
		BERTScore & Token F1 w. \mbox{token embedding} $\cos$ similarity
			& $\displaystyle \frac{2R_{ac}P_{ac}}{R_{ac}+P_{ac}}$
			& Highest target/response token pair cos. similarities give avg. token similarity for target ($R_{ac}$), avg. token similarity for response ($P_{ac}$) \\		
		
		\textbf{N-gram Level }\\
		
		BLEU & Bi-Lingual Evaluation \mbox{Understudy}
			& \citep{papineni-etal-2002-bleu}
			& $[0,1]$ score from shared target answer/response n-grams + penalty for short outputs.\\
			
		sBLEU & Sentence-BLUE
			&
			& BLEU for individual sentences.\\
		
		\forceline
		\textbf{String Level}\\
		Edit distance & \mbox{String edit} \linebreak distance 
			& \citep{edistance}
			& Operation count to convert one string to another (\eg insert, delete, replace). Can be normalized to $[0,1]$  using string lengths.\\
				
		\bottomrule
		\end{tabular}
		}
		
		\scriptsize
		\forceline		
		
		$\delta(a,b) = (a=b)$; $e(a,b) = (normalize(a) = normalize(b))$
		\linebreak
		BLEU and edit distance have many variations.
	}
	{Question Answering Metrics.  
	$a_k$: answer for question $q_k$. $A_k$: prob. distribution for possible answers to $q_k$ (\eg multiple choice). \textbf{Note:} Metrics are computed using target answers, and text measures are tokenization-dependent.}
	{tab:ametrics}

Similarity based on token sequences such as n-grams or full strings can also be used.
$BLEU$ was originally developed to measure the success of translations by comparing a translation against one or more accepted translations of a sentence  \citep{papineni-etal-2002-bleu}.
It computes the similarity of n-grams (\ie token sequences of a fixed length, for different values of $n$) between a response and target answer(s), with an additional penalty for short answers.
$sBLEU$ modifies this to compute similarity at the sentence level, rather than for complete responses.
Edit distance considers an entire token sequence, and computes the number of operations from a fixed set needed to transform one to the other (\eg insert, delete, replace \citep{edistance}).

For the text similarity measures, it is important to realize that they reflect surface structure or correlations learned by a model, and do not directly quantify \emph{semantic} similarity.
These metrics are certainly correlated with semantic similarity, but they only indirectly capture differences in \emph{information content} between responses and target answers. 
What they actually measure is how closely one string imitates the other based on tokens or token embeddings.
They are still very useful, but one has to be a bit careful about their interpretation.

\section{System comparisons and statistical tests}

Imagine that we have two math search systems returning 10 results per query for a test collection with binary relevance grades. 
We determine that the average $P@10$ is 50\% for both systems.
However:
\begin{compactenum}
\item   For the first system, 5 of 10 results are relevant for every query (\ie every query has $P@10$ of 50\%).
\item For the second system, half of the queries return \emph{no} relevant sources (\ie $P@10$ of 0\% per query) while the remaining half return only relevant sources (\ie $P@10$ of 100\% per query). 
\end{compactenum}
While the average $P@10$ scores are identical, we would probably much prefer first system because it is more consistent and avoids missing relevant answers altogether.

To capture variance in our evaluations, such as for rank metrics or differences between target and provided numeric answers to questions, we want to compare \emph{distributions} (\ie sets) of values rather than simply averages.
Statistical \emph{hypothesis tests} are used to check whether differences in average measures are likely to be stable when running additional queries. 
They include an estimate for the probability of detecting a difference incorrectly (\ie a \emph{Type-1 error}) given as the \emph{p-value}. Generally we consider a \emph{p-value} less than either 5\% or 1\% (\ie $p < 0.05$ or $p < 0.01$, chosen \emph{before} running an experiment)  to be a `statistically significant' difference suggesting that the averages are \emph{unlikely} to be the same after running a large number of additional queries/questions.\footnote{Important note: using `significantly improved' or `significantly different' without a hypothesis test is a short path to having a research paper rejected.}  Note that hypothesis tests are probabilistic estimates, and not certain answers regarding whether averages are \emph{actually} different in the limit. It is not possible to run all possible queries/questions to know for certain. Despite this limitation, statistical tests provide a more rigorous and nuanced characterization of differences in metric values than comparing average values directly.

Commonly used statistical hypothesis tests for performance metrics include the standard t-test for comparing two distributions, and the Bonferonni \emph{corrected} t-test when comparing two or more models to a single baseline system. 
The correction here adjusts computed $p$-values when multiple comparisons are made, because without correction the probability of detecting a difference increases with additional comparisons.
Many other tests and comparison types are also used.
The selection of a chosen measure or test is motivated by the goal of a comparison, variable data types, and data distribution assumptions (\eg correlation coefficients, $\chi^2$ (`chi-squared'), and Wilcoxon rank sum tests).

It is also very important when comparing two systems to check raw metric values, and to examine the specific topics where performance differs substantially. 
For example, visualizing raw metric data can reveal whether metrics are similar across queries/questions, or vary dramatically
(\eg for the $P@10$ example from the start of this section).
One simple approach is to sort the metric values and then produce a `ski jump' bar graph.
Specific queries/questions where larger differences in metric values are seen can help identify specific limitations, patterns of behavior, and information use by the models. 
Equally importantly, this also helps identify bugs in system implementations, including where computed metrics are unusually strong, but \emph{not} because the model is effective.\footnote{Based on a true story. Or three.} 
For search tasks, frameworks like PyTerrier\footnote{https://pyterrier.readthedocs.io/en/latest} provide easy access to query-specific differences between models, and can be used to compute common statistical tests.  
Additional helpful evaluation tools include {\tt trec\_eval}\footnote{\url{https://github.com/usnistgov/trec_eval}}, {\tt pytrec\_eval}\footnote{\url{https://github.com/cvangysel/pytrec_eval}} and {\tt ranx}.\footnote{\url{https://github.com/AmenRa/ranx}}
For QA tasks, frameworks such as {\tt nltk}\footnote{\url{https://www.nltk.org/}} can be used to compute standard text metrics.
Standard data matrix tools (\eg Pandas\footnote{\url{https://pandas.pydata.org}}) and statistical tools {(\eg Scipy stats\footnote{\url{https://docs.scipy.org}}) can also be used to compile descriptive statistics and compute hypothesis tests.

\chapter{Formula Search}
\label{c-formula-search} 

	

%
%


As described in \refchapter{c-sources}, the information conveyed in a formula is primarily structural, representing a hierarchy of operations over arguments.
This hierarchy can be represented in an operator tree (OPT) obtained by mapping symbol layout to an operation hierarchy.
For brevity and clarity, authors often assume that readers are familiar with common operations and variable types for the subject area they are writing upon.
The full information that a formula conveys and is associated with includes these notation conventions and related information presented in surrounding text, other formulas, and even other graphics (\eg tables or figures).
Considering these pieces of context is very helpful for formula search.

With that said, there are certainly situations where searching for isolated formulas is  helpful. This includes defining unfamiliar notation, re-finding sources using part of a formula (\eg {\tt ctrl-f} for formulas), browsing through variations of a formula (\eg loss functions using the cross-entropy loss), identifying applications in different domains (\eg medicine vs. computer science), and formula autocompletion.

\ourfigscaled
	{!t}
	{0.5}
	{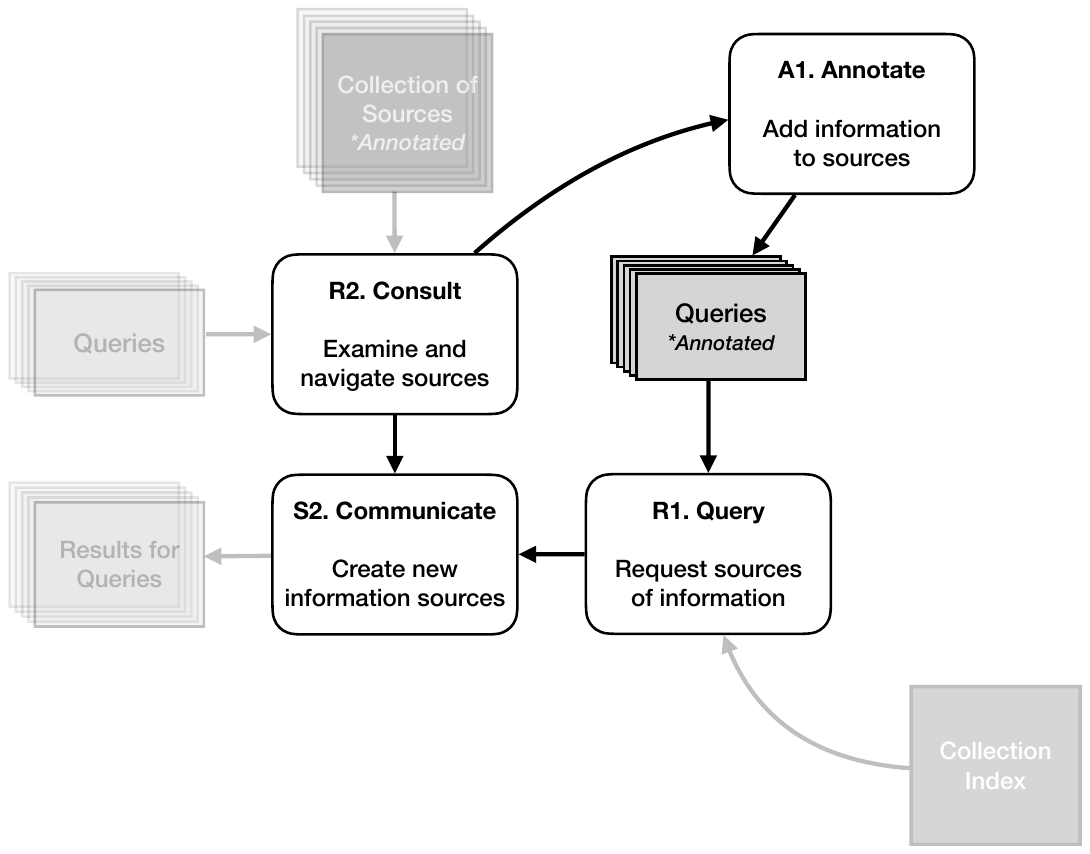}
	{Information Tasks Performed in Formula Search. Prior to search, formula patterns in a collection of sources (\eg OPT and SLT paths) are enumerated or embedded in vectors. These patterns annotate formulas and provide lookup keys  in the collection index. 
	Formulas in the index with patterns identical to the query (sparse retrieval) or similar to the query (dense retrieval) are selected, ranked, and then communicated to the user in a new source (\eg search result page).}
	{fig:ch4overview}

\reffig{fig:ch4overview} illustrates the information tasks used in formula search, using the model from \refchapter{c-inf-needs}.\footnote{Communicating formula search results is important, but little studied.} 
We assume that formulas have already been indexed using one or more representations (\eg sparse and/or dense: OPT, SLT, visual-spatial, etc.) as described in the previous chapter.


In this chapter, we first present test collections used for developing and evaluating formula search. 
We then present formula search models organized by the formula representations they use,
and then summarize their effectiveness on the test collections. 
State-of-the-art models use more than one formula representation. 
This is because dense retrieval models can more flexibly match related formulas using abstract/latent contextual patterns, while concrete patterns (\eg SLT and OPT paths) are better for retrieving highly similar formulas using specific symbols and structures.

\section{Test collections for formula search} 
\label{FormulaSearchEval}

%
%
%
%
%

Relevance definitions for formula search differ based on information needs, and formula search relevance definitions have evolved over time in test collections. There are two basic formula search tasks that have been explored, which differ in their consideration of context.
\begin{compactenum}
\item \textbf{Isolated:} structural similarity of query vs. candidate formulas (SLT and/or OPT), sometimes with optional wildcard symbols
\item \textbf{Contextual:} formula relevance depends upon text where query and candidate formulas appear  
\end{compactenum}
Both tasks are illustrated in \refexample{eg:ftasks} from the previous chapter.

\ourtable
	{!t}
	{
		\vspace{-0.2in}
		\scriptsize	
		\begin{tabular}{p{1.05in}   p{0.7in} p{0.75in} p{0.7in} p{0.7in} p{0.6in} }
		
		\toprule
		\textbf{Test Collection}	& \textbf{Sources} & \textbf{Queries} & \textbf{Results} & \textbf{Metrics}\\
		\hline
		

		\textbf{NTCIR-10} \hfill(2013)	& 	arXiv papers 	 &  Organizers$^w$ & 	F in paper & mAP, P@\{5,10,hit\}\\
		NTCIR-11 \hfill(2014)	& 	Wikipedia		 &  Individual Wiki \linebreak F (known item) 	& F in articles & mRR\\
		NTCIR-12 \hfill(2016)	& 	Wikipedia$'$ & \mbox{Organizers}$^w$ &  F in articles 		& P@\{5,10,15,20\}, Bpref\\

		\textbf{ARQMath-1} \hfill(2020)	& 2018 MSE As & 2019 MSE Qs 	& F in MSE As & nDCG$'$, mAP$'$, P$'$@10 \\
		ARQMath-2 \hfill(2021)	&  	& 2020 MSE Qs 	& 	 &  \\
		ARQMath-3 \hfill(2022)	& 	 & 2021 MSE Qs 	& 	 &  \\
		\textbf{AccessMath}$^\dagger$ \hfill (2018) & \mbox{Videos +} \LaTeX{}~notes & \mbox{Image~region} \linebreak or~\LaTeX{} & Image regions or \LaTeX{} form. & R@10, mRR@10\\
		\midrule
		
		
		\end{tabular}\\
		\scriptsize
		\textbf{F:} Formula; \textbf{As}: Answers; \textbf{Qs}: Questions; \textbf{MSE:} Math Stack Exchange
		\linebreak
		$^w${\textbf{Wildcard symbols} in at least some query formulas} 
		\linebreak
		 $^\dagger$Cross-modal or cross-language retrieval; \textbf{frags.}: Fragments (roughly paragraphs)
   
	}
	{Formula Search Test Collections}
	{tab:tasks-sources}

\textbf{NTCIR:} NTCIR-10 was the first shared math-aware search task \citep{aizawa_ntcir-10_2013}.
It had a formula search subtask, in which systems needed to retrieve formulas similar to a given formula query. 
The collection included 100,000 technical papers from arXiv (mathematics, physics, and computer science) with 35.5 million formulae.

In NTCIR-11 the formula search task 
 was a known-item retrieval task~\citep{aizawa_ntcir-11_2014}.
 The query was either identical to a specific formula instance  in a Wikipedia article, or a version with wildcard replacements for subexpressions. 
 Systems were evaluated based on ranks for target formula instances.
  A Wikipedia page collection with mathematical formulas was used, which was much smaller than the NCTIR-10 arXiv collection.

 NTCIR-12 introduced the Wikipedia Formula Browsing (WFB) task that is similar to NTCIR-10:  retrieve relevant formulas for a formula query
 ~\citep{zanibbi_ntcir-12_2016}.
 NTCIR-12 uses 319,689 articles from English Wikipedia with over 590,000 formulae in the corpus. 

All three test collections use lab-generated topics. In NTCIR-10, 21 formula queries were chosen by the organizers for arXiv papers, of which 18 queries included wildcards and 3 were concrete queries. 
NTCIR-11 had 100 queries, with 59 including wildcards, and 41 concrete queries without wildcards. 
Queries were randomly sampled from Wikipedia pages and then modified to include query variables.
The NTCIR-12  task had 40 queries, divided into 20 concrete queries and 20 wildcard queries.
The wildcard queries are created by replacing one or more sub-expressions in each concrete formula query with wildcards.
The intent was to observe differences in retrieval behavior when wildcards were added to queries.  

Different pooling processes are used in each NTCIR collection \cite{evalarxiv}.
No pooling was needed for
%
NTCIR-11, because retrieval targets were specific formulas.
In contrast,  every formula instance was treated as a separate source for the NTCIR-12 WFB.
This led to limited diversity in the judgement pools after selecting the top-20 instances from each submitted run. 
For example, for the query $\beta$ (a short formula consisting of a single symbol), every formula instance in the pool of formulas to be judged was $\beta$.

Both the NTCIR-10 and -12 test collections use graded relevance (0-2): 2: Relevant (R), 1: Partially Relevant (P), or 0: Non-relevant (N). 
For NTCIR-10, the assessors were mathematicians or math students who viewed each formula instance from the judgment pool in isolation, considering the query-specific scenario and judgment criteria specified for the query. For the NTCIR-12 task there were two groups of assessors, with each group independently judging pooled formulas. One group was computer science graduate students, and the other was computer science undergraduates. Pooled formula instances were shown to the assessors in context by highlighting them in sources, but assessors were not asked to interpret the pooled formula in that specific context. 
Instead, the assessment was to done based on the pooled formula alone.

For each topic in NTCIR-11, the single Relevant (R) formula instance was defined as the formula instance that had been used as the formula query.  Note that there may have been other instances of the same or similar formulas in the collection, but like all instances of other formulas, they would be scored as Non-relevant (N).
NTCIR-11 used mean reciprocal rank (mRR), which is appropriate for single retrieval targets per topic.

For assessment, NTCIR-10 and -12 combined the judgments from two assessors to form a 5-level ``Aggregate'' relevance score. This was done by summing the two scores from assessors for each pooled formula.  Relevance scores ranged from 0 (both assessors judged N) to 4 (both assessors judged R). To compute evaluation measures the 4 level-relevance is binarized, by treating scores of 0-2 as non-relevant, scores of 3-4 as relevant. 

NTCIR-10 reported P@5, P@10, P@hit (i.e., for all returned results), and MAP. NTCIR-12 uses P@k for k = \{5, 10, 15, 20\}. Later, researchers used Bpref \citep{buckley_retrieval_2004} to avoid penalization for unevaluated formulas. 

\textbf{ARQMath:}  The Answer Retrieval for Question on Math (ARQMath) lab introduced a \textit{contextualized} formula search task illustrated in \refexample{eg:ftasks}. The test collection was developed over three years, generating test collections referred to as ARQMath-1 (2020), ARQMath-2 (2021), and ARQMath-3 (2022). 

ARQMath's collection consists of question and answer posts from a math community question-answering website, Math Stack Exchange (MSE). These question posts provide a diversity in subject areas and required mathematical expertise, ranging from simple questions from high school to advanced topics. 
Formula queries are taken from question posts, and the task is to find relevant formulae inside other question and answer posts. 
All MSE questions and answers posted from 2010 to 2018 are used as the collection of sources. 
Formula queries for topics were selected from questions posted in 2019 (ARQMath-1), 2020 (ARQMath-2), and 2021 (ARQMath-3). 
Additional training topics are provided in the test collection.

To make topics diverse, ARQMath attached a complexity label to topic formulas, dividing them into low, medium, and high-complexity topics.  
Additional details on topic selection can be found elsewhere \citep{evalarxiv}.
To avoid the lack of diversity in pooled formulas seen in NTCIR-12, ARQMath pooling selects \emph{visually distinct} formulas:
 two formulas are \emph{visually distinct} if their Symbol Layout Trees differ.
The canonicalized SLT representation from Tangent-S~\citep{davila_layout_2017} was used to identify visually distinct formulas when two formulas are parseable, or had identical \LaTeX~strings otherwise. 

For each visually distinct pooled formula, up to five instances of that formula were shown to the assessors. \refexample{eg:turkle} shows the Turkle\footnote{\url{https://github.com/hltcoe/turkle}} interface used for assessment. As shown in the left panel of the figure, the formula query $\sum_{k=0}^{n} \binom{n}{k} k$ is highlighted in yellow. The assessors can use the question post to understand the user's information need. In the right panel, two instances of one visually-distinct formula, $\sum_{k=0}^{n} \binom{n+k}{k}$, are shown in two different posts. 
For each instance, the assessor could consider the post in which the instance appeared when deciding the relevance degree. 
The final relevance score for a formula is the \emph{maximum} relevance score for any judged instance of that formula.  

While the official evaluation using visually distinct formula pools, ARQMath introduced the use of ``Big Qrel Files'', where nearly all assessment data is provided. This includes assessment for each individual formula instance, along with assessor ID. 
This can be used to study effectiveness of formula search models, under assessment of different people, and to \emph{change} how final relevance scores are defined (\eg using average rather than maximum relevance scores).   

ARQMath organizers hired undergraduate and graduate in mathematics or with strong mathematical backgrounds to act as assessors.
Each year, the assessors were trained by a math professor during three training sessions.
The sessions included discussing relevance ratings for practice topics, with the goal of reducing variation in ratings across assessors, and minimizing assessment errors. 
After some discussion between organizers and assessors in ARQMath-1, relevance for retrieved formulas was defined as follows:

\textit{For a formula query, if a search engine retrieved one or more instances of this retrieved formula, would that have been expected to be useful for the task that the searcher was attempting to accomplish?}

\ourtable
{!t}
{
\vspace{-0.2in}
\begin{small}
{
\resizebox{\columnwidth}{!}{
\begin{tabular}{c l l }
\toprule
\sc Score~ & \sc Rating & \sc Definition\\
\midrule
3 & High & Just as good as finding an exact match to the formula query would be\\
2 & Medium & Useful but not as good as the original formula would be \\
1 & Low & There is some chance of finding something useful\\
0 & Not Relevant ~~& Not expected to be useful    \\
\bottomrule
\end{tabular}
}
}
\end{small}
}
{Relevance scores and definitions for ARQMath Formula Search task.}
{tab:arqmath-task2-relevance}

Assessors 
assigned each formula instance in the judgment pool one of four scores as defined in Table \ref{tab:arqmath-task2-relevance}. 
For example, if the formula query was ${\sum{\frac{1}{n^{2+\cos{n}}}}}$, and the formula instance to be judged is ${\sum_{n-1}^\infty {1 \over n^2}}$, the assessors would decide whether finding the second formula rather than the first would be expected to yield good results. To do this, they would review the question post containing the query (and, optionally, the thread containing that question post) to understand the searcher's information need. Here the question post fills a role akin to Borlund's simulated work task~\citep{pia2003iir}, although here the title, body, and tags from the question post are included in the topic and may be used by retrieval systems. 
Assessors also consult the posts where retrieved formula instances come from (these may be question or answer posts), along with the associated thread to see whether the formula would have been a useful basis for a search, \ie
how likely useful content would be found if this or other instances of the retrieved formula were returned by a search engine.

The ARQMath organizers did make one 
change to the way this relevance definition was interpreted for ARQMath-2 and -3. 
ARQMath-1 assessors were instructed during training that if the query and candidate formulas had the same appearance, 
then the candidate was highly relevant. 
For ARQMath-2 and -3, the interpretation of `exact match' was clarified to take the formula semantics and context into account.
 For example, variables of different types would not be considered the same, even if variable names are identical. 
 This means that an exact match with the formula query may be considered not relevant. 
On the other hand, formulas that do not share the same appearance or syntax as the query might be considered relevant. This is usually the case where both formulas refer to the same concept. 
For the formula query $\frac{S}{n} \ge \sqrt[n]{P}$ (ARQMath query B.277), formula $\frac{1+2+3+...+n}{n}\geq \sqrt[n]{n!}$ has medium relevance. Both formulas are referring to the AM-GM inequality (of Arithmetic and Geometric Means).

System evaluation is performed after removing duplicate instances of visually identical formulas from search results, and then calculating effectiveness measures over the ranked series of visually distinct formulas. This is done by replacing each formula instance with its associated visually distinct formula id, and then removing duplicates starting from the top of the ranking. 
%
To avoid earlier issues with unevaluated hits for people using the test collection assessment was complete, the organizers chose the nDCG$^\prime$ measure (read as ``nDCG-prime'') introduced by Sakai~\citep{sakai_alternatives_2007} as the primary measure. The nDCG measure on which nDCG$^\prime$ is based is widely used when graded relevance judgments are available. ARQMath also uses two other measures: Mean Average Precision (MAP$^\prime$), and Precision at 10 (P$^\prime$@10), after removing the unjudged hits. For MAP$^\prime$ and P$^\prime$@10 High+Medium binarization is used, meaning only the medium and high relevance ratings  (2 and 3) were considered relevant.  


\textbf{AccessMath.}
As an example of a very different (albeit small) test collection,
the AccessMath system described in the first chapter \citep{DBLP:conf/icfhr/DavilaZ18} was developed using lecture videos and \LaTeX{} lecture notes produced for those lecture videos.\footnote{Notes: \url{https://www.cs.rit.edu/~dprl/files/TangentV-data_results.zip},\\ Videos: \url{https://www.cs.rit.edu/~accessmath/am_videos}} 

\section{Formula retrieval models}

We have organized formula retrieval models by the formula representation they use for search.
Some of the representations imply sparse vs. dense representations as noted below. A more detailed discussion of formula representations can be found in Chapter 2. The formula representation types we distinguish here are:
\begin{compactdesc}
    \item \textbf{Text-Based:} Formulas represented by tokens in text encodings (\eg \LaTeX{} tokens). Early systems used this with traditional sparse (\ie inverted index-based) retrieval models such as TF-IDF.
  
    \item \textbf{Tree-Based:} Use formula tree representations (\eg SLT and OPT). Retrieval is performed over sparse tuple indexes for substructures (\eg paths, subexpressions) and/or (re-)ranking by tree edit distance or graph alignment.
  
    \item \textbf{Visual-Spatial:} Captures formula appearance symbolically without writing lines (\eg in SLTs) or operation-argument relationships (\eg in OPTs).
  
    \item \textbf{Embedding:} Text, tree, or visual-spatial representations for formulas and/or subexpressions are embedded in vector spaces.  Nearest-neighbor search using vector similarity identify candidates. 
    
    \item \textbf{Other:} Use images or other representations not described above.
\end{compactdesc}

\begin{myexample}{Formula comparison in different representations.}{ex:overviewFormulaSearch}
\centering
\scalebox{0.56}
	{
		\hspace{-0.3in}
		\includegraphics{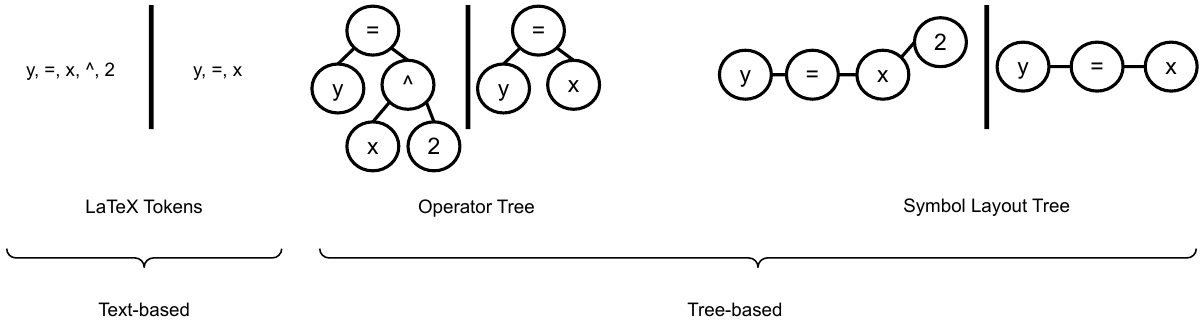} 
	}
	~\\
\scalebox{0.6}
	{
		\hspace{-0.3in}
		\includegraphics{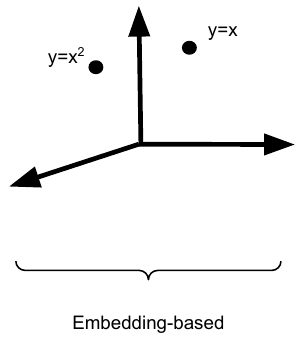} 
	}
\end{myexample}


 \refexample{ex:overviewFormulaSearch} compares the formulas  $y=x^2$ and $y=x$ in different representations. 
At top-left we see a text based representations, where we have two lists of tokens produced by linearizing SLTs represented in  \LaTeX{}. 
At top-right  we have operator and symbol layout trees at left and right respectively. 
Finally, at bottom we see the formulas represented as vectors (points) in a 3d embedding space. Such a representation can be created from text, tree, or other representation. 
The specific positions of vectors depend upon training data and the training tasks, learning algorithms, and loss functions used for embedding.


Table \ref{tab:sampleResults} show the first formula returned by models using different formula representations.
Each model has a `reasonable' first hit. 
While it is helpful for identical or near-identical formulas to be highly ranked, 
matching identical/nearly-identical formulas is easy for reasonably expressive representations and indexing patterns.

Effectiveness-wise, the ability to capture relevance for formulas that are progressively more distinct from the query is  what differentiates most formula search  models.
This is one of the reasons that the ARQMath test collections used full rank metrics for ranking formula search systems (\ie $nDCG'$, with $mAP'$ added for comparison), in addition to observing metrics focused on the top of a ranking (\eg $P'@10$, $mRR$).


A summary of formula search models is provided in Table \ref{tab:formulasearchmodels}. In the remainder of this chapter, we will discuss the families of formula search models based on their representation types.

\ourtable
{!t}
{
\vspace{-0.2in}
\begin{small}
\resizebox{1\textwidth}{!}
{
\begin{tabular}{l  l |  l l l}
    \toprule
    & & \multicolumn{3}{c}{\textbf{Query}}\\
    & &\\
    \bf Model & \bf Repr. & \boldmath{$Cov(x,y) = 0$} & \boldmath{$\int_{0}^{1}\frac{\sin^{-1}(x)}{x}$} & \boldmath{$x! = \sqrt{2\pi x} * (\frac{x}{e})^x$} \\  
    \midrule 
    \textbf{MathDowsers} &Tree paths & $COV (X,Y) = 0$&$J=\int_0^a \frac{\sin ^{-1}(x)}{x}\,dx$ & $ x! \approx \sqrt{2 \pi x} \cdot \left(\dfrac{x}{e}\right)^x$\\ 
    & (SLT) &\\
    \textbf{Approach0} & Tree paths & $Cov(x,y) = 0$&$\int_{0}^{1} \frac{\sin^{-1}(x)}{x} dx = \frac{\pi}{2}\ln2
$ & $ \large x! \sim \sqrt{2\pi x}(\frac{x}{e})^x
$\\
	& (OPT) & \\
    \textbf{Tangent-CFT} & Tree embed. & cov(y,x) = 0&$\displaystyle\int_1^\infty\dfrac{\sin^2(x)}{x}$& $ n!=\sqrt{2\pi x} \left({\frac{x}{e}}\right)^x$\\
    & & \\
    \textbf{XY-PHOC} & Visual-spatial& $Cov (x,y) = 0$&$\int_{0}^{1} \frac{\sin(x)dx}{x}$&$ n!=\sqrt{2\pi x} \left({\frac{x}{e}}\right)^x$\\
  \bottomrule
\end{tabular}
}
\end{small}
}
{The first formula returned by different search engines for three queries of increasing structural complexity.}
{tab:sampleResults}


\ourtable
{!t}
{
\vspace{-0.2in}
\begin{small}
\resizebox{\textwidth}{!}
{
\begin{tabular}{l   c   c  c  c | c  c  |   l | l}
\toprule
& \multicolumn{4}{c}{\textbf{Representation}} & \multicolumn{2}{c}{\textbf{Canonicalization}} & \\
 \textbf{Model} &    SLT &  OPT & Others & Context & Unif. & Norm. & \textbf{Rank}& \textbf{References}\\

\hline
~\\
\multicolumn{8}{l}{ \bf Text}\\
\hline

DLMF & &  &  \checkmark&&&\checkmark& TF-IDF & \cite{miller_technical_nodate}\\
ActiveMath&  & &\checkmark &\checkmark&& & Tokens& \cite{libbrecht_methods_2006}\\
MathDex  & \checkmark &  & & & \checkmark& \checkmark&TF-IDF& \cite{miner2007approach}\\
EgoMath   & \checkmark&  &  &\checkmark& \checkmark& \checkmark&TF-IDF&\cite{misutka_extending_nodate}\\
MIaS & \checkmark &&&\checkmark&\checkmark&\checkmark&TF-IDF&\cite{sojka_indexing_2011}\\
LCS & & & \checkmark & & \checkmark&\checkmark& LCS.&\cite{pavan_kumar_structure_2012}\\

\hline
~\\
\multicolumn{8}{l}{ \bf Tree}\\
\hline

MathWebSearch&&\checkmark&&&&\checkmark&Paths&\cite{kohlhase_search_2006}\\
WikiMirs&  & \checkmark & \checkmark&  \checkmark&\checkmark&\checkmark&TF-IDF&\cite{hu_wikimirs_2013}\\
SimSearch &\checkmark &&& &  & &TED&\cite{kamali_retrieving_2013}\\
MCAT &\checkmark& \checkmark&&\checkmark&\checkmark& \checkmark&Paths&\cite{kristianto_mcat_2016}\\

Tangent-3 &  \checkmark& & &\checkmark& & &Paths&\cite{zanibbi_multi-stage_2016}\\
Tangent-S &  \checkmark& \checkmark& &\checkmark& & &Paths&\cite{davila_layout_2017}\\
Tangent-L&\checkmark&&&\checkmark&&&BM25+&\cite{fraser_choosing_2018}\\
Approach0 & & \checkmark& &\checkmark& \checkmark& \checkmark&Paths&\cite{zhong_structural_2019}\\ 
MathDowsers&\checkmark&&&\checkmark&\checkmark&\checkmark&BM25+&\cite{ng_dowsing_nodate}\\
Tangent-CFTED & \checkmark & \checkmark & & && &TED& \cite{mansouri_dprl_2020}\\
\hline
~\\
 \multicolumn{8}{l}{ \bf Embedding}\\
\hline
SMSG5 &\checkmark&&\checkmark&\checkmark&\checkmark& &Cosine&\cite{thanda_document_2016}\\
Formula2vec &  & &\checkmark&\checkmark&& &Cosine&\cite{gao_preliminary_2017}\\
EqEmb.& \checkmark & & &\checkmark& & &Cosine&\cite{krstovski_equation_2018}\\
Tangent-CFT & \checkmark & \checkmark & &  & \checkmark& &Cosine&\cite{mansouri_tangent-cft_2019}\\
NTFEM & & & \checkmark& && &Cosine&\cite{dai_n-ary_2020}\\
Semantic Search&\checkmark&&&&&&Cosine&\cite{pfahler_semantic_2020}\\
Forte & & \checkmark & &  && &Cosine&\cite{9671942}\\
MathEmb & &\checkmark &&\checkmark&\checkmark &\checkmark&Cosine&\cite{song_searching_2021}\\
MathBERT & & \checkmark & & \checkmark& & &Cosine&\cite{peng_mathbert_2021}\\ 
MathAMR & & \checkmark & &\checkmark& & &Cosine&\cite{10.1145/3511808.3557567}\\
\hline
~\\
\multicolumn{8}{l}{ \bf Visual-Spatial}\\
\hline
Tangent-V& &  & \checkmark& & & &Tokens&\cite{davila_tangent-v_2019}\\
XY-PHOC& &  & \checkmark& & & &Cosine&\cite{avenoso_spatial_nodate}\\
EARN& \checkmark& \checkmark & \checkmark&& && K-NN&\cite{ahmed_equation_2021}\\
\hline
~\\
\multicolumn{8}{l}{ \bf Other}\\
\hline
TanAPP&\checkmark &  \checkmark&&&\checkmark&\checkmark& Ens.&\cite{mansouri_tangent-cft_2019} \\
Math-L2R&\checkmark &  \checkmark& & &\checkmark&\checkmark&SVM$^{rank}$& \cite{mansouri_learning_2021}\\
MathAPP&&  \checkmark&&\checkmark& \checkmark&\checkmark&Ens.&\cite{peng_mathbert_2021}\\
FORTEAPP&&  \checkmark&&& \checkmark&\checkmark&Ens.&\cite{9671942}\\
\bottomrule
\end{tabular} 
}
\end{small}
}
{Formula Search Models. `Others' representations includes images.}
{tab:formulasearchmodels}


\section{Text-based and tree-based models}
It is common to use traditional sparse retrieval models for more complex domains such as math.
Particularly 
in the early days of formula search, traditional token-based sparse models such as TF-IDF were used. 
An example is one of the earliest formula large-scale formula search engines created for the Digital Library of Mathematical Functions (DLMF) \citep{miller_technical_nodate}. 
\LaTeX~is parsed into an SLT-type tree,
which 
is then linearized after normalizing token symbols.
Normalizations  include converting symbols to text tokens, and mapping non-alphanumeric characters to alphanumeric strings. 
For instance, $x^{t-2}=1$ given as  `{\tt x$\wedge$\{t-2\}=1}' is converted to the token sequence: 
$$x,~~BeginExponent,~~t,~~minus,~~2,~~EndExponent,~~Equal,~~1.$$ 
A second normalization is canonical orderings: for commutative operations where argument order is unimportant (\eg multiplication and addition), a fixed ordering is produced using the lexicographic order of argument tokens. 
After linearization, DLMF creates an inverted index used with TF-IDF scoring of query tokens in the same manner as text.\footnote{text and formula tokens are stored and retrieved from the same index, using a unified token representation.}

Approaches like DLMF later included additional canonicalization steps. 
For example, EgoMath \citep{misutka_extending_nodate} canonicalizes argument ordering, and enumerates variables and constants, using identical symbols to capture variable repetitions.\footnote{See Chapter 2 for discussion of symbol enumeration.} 
For constants, formula $74+a^2+b^2$ is also indexed as $const+a^{const}+b^{const}$. With variable normalization, formula $a-b$ is also indexed as $id_1-id_2$. Other normalizations such as removing brackets using distributivity rules are also applied. The goal in these normalizations is to increase recall by increasing the number of formulas with similar token representations.

Aside from sparse retrieval, some other approaches such as using the Longest Common Subsequence (LCS) of a string \citep{pavan_kumar_structure_2012} have been used to produce similarity scores.
As before, formulas are canonicalized before applying LCS so that each function, variable, and number is mapped to a unique token,  and constants and variables are enumerated. 

\paragraph{Tree-based models.}
As discussed in Chapter 2, we normally use graphs to represent structured data, and specifically for formulas, trees to capture a hierarchy of writing lines in SLTs, and a hierarchy of mathematical operations and arguments in OPTs. 

Tree-based approaches can be categorized into two main groups: part-based, and full-tree matching. One of the earliest part-based models is MathWebSearch \citep{kohlhase_search_2006} that relies on subexpression indexing used originally to unify terms in theorem provers (substitution indexing trees \citep{10.1007/3-540-59200-8_52}).
Using operator trees, relationships between progressing more concrete formulas are presented by a series of variable substitutions. 
A search for expressions with similar operator structures and operands starts from the lowest-precedence operators. 
Nodes in the substitution indexing tree correspond to expressions with common structures at the top of their operator trees. 
Moving from the root to the leaves of the substitution tree yields increasingly concrete expressions (\ie after more variable replacements).

Another category of tree-based models represent formula tree substructures. 
The Math Indexer and Searcher (MIaS) 
\citep{sojka_indexing_2011,  ruuvzivcka2016math} 
system uses Presentation MathML, encoding subtrees as compact strings. 
For example, $a+b$ is represented by $math~(mi(a)~mo(+)~mi(b))$.
A similar approach that uses subtrees of differing structural complexity is
WikiMirs \citep{hu_wikimirs_2013}.
WikiMirs creates  terms (patterns) for search from SLTs by recursively replacing subexpressions with wildcards.
For example, the formula $(x+3) \times \frac{a}{b}$, is tokenized into 4 concrete terms, and four generalized terms with wildcards for argument subexpressions:
\begin{equation*}
\begin{array}{l l l l l}
 \mathbf{Concrete~terms:} & \{~ (x+3) \times \frac{a}{b}, & (x+3), &  \frac{a}{b}, & x+3 ~\} \\
 \mathbf{Generalized~terms:} & \{~ (*) \times *, 	& (*), 	& \frac{*}{*},   & *+* ~\}
 \end{array}
 \end{equation*}
Term construction is performed recursively 
until no new terms can be produced. 
Unique tokens are enumerated, and then used to create an inverted index that is searched using TF-IDF.
This system was later extended, incorporating text keywords and using operator trees \citep{gao2016math}. 


    

MCAT \citep{kristianto_mcat_2016} improved part-based retrieval by encoding path and sibling information in symbol layout and operator trees. Tuples capturing tree paths are used for retrieval patterns in an inverted index. In addition to the path-based lookup, this model also uses a hashing-based formula structure encoding scheme, and also includes text at three levels of granularity. The first level considers words around a formula within a context window of size 10, along with descriptions and noun phrases in the same sentence as the formula. The second level includes all words from the paragraph where the formula appeared. At the third level, the title, abstract, keywords in the document, descriptions of all the formulas, noun phrases, and all words in the document are considered.
A formula query combines lookup up in multiple inverted indexes for both formula and text representations.
This was perhaps the first model to capture surrounding context for formulas in a detailed manner.

\textbf{Tangent.}
Tangent-3 \citep{zanibbi_multi-stage_2016} is a two-stage part-based retrieval model. 
From an SLT, path tuples are generated in the form of ($s_1$, $s_2$, R, \#) with parent symbol $s_1$, child symbol $s_2$, the spatial relationship sequence R from s1 to s2, and a count used to capture repetitions (\#).  These tuples are used to identify an initial set of top-k candidates using a sparse bag-of-words model, scoring by F1 (\ie harmonic mean of tuples matched on the query and a candidate formula, also known as the \emph{dice coefficient}). 
Top candidates are then re-ranked using full-tree matching, aligning the query SLT to each candidate SLT. 
After alignment, each top-k candidate is scored using the harmonic mean of symbol and relationship recall (the \emph{Maximum Subtree Similarity (MSS)}) and two tie-breakers: symbol precision after unification, and symbol recall without unification.

The symbol layout tree representation developed for Tangent-3 has been used in a number of retrieval models.
The model includes a container object for matrices, tabular structures, and parenthesized expressions, as well as explicit whitespace, and variable and operation types attached to names (\eg $N!x$ for the number $x$).\footnote{See \citep{zanibbi_multi-stage_2016} for details.}
The Tangent-S model later included retrieval using both symbol layout and operator trees (Tangent-S  \citep{davila_layout_2017}). 
In operator trees, commutative and non-commutative operators have node type (U!) and (O!) for unordered and ordered operations, respectively.
Tangent-L \citep{fraser_choosing_2018} improved retrieval results further through richer indexing patterns/features, and scoring with language statistics using the BM25+ \citep{lv2011lower} model.

The Tangent-L tuple generator was later re-implemented in the MathDowsers system \citep{ng_dowsing_nodate, ng2021dowsing, kane2022dowsing}.
The new generator adds additional patterns for repeated symbols, and additional normalizations. 
Normalization rules are defined to support operation (`semantic') matches. 
For example, for commutative operators ($A+B$, $B+A$) and symmetry ($A=B$, $B=A$) the order of adjacent symbols is ignored. Using a canonical symbol for operator equivalence classes, the model also canonicalizes alternative notations ($A \times B$, $AB$), operator unification ($A \prec B$, $A < B$), and inequality equivalence ($A \leq B$, $B \geq A$).   
This captures OPT-type relationships in an SLT representation.

 \textbf{Approach0.}
Approach0 is a state-of-the-art formula retrieval model that uses OPT leaf-root paths in an inverted index within a two-stage model for retrieving operator trees \citep{zhong_structural_2019}. 
An illustration of OPT leaf-root paths is shown in \refexample{eg:paths}.
 OPTs are generated from \LaTeX~using a small but robust expression grammar.
 To boost recall, variable enumeration is applied. 
 Like Tangent-3/-S, retrieval is performed in two steps. 
 Candidates are first retrieved using matching leaf-root paths in a sparse index, and then
re-ranked using matches of up to three largest common subtrees identified via dynamic programming.
 Similarity is scored by a weighted sum of matched leaves (operands) and operators from the common subtrees. 
In the later version of this system text context is used \citep{zhong2021approach}. 
A textual similarity score is produced using Lucene BM25, and formula structure-based scoring uses the IDF of paths and symbol similarity.
These scores are combined in a linear combination.

\textbf{Full tree matching and tree-edit distance (TED).}
In addition to tree alignments used for reranking in Tangent-3/-S and Approach0,
full-tree matching from tree-edit distances (TED) have been used.
Tree edit distance generalizes string edit distance, defined by the number of operations needed to convert one tree to the other. 
The SimSearch model uses tree-edit distance (TED) on SLTs directly as the similarity measure \citep{kamali_retrieving_2013}.
Three editing operations are used: insertion, deletion, and substitution.
\refexample{ex:treeedit} shows operations converting the SLT for $x^2-y$ to $x+y^2$. 
Accelerations such as cost-based pruning of candidates and caching sub-trees can be used.
In SimSearch operation costs are defined using the similarity of node labels, a node's parent’s label, and whether they are leaf nodes. 
The final ranking is the inverse edit distance normalized by tree sizes, as given in Equation \ref{eq_treeedit}.
\begin{equation}
    sim(E_1, E_2) = 1 - \frac{dist(T_1, T_2)}{|T_1|+{|T_2|}}
    \label{eq_treeedit}
\end{equation}
    
\begin{myexample}{Converting SLT $x^2-y$ to $x+y^2$ in three edits:\\ Delete $2$, replace $-$ by $+$, and add $2$ as superscript of $y$.}{ex:treeedit}
\centering
\scalebox{0.9}
	{
		\hspace{-0.0in}
		\includegraphics{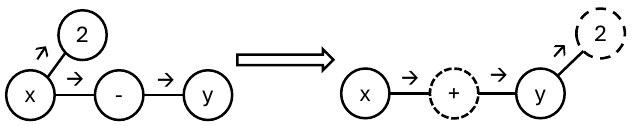}
	}
\end{myexample}

    
Tangent-CFTED  reranks results from a path-based dense retrieval model using tree-edit distances \citep{mansouri_dprl_2020}. 
Unlike SimSearch where edit operation weights were defined using heuristics, here weights are learned for each edit operation. 
The model uses inverse edit distances for scoring, as shown in Equation \ref{eq_treeedit2}.\footnote{1 is added to the denominator to avoid division by zero.} 
Tangent-CFTED uses both symbol layout and operator trees, and the final ranking score is a weighted combination of individual rank scores. 
\begin{equation}
    sim(E_1, E_2) = \frac{1}{TED(T_1,T_2)+1}
    \label{eq_treeedit2}
\end{equation}

\section{Dense retrieval with formula tree embeddings}
As for information retrieval and natural language processing in general, researchers working in math IR turned to embedding models to avoid the types of \emph{vocabulary problems} that traditional sparse model have, and to make greater use of context in patterns used for matching, as discussed earlier.
 Early text embedding models such as Word2Vec \citep{mikolov2013efficient} produced revolutionary results for text problems. These models were then extended to graph data types. The earliest approaches to graph embddings were simple: linearizing a graph using different traversals, treat nodes as tokens, and then apply Word2Vec. 
 Graph embedding models of this type include Node2Vec \citep{grover2016node2vec} and DeepWalk \citep{perozzi2014deepwalk}.

SMSG5 was the first known embedding model for math formulas, and used for re-ranking text-based sparse retrieval results \citep{thanda_document_2016}. 
For first-stage retrieval formulas in Presentation MathML (SLTs) are linearized and indexed as keywords in a sparse index using ElasticSearch. 
For re-ranking the doc2vec embedding model was used \citep{pmlr-v32-le14} to covert
binarized expression trees into real-valued vectors. 
Each operator and its operands are treated as tokens for the doc2vec model, and linearized using an in-order traversal. 
If the operand is a subexpression rather than a symbol, a token identifier for the subexpression is used. 
\refexample{ex:formula_SMSG5} shows tokens extracted for formula $x^3+(1+x)^2$. 
The tuples/patterns produced are similar to those used for other tree part-based retrieval models described above,
and includes variable and subexpression enumeration.  
 The final similarity score for reranked formulas is the cosine similarity of the query and candidate vectors.

\begin{myexample}{Binary OPT for $x^3+(1+x)^2$ with tokens generated for operator nodes. MiAS, SMSG, WikiMIRS and other models index similar patterns.}{ex:formula_SMSG5}
\centering
\scalebox{0.65}
	{
		\hspace{-0.0in}
		\includegraphics{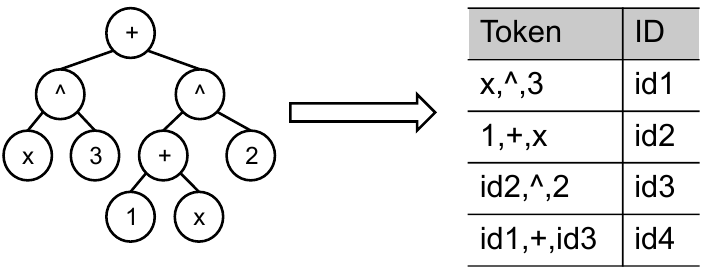}
	}
\end{myexample}


Tangent-CFT \citep{mansouri_tangent-cft_2019} was the first dense retrieval model to use both symbol layout and operator trees for formula embeddings, using an approach similar to SMSG5. 
Tangent-S is used to generate OPT and SLT path tuples.
Canonicalization is performed using enumeration of variables and constants, and in later versions, operator types.
The modified tuples are enumerated and  grouped into n-grams, which are then embedded individually. 
The novelty of this model lies in using an n-gram embedding model, fastText \citep{bojanowski_enriching_2017}.
This approach is better suited for queries not seen in the collection, as it represents formulas using subexpressions.
n-gram vector representations for formulas were averaged to obtain the final embeddings used for retrieval, with ranking by cosine
similarity.
 In the early version of this model, the vectors of different representations (SLT, OPT, Unified SLT) were averaged to get the final vector for a formula. Later, this was converted to combining retrieval results from each representation using a modified Reciprocal Rank \citep{mansouri_dprl_2020}. 
  A similar approach was applied on other representations such as N-ary trees  in the N-ary Tree-based Formula Embedding Model (NTFEM \citep{dai_n-ary_2020}).

\ourfigscaled
{!t}
{0.55}
{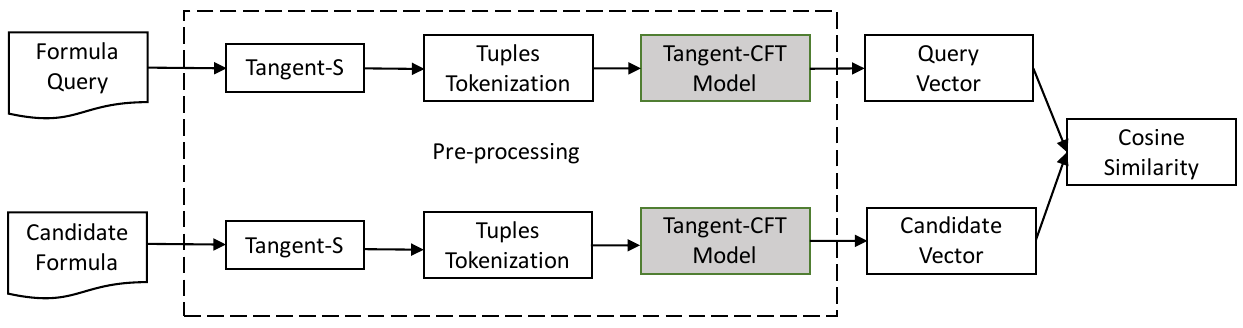}
{Retrieval with Tangent-CFT model. Query and candidate formulas are passed to the same pre-processing pipeline to extract their vector representations. The cosine similarity between these two vectors is the similarity score.}
{fig:formula_cft}

The approaches seen so far for embedding use linearized tree representations and apply sequence embedding models. As formulas are more naturally represented as trees, graph convolutional neural networks are well-suited to formula embedding. The Semantic Search model generates graph representations from Presentation MathML (SLTs), and derives context features from tags, attributes, and text \citep{pfahler_semantic_2020}. These features were then used to represent nodes as one-hot encoded vectors. A graph convolutional neural was trained using two unsupervised tasks: 1) a contextual similarity task where labels are generated from the surrounding contexts of mathematical expressions, 2) a self-supervised masking task. Other graph-embedding approaches have considered using symbol layout and operator trees. For example, MathEmb  uses operator trees with a Graph Convolution Network, Graph SAmple and aggreGatE (GraphSAGE), and a Graph Isomorphism Network \citep{song_searching_2021}.

EARN \citep{ahmed_equation_2021} is a multimodal embedding model that takes advantage of both image and graph formula representations.  An image encoder uses a formula image rendered from \LaTeX{} passed to a ResNet \citep{He_2016_CVPR} model, followed by a Bi-LSTM to produce an image embedding. For the graph representations, a message-passing-based graph encoder is used. The distances between graph-based and image-based embeddings are used as patterns for retrieval. The visual and graph-based similarities are combined using a  linear combination in a manner similar to the Tangent-S system.

Encoder-decoder architectures have also been used for formula search. Similar to NLP tasks, reconstruction  (also known as the `fake task') where a formula must be decoded from an embedded vector  can be used to train these architectures.
After training, only the formula encoder is needed for embedding formulas. 
FOrmula Representation learning via Tree Embeddings (FORTE)  uses this architecture by taking an operator tree as input, generating the vector embedding, and then reconstructing the formula in the decoder \citep{9671942}. 
The encoding process is shown in \refexample{ex:forte}.
On the encoder side, trees are traversed depth-first, with each node represented by an embedding. 
To preserve formula structure, a positional encoding in a fixed-length vector is concatenated to each node embedding.
The positional vector represents the binary branching path from the root to a node in the tree. 
On the decoder side, this model uses a novel tree beam search generation algorithm to reconstruct a slightly different version of the input tree, with attached `end' nodes.

\begin{myexample}{FORTE encoding process for formula $x=2x-4$.}{ex:forte}
\centering
\scalebox{0.465}
	{
		\hspace{-0.47in}
		\includegraphics{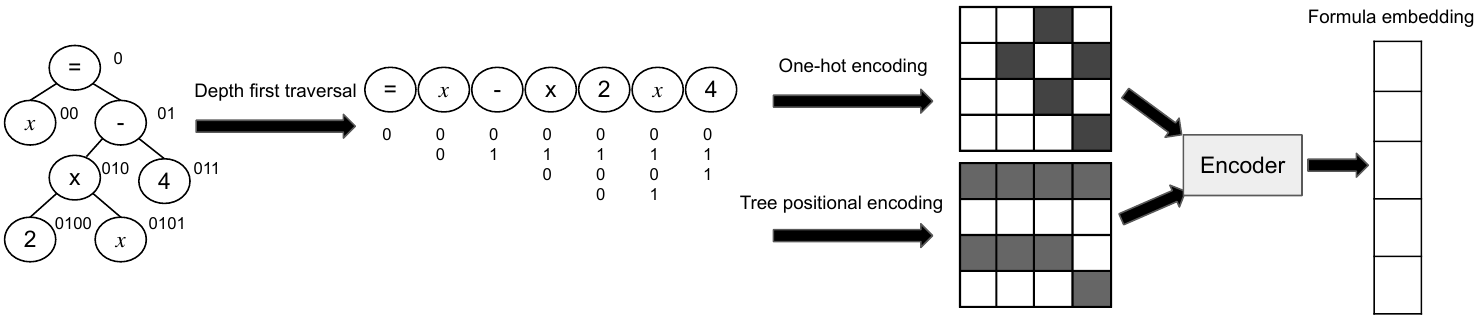}
	}
\end{myexample}

%
Some embedding models include the textual context surrounding a formula. 
Early attempts include an embedding model  generating embeddings for words and formulae using textualized formulas \citep{krstovski_equation_2018}. Linearization is done using SLT tuples from the Tangent-3 system, after which
Word2Vec is used within a larger context window size for formulas than words. 
 A similar idea was adopted in the early days of  BERT transformer models
 \citep{devlin-etal-2019-bert}, which were trained on general text. 
Later models brought attention to the need for specific tokenizers for math.
Note that directly fine-tuning a BERT-based model for formula search is not ideal, as their tokenizers (byte pair encoding or WordPiece) are trained using general text, and may not handle formulae correctly (\eg by splitting \LaTeX{} commands). 

MathBERT pre-trains a BERT model using two tasks: Masked Language Modeling and Context Correspondence Prediction \citep{peng_mathbert_2021}. 
For formula search, they use a Masked Substructure Prediction task as their masking task, with masked structures representing an operator along with its parent node and child nodes in an operator tree. 
During training, the input to MathBERT includes formula \LaTeX~tokens, context, and operators:
\begin{quote}
    \begin{center}
   {\tt  [CLS]~~\LaTeX~~[SEP]~~Context~~[SEP]~~(OPT Nodes)}
    \end{center}
\end{quote}    
with [CLS] and [SEP] defined as special tokens. To further incorporate structural information from the operator tree, this model modifies the attention mask matrix, leveraging the edges between nodes in the operator tree.

Rather than linearize formulas to produce a unified formula and text representation,
MathAMR \citep{10.1145/3511808.3557567} uses Abstract Meaning Representation (AMR) graphs \citep{banarescu2013abstract}to produce a structured unified representation. The process to get the unified tree representation is shown in \refexample{eg:amr}. First math formulae are enumerated and replaced with a special token `EQ:ID' (where ID is an enumeration). Then, an AMR parser produces an AMR tree, after which the root of the formula OPT is inserted at the enumerated formula node. OPT edge labels are modified to be consistent with AMR conventions. A special edge label `math' is added to the set of AMR edge labels to indicate a math formula. 
In the first version, despite having a unified tree representation, the AMR tree is linearized and then used to fine-tune a Sentence-BERT model. Unfortunately the linearization loses structural information, and Sentence-BERT tokenizer may not be well-suited to the AMR annotations.



\section{Visual-spatial models and formula autocompletion}

Many math formulas in digital libraries are represented in PDF documents. Even for online resources such as Wikipedia, formulae are often represented by images rather than Presentation MathML or \LaTeX{}. 
Formulae are also often represented as images in videos, handwriting, and slide decks. 
Formulas may be recognized and converted to \LaTeX{} for use with retrieval models that we have discussed.
An alternative approach is to search using visual-spatial representations of formulae that require symbols but no representation of writing lines or operation hierarchies. 

Tangent-V  retrieves mathematical formulas and other graphics in PDF and PNG images \citep{davila_tangent-v_2019}. Built on top of the Tangent-S system, Tangent-V utilizes symbol pairs extracted directly from images: for PDFs, symbols are taken directly from the file, and for PNGs, symbols are identified using an open-source OCR system \citep{davila2014using}. 
Line-of-sight graphs are created to capture which pairs of symbols are unblocked by other symbols. 
The visible symbol pairs are indexed with their relative angles in a 2$\frac{1}{2}$D representation to capture symbols inside square roots and other containers. 
For search, candidates with shared symbol pairs are retrieved from a sparse index, and formulas with large differences in displacement angles and/or symbol size ratios relative to the query pairs are filtered.\footnote{This is a Boolean query constraining symbol angles and relative sizes.} 
A re-ranking step aligns matched pairs one-to-one with the query, and then the Tangent-S Maximum Subgraph Similarity (MSS) from Tangent-S scores candidates by the harmonic mean of query node and edge match percentages (F1). 

Another visual-spatial representation is 2d histograms of symbols. 
XY-PHOC  uses a sparse visual-spatial representation for retrieval (see \refexample{eg:appear} \citep{avenoso_spatial_nodate}). 
This representation generalizes a one-dimensional spatial encoding previously used for word spotting in handwritten document images, the Pyramidal Histogram of Characters (PHOC) \citep{sudholt2016phocnet}.
Scoring is done by cosine similarity of the PHOC embedding vectors. 
Formulas are represented by a bag of symbols; for each symbol, a binary vector of 29 elements is generated, where each element corresponds to a region, and 1 represents the existence of that symbol in that region.\footnote{PHOC may be a 2d generalization and/or variation of an unweighted binary independence term model (see \citep{DBLP:books/daglib/0022709}).}

Later work found that using concentric rectangles improved PHOC-based retrieval, and that similar effectiveness is obtained using fewer region partitions, \eg only odd-numbered partitions (\ie 1, 3, 5, etc.) \citep{phocrectangle}. 
As seen in other formula retrieval models, using PHOC for part-based rather than whole formula matching can also improve retrieval effectiveness 
\citep{qproject}. 
The model is surprisingly effective for formula search despite its simplicity; models that incorporate IDF-like information (\eg BM25), SPLADE-like token expansions and dense retrieval have not been properly explored with this representation yet.
Also, because PHOC is domain-agnostic, requiring only a symbol vocabulary, it might provide a simple but effective unified representation for visual-spatial search of text, formulas, and other graphics.

\textbf{Formula autocompletion.}
Query auto-completion (QAC) can help users input queries more quickly, and with formulating queries when they have a specific intent but lack a clear way to express it in words. 
For text queries this helps prevent spelling errors, particularly on devices with small screens. 
It was  reported in 2014 that for English queries, using QAC by selecting suggested completions saved over 50\% of keystrokes for global Yahoo! searchers \citep{zhang2015adaqac}.

Formula auto-completion is employed in search engines like WolframAlpha. This system employs prefix matching for retrieving candidates. Consequently, mathematical expressions that are reordered around commutative operators (e.g., a + b = b + a) or use different symbols than the query are not presented as candidate completions.

%
	
Despite extensive research in general query autocompletion, formula autocompletion remains underexplored. Rohtagi et al. \citep{rohatgi2019query} proposed an approach that uses \LaTeX~strings and considers three methods: exact matching, prefix matching, and pattern matching. MathDeck \citep{diaz_mathdeck_2021} uses TangentCFT \citep{mansouri_tangent-cft_2019} to search a small collection of indexed formulae online as a user inputs a formula, displaying similar formulas.
Both approaches complete the right side of a query assuming that the left side has been entered. For math formulas, entry is not always left-to-right: for example, when writing fractions or integrals.

For autocompletion, XY-PHOC has been used with conjunctive queries where all query symbols must be present in a candidate. 
An additional boolean constraint is also added: that returned formulas must contain no fewer symbols than the query. This is the first known model to allow symbols to be inserted in any order for formula autocompletion, because the XY-PHOC is a spatial representation rather than a tree-based one.
\begin{myexample}{Different entry orders for three symbols in $\int_{0}^{\infty} \frac{sin(x)}{x} dx$ a) left-right, b) right-left, c) outside-in, d) middle-out.}{ex:XYPHOCAutoComplete}
\centering
\scalebox{0.625}
	{
		\hspace{-0.0in}
		\includegraphics{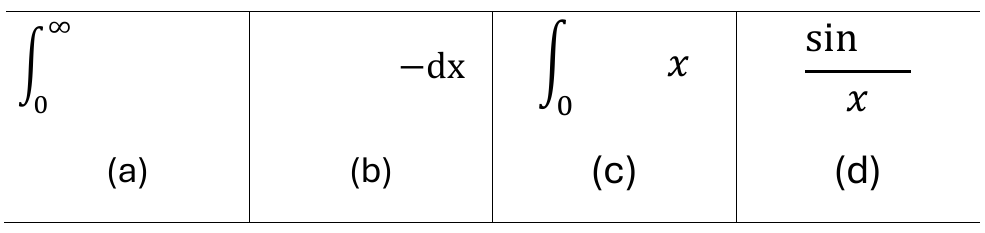}
	}
\end{myexample}
For evaluation,  formula search test collections were used.
Four different symbol entry orders for XY-PHOC were compared, as illustrated in \refexample{ex:XYPHOCAutoComplete}. 
Experiments confirmed the outside-in ordering constrains formula completions most quickly, raising target formulas to the top of the completion list using fewer symbols on average.

\section{Retrieval effectiveness and combining representations} 

In this section we summarize the effectiveness seen to-date for the different retrieval model families presented in this chapter,
using standard formula test collections (see \reftab{tab:tasks-sources}).

While easier to implement, text-based models are generally less effective than other representations for formula search. 
This is because these models do not capture the hierarchical structure of formulas. 
In contrast,
the strongest models from the most recent ARQMath formula search tasks are tree-based models.
Full-tree matching approaches, \eg using tree-edit distance can be time-consuming and seem to be better suited for re-ranking. Also, these models are stricter than path-based models as they match full tree representations. 

Results from the ARQMath-3 formula search task show tree-based approaches obtaining the strongest formula retrieval results. 
Approach0 obtains the highest nDCG$'$ of 0.72.
Interestingly, Approach0 uses only OPTs for its formula representation.
Tangent-CFTED obtains an nDCG$'$ of 0.69 using both SLT and OPT representations.\footnote{Tangent-CFTED is actually a two-stage model using dense embeddings of tree paths for first-stage retrieval, and tree edit distance for reranking. }
MathDowsers uses an SLT-based representation, and also obtains a high nDCG$'$ score of 0.64.
There is also evidence that multi-modal representations can be helpful. For example, the multimodal image + SLT dense model EARN obtained higher Bpref scores compared to the tree-based Tangent-S model that it extends (0.69 vs. 0.64 on the NTCIR-12 Wikipedia Formula Browsing task).

Experimental results suggest that despite providing rich contextual features, dense formula retrieval models may be  better suited for finding similar or partially-relevant formulae than very similar or fully relevant formulas.
The embedding vectors do well at capturing shared contexts, but not necessarily specific symbols and structures in current models.
Given their ability to match similar formulas well, albeit not rank them in the ideal order,
many approaches use embedding-based models to select candidates, and then re-rank the results using similarity scores from tree-based models. Looking at the ARQMath-3 formula search task, Tangent-CFT's first stage dense retrieval obtains an nDCG$'$ of 0.64; after re-ranking with tree-edit distance this increases to 0.69. A similar pattern is seen for MathBERT on the NTCIR-12 formula browsing task: for partial matching the Bpref score is higher than that for reported tree-based models (0.74), but drops to 0.61 for full relevance matching.

For visual-spatial models, the original XY-PHOC had much lower effectiveness than other approaches at ARQMath-3 for full rank metrics (nDCG$'$ of 0.47). However, it was surprisingly competitive in metrics focused on the top of rankings (\eg $P'@10$). Later refinements including using additional levels and rectangular regions increased $nDCG'$ to 0.623, and $P'@10$ to 60.9\%. These measures were respectively 10\% and 8\% lower than the best performing tree-based model at ARQMath-3 (Approach0).
This is interesting because PHOC models use less complex representations and a simple sparse retrieval model, and do not employ statistical weighting or machine learning \citep{phocrectangle}.


\paragraph{Ensembling and learning-to-rank.}
Ensembling and learning-to-rank approaches have been used to combine the benefits of  different representations, and to combine dense and sparse retrieval models. 
TanApp \citep{mansouri_tangent-cft_2019}, FORTE-App \citep{9671942}, and MathApp \citep{peng2021mathbert} are models that use a linear combination of relevance scores from dense embeddings (Tangent-CFT, FORTE, and MathBERT, respectively) with tree-based sparse retrieval and re-ranking (Approach0). These ensemble models provide better effectiveness compared to each of the individual models. 
MathApp and TanApp both obtained higher Bpref values on the NTCIR-12 formula browsing task compared to their component systems.

Using learning to rank models for formula search is underexplored. As mentioned earlier, an early attempt at this was using RankBoost with formula and text features \citep{gao2016math}. A more recent study used only formula features, combining similarity features from tree-based and embedding-based models and training an SVM-Rank model \citep{joachims2006training,mansouri_learning_2021}. The input features in this approach include similarity features from tree paths, full-tree matching, and embeddings. The SVM-Rank weights allowed observing feature importance. Consistent with results seen for the best-performing tree-based retrieval models, it was found that full-tree matching features on both symbol layout and operator trees were among the important features.

\chapter{Math-Aware Search}
\label{c-formula-text-search} 




When we have math information needs for topics that include text,  or are too complex to be addressed by formula search alone, we can use 
\emph{math-aware} search engines supporting queries with both formulas and text. 
Math-aware search tasks range from the simple ad-hoc query ``$a^2+b^2=c^2$ proof'' to complete questions expressed with formulas and text. 
Math-aware search can be understand as a multi-modal extension of traditional text-based search models.

In this chapter we first present test collections for math-aware search.
We then present multi-modal math-aware search models, which either (1) retrieve text and formulas separately and combine relevance scores for individual sources,  or (2) use a unified formula + text representation to retrieve sources directly.
Large Language Models (LLMs) have recently been applied to math-aware search,
including a state-of-the-art model that transforms the problem of retrieving question answers. 
A question answer in text/\LaTeX{} is generated using an LLM, and the LLM answer is then embedded using a unified formula/text representation to search embedded answers rather than the original question. 
The chapter closes with some additional insights related to math-aware search and LLMs.

\section{Test collections for math-aware search}

\ourtable
	{!t}
	{
		\vspace{-0.2in}
		\scriptsize	
		\begin{tabular}{p{1in}   p{0.7in} p{0.7in} p{0.7in} p{0.75in} p{0.5in} }
		
		\toprule
		\textbf{Test Collection}	& \textbf{Sources} & \textbf{Queries} & \textbf{Results} & \textbf{Metrics}\\
		\hline
		


		
		MREC 	\hfill(2011) 	& arXiv papers 		& $--$ & $--$ & $--$\\
		CUMTC 	\hfill(2015) 	& \mbox{arXiv papers} \linebreak (MREC data) & MathOverflow Qs & Papers & mAP\\
		\textbf{NTCIR-10} \hfill(2013)	& arXiv papers & Organizers$^w$ & Papers & mAP, P@\{5,10,hit\}\\
		NTCIR-11 \hfill(2014)	& \mbox{arXiv frags.} & Organizers$^w$ & Paper frags. & mAP, P@\{5,10,hit\}, Bpref\\
		NTCIR-12 \hfill(2016)	& \mbox{1. arXiv frags.} \linebreak 2. Wikipedia$'$ & \mbox{1. Organizers$^w$} \linebreak \mbox{2. Organizers$^w$} & \mbox{1. Paper frags.} \linebreak \mbox{2. Wiki articles} & P@\{5,10,15,20\}\\
		\textbf{ARQMath-1} \hfill(2020)	& 2018 MSE As & 2019 MSE Qs & F in MSE As &  nDCG$'$, mAP$'$, P$'$@10 \\
		ARQMath-2 \hfill(2021)	&  & 2020 MSE Qs & 	 &  \\
		ARQMath-3 \hfill(2022)	&  & 2021 MSE Qs &  	&  \\
		Cross-Math$^\dagger$ \hfill (2024)	&  & \mbox{ARQMath 1-3 Qs} (4 languages) & 	  & P$'$@10, nDCG$'$@10\\
		\midrule
		
		\end{tabular}\\
		\scriptsize
		\textbf{F:} Formula; \textbf{As}: Answers; \textbf{Qs}: Questions; \textbf{MSE:} Math Stack Exchange
		\linebreak
		$^w${\textbf{Wildcard symbols} in at least some query formulas} 
		\linebreak
		 $^\dagger$Cross-modal or cross-language retrieval; \textbf{frags.}: Fragments (roughly paragraphs)
	}
	{Math-Aware Search Test Collections}
	{tab:mas-colls}

As with formula search, currently the primary standard test collections are NTCIR and ARQMath.\footnote{Additional details and comparison of these collections are available \citep{evalarxiv}.}
Test collections for math-aware search are summarized in \reftab{tab:tasks-sources}.

\textbf{MREC and CUMTC:} The first math-aware search collection with annotated formulas we are aware of is the Mathematical REtrieval Collection\footnote{\url{https://mir.fi.muni.cz/MREC/index.html}} (MREC) \citep{livska2011web} data set was introduced. MREC consists of 439,423 scientific documents from arXiv with more than 158 million formulae with MathML annotations. 
Four years later, the Cambridge University MathIR Test Collection (CUMTC) \citep{stathopoulos2015retrieval} built on MREC, adding 160 test topics derived from 120 MathOverflow discussion threads. This was one of the first attempts to use math community question-answering websites for producing real-world topics rather than topics created by shared task organizers. CUMTC topics were selected from question excerpts from 120 threads. These threads have at least one citation to the MREC collection in their accepted answer.\footnote{Answer accepted by the user who posted the question.} The majority of topics (81\%) have only one relevant document, and 17.5\% have two relevant documents. 

\textbf{NTCIR:} NTCIR-10, -11, and -12 used largely the same collections as for formula search tasks described in the previous chapter, consisting of arXiv papers and Wikipedia articles. Sources for NTCIR-10 were complete technical documents, which makes assessment challenging. For NTCIR-11 collection sources were reduced to excerpts (roughly paragraphs) resulting in 8,301,578 search units. NTCIR-12 uses the NTCIR-11 arXiv collection, along with a collection of (full) Wikipedia articles.

The NTCIR collections contained an increasing number of math-aware (formula+text) search topics with assessments for each lab (NTCIR-10: 15, NTCIR-11: 50, NTCIR-12: 29 (arXiv collection) + 30 (Wikipedia collection)).
In NTCIR-11, topics (queries) all had at least one keyword and one formula.
In NTCIR-12, topics were developed for two different collections (arXiv and Wikipedia), 
and all topics contained at least one formula, however 5 arXiv and 3 Wikipedia topic had no keywords. 
All NTCIR math-aware search topics are lab-generated, and only the query is provided with no additional description of information needs and search scenario.
Pooling methods also differ between the different collections \citep{evalarxiv}.


The assessment process in NTCIR-10 for text+formula searches was similar to the formula search task, with the same assessors.  
Relevance was decided from retrieved formulas rather than documents due to their size and complexity.
For each formula, assessors used a graded 0-2 scale, to represent 0: non-relevant (N), 1: partially relevant (PR) or 2: relevant (R) judgements.
Each formula was assessed by one or two assessors. 

In NTCIR-11, assessors were shown the title of the topic, the relevance description, and an example hit (if any) as supplementary information. For this collection, relevance was determined using roughly paragraph-sized sources rather than individual formulas. 
The assessors were undergraduate and students in mathematics for the arXiv topics, and computer science for the Wikipedia topics.
Each hit was evaluated by two students, with their judgements combined. 
Relevance levels were defined the same as for NTCIR-10.
The final relevance score from two assessments assigned R/R and R/PR to relevant (2), PR/PR, R/N, PR/N to partially-relevant (1), and N/N to not relevant (0). 

To evaluate effectiveness, NTCIR-10 uses MAP and P@\{5,10, hit\}. 
In NTCIR-11 used MAP, P@\{5, 10\}, and bpref to accommodate unjudged instances. 
NTCIR-12 reported P@\{5, 10, 15, 20\}. 
In all cases, relevance judgments for 
sources missing from the pools (as can happen for P@hit and MAP) were treated as not relevant.

\textbf{ARQMath:} ARQMath's main task is Answer Retrieval, where Math Stack Exchange (MSE) question posts containing text and formulas are used to search MSE answer posts. 
ARQMath's topics and collection were built as shown in \reffig{fig:ARQMath}. 
All questions and their related answers posted from 2010 to 2018 are provided for training, including roughly 1 million questions and 28 million formulas. ARQMath topics were selected from new questions posted in 2019, 2020, and 2021. 

The answer retrieval task was motivated by three things.
First, short answer posts are easier to assess, as they usually contain at most a few paragraphs, and are organized within question threads.
Second, a query log analysis  showed that the number of question queries was almost 10\% higher for math searches compared to searches on other topics \citep{mansouri_characterizing_2019}.  
Third, question posts act as both queries and information need descriptions.

There are 226 assessed test topics in total (ARQMath-1: 77,  ARQMath-2: 71, ARQMath-3: 78), along with
assessments for additional training topics.
ARQMath also provides all system runs used for pooling.
This provides a way to study different approaches and understand their behavior in greater detail. 
Assessors use four relevance ratings, as defined in Table \ref{tab:ARQMathTask1Relevance}. All relevance ratings organized by topic and assessor may be found in a `big' qrels file available with the test collection.


\ourfigscaled
	{!t}
	{0.7}
	{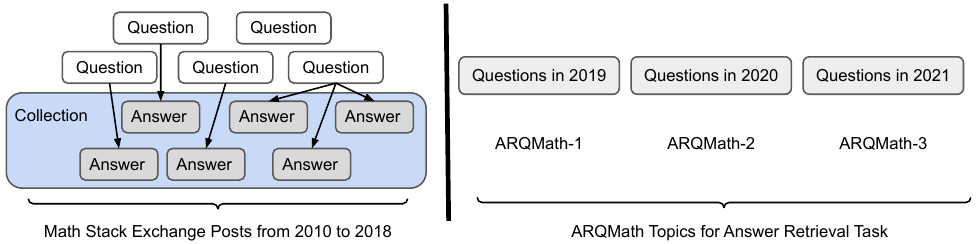}
	{Topics and Collection for ARQMath Answer Retrieval Task. Math Stack Exchange (MSE) question answers from 2010--2018 are the collection searched, while topic questions were posted in 2019 or later. Question and answer posts from 2010--2018 are also provided for training (\ie all posts shown at left).}
	{fig:ARQMath}

To be selected as a topic, a question needs to contain at least one formula.
To diversity topics, 3 categorizations were assigned to candidate question posts, and a stratified
sampling strategy was used to select the final topic sets.
These categories are: 
\begin{compactenum}
\item \textbf{Topic type}: \textit{computation}, \textit{concept} or \textit{proof}
\item  \textbf{Difficulty}: \textit{low}, \textit{medium}, and \textit{high}
\item  \textbf{Representation dependency}: \textit{text}, \textit{formulas}, or \textit{both}
\end{compactenum}


Assessors were selected from students in mathematics, similar to the ARQMath formula search task. 
For each edition, there were 2-3 training sessions with a math professor to introduce the task and train the assessors.
Some questions might offer clues as to the level of mathematical knowledge on the part of the person posing the question; others might not.
To avoid assessors having to guess the level of mathematical knowledge available to the person posing the question, we asked assessors to base their judgments on the degree of usefulness for an expert (modeled in this case as a math professor) who might then try to use that answer to help the person who had asked the original question. 
Finally, for evaluation, the same evaluation measures as the formula search task were used: nDCG$^\prime$, MAP$^\prime$ and P$^\prime$@10. Prime metrics avoid issues with unevaluated hits (see \refchapter{c-eval}).

\ourtable
{!b}
{
\vspace{-0.2in}
\begin{scriptsize}
{
\resizebox{\columnwidth}{!}{
\begin{tabular}{c l p{3.6in} }
\toprule
\sc Score~ & \sc Rating & \sc Definition\\
\midrule
3 & High & Sufficient to answer the complete question on its own \\

2 & Medium & Provides some path towards the solution. This path might come from clarifying the question, or identifying steps towards a solution \\

1 & Low & Provides information that could be useful for finding or interpreting an answer, or interpreting the question  \\

0 & Not Relevant & Provides no information pertinent to the question or its answers. A post that restates the question without providing any new information is considered non-relevant \\
\bottomrule
\end{tabular}
}
}
\end{scriptsize}
}
{ARQMath Answer Retrieval: Relevance Assessment Criteria}
{tab:ARQMathTask1Relevance}




\textbf{Cross-lingual math information retrieval:} 
Current test collections for math-aware search are primarily developed for the English language, limiting their accessibility and inclusivity. Cross-lingual math information retrieval (CLMIR) is a new task, focusing on retrieving mathematical information across languages. CLMIR has been explored for math-word problems \citep{tan-etal-2022-investigating}, where existing datasets were translated into Chinese using online machine translators, and manually refined the translations. CrossMath~\citep{gore2024crossmath} is a novel CLMIR test collection comprised of  ARQMath Answer Retrieval task topics that have been manually translated into four languages (Croatian, Czech, Persian, and Spanish). This collection helps address a research gap in need of filling.
Some approaches proposed for machine translation of mathematical text have been proposed~\citep{9369381_1, petersen-etal-2023-neural}, and we expect more in the future.

\section{Searching with Formulas and Text}
Unlike text or formula search where queries and sources have one representation, for math-aware search queries and sources combine text with  one or more formula representations (\eg SLT, OPT, or \LaTeX{} tokens). 
Because of the different representations (\ie \emph{modalities}), this is a math-specific variation of \emph{multimodal} information retrieval \citep{DBLP:journals/tkde/ZhuZGLYS24,DBLP:journals/mta/ShirahamaG16}.

A key challenge in multi-modal search is bridging the gap between diverse data types such as text, images, videos, and audio \citep{Bozzon2010}. 
There are two main approaches for searching multiple representations: searching modalities separately and then combining rank scores for individual sources  (\ie \emph{federated search}), or by searching a single unified representation for all modalities. 
Below is a summary of approaches for combining formula and text search in sparse and dense search indexes.



\begin{compactdesc}
    \item  \textbf{Sparse retrieval:}\\Formulas are represented using text tokens (\eg in \LaTeX{}) or token sequences annotated on formulas after traversing nodes of formula trees (\eg depth-first traversal of OPT or SLT). These formula tokens may represent individual symbols, or tuples, \eg `+' node in OPT for $x+1$ as operator-prefix tuple $(+,x,1)$, or $(x, 1, \rightarrow \rightarrow)$ for the SLT path from x to 1.
    
    \begin{compactdesc}
    \item \textbf{Federated:} text and each formula representation have their own inverted index. Query text and formula representations are separated before searches are run, and relevance scores are combined to score individual sources. 
    
    \item \textbf{Unified:} text and formula tokens belong to one vocabulary, and a `traditional' inverted index is used to search both together (see Ch. 1). Linearized formula tokens are inserted in source text before indexing, and text/math tokens produced for queries are looked up in the unified inverted index.
    \end{compactdesc}
    
    \item \textbf{Dense retrieval:} \\
    Formulas and/or subexpressions are annotated with vectors in embedding spaces (see Ch. 2). Before embedding, formula representations may be linearized tokens (\eg \LaTeX{}), trees (\eg OPT, SLT) or other representations (\eg PHOC).
    
    \begin{compactdesc}
    \item \textbf{Federated:} each text granularity (\eg token vs. sentence) and formula representation have separate embeddings.  Query text and formula vectors return their nearest-neighbors, and similarity scores are combined to score individual sources.
    
    \item \textbf{Unified:} text and formula elements for sources and queries are embedded in the same dense vector space. Multiple space may be used for different granularity (\eg text+math in passages, vs. individual math/text token embeddings). Vector(s) for sources close to queries in the unified embedding space(s) are used to score sources.
    
    \end{compactdesc}
\end{compactdesc}

\section{Federated search: Combining formula and text results}




With several sparse and dense formula search models available, one approach to math-aware search is combining text search with one or more separate formula searches, and then combining the results to produce scores for matches sources. 
Techniques for combining results include boolean constraints, linearly combining formula and text scores, learning-to-rank, and voting methods. 

\reffig{fig:fusedmathawaresearch} shows a math-aware search model that uses independent searches for text and formulas identify relevant sources (\eg documents or passages).
At indexing time, the \emph{extractor}  is used to separate the formulas and text of sources into two separate search indexes, and formulas are annotated with token sequences for sparse models, or embedding vectors for dense models.
At query time, the same extractor splits a query into sub-queries for text and formulas, and annotates formulas with a token sequence or vector.
The results from both searches are combined into the final score for retrieved sources, and the final result is communicated to the user.

\ourfigscaled
	{!t}
	{0.55}
	{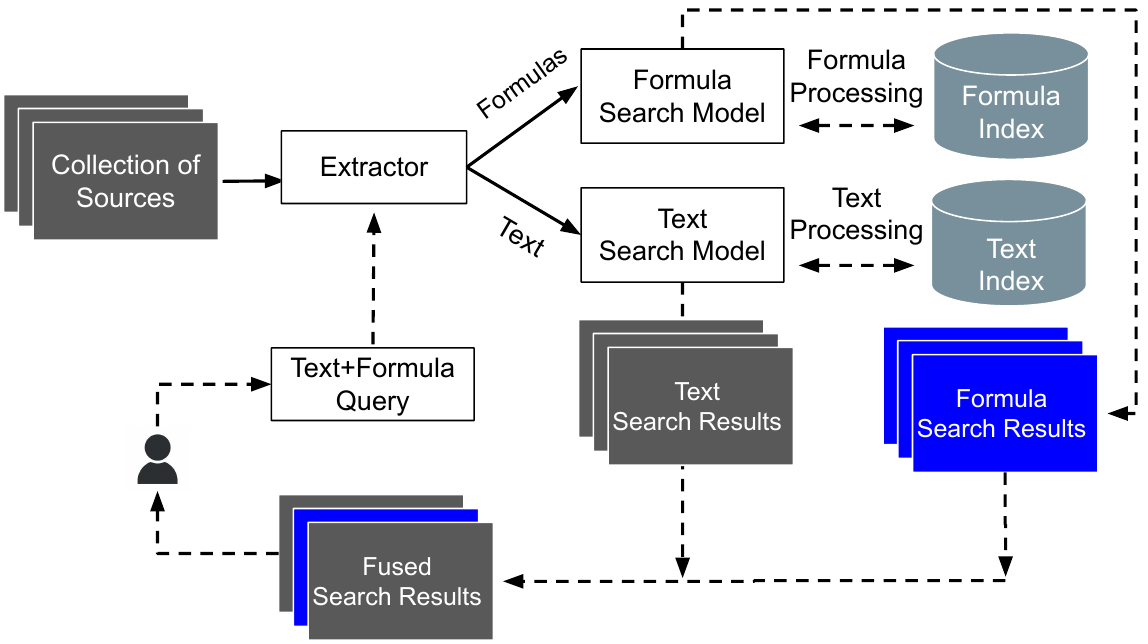}
	{Federated Search for Formulas and Text. Formulas and text are first retrieved and scored independently. Multiple relevance scores for formulas and/or text passages from individual sources are then fused before producing a final ranking. }
	{fig:fusedmathawaresearch}
 
The first approach that we'll consider for fusing independent formula and text searches is using Boolean queries to filter sources that do not contain both formula and text matches.
MathWebSearch \citep{hambasan_mathwebsearch_2014} 
represents formulas as tokens for OPT subexpressions stored in a separate inverted index for formulas  (implemented in ElasticSearch).
Text search results are used to define Boolean queries of the form $(formula_1 \vee ... \vee formula_n) \wedge (term_1 \vee...\vee term_n)$, 
requiring at least one formula and one text token from the query to match a source.
Sources without text token matches are removed from the formula rank score list, which produces the final ranking.
Despite the simplicity of this approach, the system achieved a $P@5$ of 0.79 for the NTCIR-11 math-aware search task (arXiv paper collection). 


The same boolean constraint is used to filter sources that do not contain both formula and text matches in the MIaS system \citep{sojka_indexing_2011}, 
which uses canonicalized tuples as formula tokens (\eg with variable unification; implemented in Apache Lucene).
Rank scores for sub-queries generated from different combinations of text and formula tokens are used to re-rank the remaining candidates multiple ways.
Results from this rankings set were interleaved in the final result to improve the diversity of returned sources \citep{sojka_mias_2018}.


Simple averaging and linear combinations of rank scores have also been used. For the ARQMath Answer Retrieval task three baselines methods were used for answer retrieval:
\begin{compactenum}
\item Tangent-S formula search (linearly combined OPT and SLT scores), 
\item TF-IDF text search, and 
\item average of normalized ($[0,1]$) Tangent-S and TF-IDF scores.
\end{compactenum}
For Tangent-S, the largest formula (SLT) in the question's title was selected, and if no formula was used in the title, the largest formula in the question body was used. The combined model was more effective then the formula or text model in isolation.

Linearly combining formula and text rank scores was used in the MathDowsers \citep{ng_dowsing_nodate} system, where Tangent-L results for formula search are combined with BM25$^+$ text search by scaling and adding relevance scores for sources. For text search, a keyword extraction model was used to select tokens in question answer posts for use in text queries. 
MCAT \citep{kristianto_mcat_2016} also linearly combines formula and text rank scores, but for multiple formula (OPT, SLT) and  text indexes. 
Text is indexed separately at the paragraph and document (title, abstract, keywords, …) levels. 
This model was the most effective for participating teams in the NTCIR-12 math-aware search task (arXiv collection), in part due to the rich variety of formula and text representations.




The WikiMirs system \citep{gao2016math} uses a learning-to-rank approach to combining formula and text scores.
As described in the previous chapter, OPT-like tokens are generated from the SLT representation for a \LaTeX{} string,
using two representations with (1) concrete symbols, and (2) subexpressions replaced by wildcards ($*$).
 Sources are retrieved using inverted indexes for text and formula tokens. These candidates are then re-reranked using RankBoost \citep{freund2003efficient} applied to features focused on  formulas. This system achieved the highest P@K values among the participating teams in the NTCIR-12 math-aware search task (Wikipedia collection). 
 


MaRec (Math answer Recommender)  \citep{10.1145/3624918.3625337} uses the borda count to combine formulas and text scores for answer retrieval.
A Naive Bayes classifier is used to categorize topics in answers and queries/questions  (\eg algebra, geometry, etc.).
 At query time, answers from topics inferred for the query question are selected, and then ranked separately by text and formula similarity.
Text similarity is scores by the Kullback-Leibler (KL) divergence between question/answer token frequency distributions, using
a vocabulary chosen from terms characterizing topics as detected via Dirichlet Allocation (LDA \citep{DBLP:journals/jmlr/BleiNJ03}).
Formula similarity is computed from average SLT tree-edit distances, and a depth score based on the sum of leaf-root OPT path lengths.
The final rank score uses the Borda Count, adding the number of answer posts that rank lower than a source in the formula and text rankings. 








Most systems that combine independent formula and text searches to date use sparse retrieval (\ie inverted indexes) to produce the initial retrieval results.
However some models combine sparse and dense retrieval models.
For example, ColBERT \citep{10.1145/3397271.3401075} has been used for text search, and the similarity scores linearly combined with formula search results produced using from Approach0 (described in the previous chapter).
The MSM team at ARQMath-3 use Reciprocal Rank Fusion (RRF) to combine  sparse TF-IDF and BM25 retrieval models with a 
RoBERTa dense retrieval model.
Final rank scores for sources are computed using Equation ~\ref{eq:rrf}.
In the RRF equation, $R$ is the set of formula and text rankings, and $r(d)$ is the rank of document $d$. 
MSM achieved nearly the same nDCG$'$ as the top participating system at the ARQMath-3 answer retrieval task (0.504 vs. 0.508).

\begin{equation} \label{eq:rrf}
RRF(d) = \sum_{r \in R}{\dfrac{1}{60 + r(d)}}
\end{equation}

\section{Unified formula + text representations}

Combining results from separate indexes for different representations can often produce useful results quickly, especially for sparse retrieval models. However, the text-notation interactions described in Ch. 2 provide important context that is missing when formulas and text are indexed separately. We next consider searching unified representations for formulas and text.

DLMF was the earliest unified sparse retrieval model indexed formula and text tokens together, using a variation of TF-IDF for scoring \citep{ miller_technical_nodate}.
More recently Latent Dirichlet Allocation (LDA) has been used to weight formula and text tokens. 
In these sparse models, formulas are represented as~\LaTeX~tokens \citep{yasunaga_topiceq_2019} or linearized tree tokens \citep{thanda_document_2016}, and a single inverted index is used to retrieve and score sources.
MIaS used a similar approach~\citep{sojka_mias_2018}.

The introduction of the ARQMath answer retrieval task coincided with the emergence of transformer-based dense retrieval models.
A common technique is using a pre-trained transformer (\eg BERT variants) that is fine-tuned using pairs of MathSE questions and answers with their associated assessor relevance ratings.
Reusch \citep{10.1007/978-3-031-13643-6_14} studied dense retrieval using ColBERT and ALBERT models. 
To fine-tune ALBERT, 1.9M triples containing questions with one relevant and one non-relevant answer were fed to the model. 
The model attempts to match assessor scores by classifying answers using the learned vector embedding for the \textit{[CLS]} token that starts each token sequence. 
A similar approach was used to fine-tune ColBERT, but using more relevant and non-relevant answers. 
Somewhat surprisingly, both models proved less effective than sparse retrieval models that participated in the ARQMath-3 shared task. 

In subsequent work, Reusch et al. explored how mathematical formulas affect a transformer model's training \citep{10.1007/978-3-031-56027-9_15}. 
They found that transformer models consider formulas when scoring relevance for answers to a given math question,
but a study of the transformer attention weights for formula and text tokens suggests that structural relationships between formula tokens are lost, and that the attention maps do not capture variables appearing between in both questions and answers. 

This finding motivates creating transformer-based models that can better capture formula structure and interactions.
One approach is adding additional tokens. The MathPredictor model \citep{jo-etal-2021-modeling-mathematical} extends BERT's tokenizers to support  2,651 new tokens. This addresses the BERT WordPiece tokenizer's oversegmention of \LaTeX{} commands such as `\textbackslash overline', which is split into three tokens: \{\textbackslash, over, \#\#line\}. This allows formulas such as $\overline h$ (in~\LaTeX~expressed as \verb|$\overline h$|) to be correctly tokenized as \{\$, \textbackslash overline, h , \$ \}. 
The model was then fine-tuned using masked tokens in formulas.

A hybrid approach that incorporates a unified representation with independent formula and text searches is taken in the MABOWDOR system \citep{10.1145/3539618.3591746}. The PyA0 toolkit is used for preprocessing mathematical formulas before tokenization by WordPiece. 
PyA0 canonicalizes math tokens by merging those likely to be semantically identical, such as \verb|\emptyset|, \verb|\empty|, and \verb|\varnothing|. 1,000 new math tokens are added to the token vocabulary, and sources are then annotated with the formula math tokens, which are  treated the same as regular text in dense embeddings.
The final search combines dense and sparse retrieval, and uses both independent and unified formula+text representations. 
A unified single-vector dense retriever is used for passage-level representations of formulas and text (DPR \citep{karpukhin-etal-2020-dense}).
To take advantage of precise formula symbol and structure matching, formulas are searched independently using Approach0 (combining path-based sparse retrieval with structural alignment). A dense embedding-augmented sparse retriever (SPLADE \citep{splade}) is also used to search the text independently, and these different retrieval results are combined. 

For the MABOWDOR unified representation dense passage retriever, a new pre-training dataset for the math domain was created using Coco-MAE, a retriever architecture and pretraining scheme. 
The pretraining task is Masked Auto-Encoding (MAE), which is similar to masked token pre-training but attempts to decode whole input passages with masked tokens using a decoder. For the final ranking, search results from each component are merged using a convex linear interpolation. This system currently archives the highest P$'$@10 for ARQMath's answer retrieval task.



In an alternative approach, MathBERT \citep{peng2021mathbert} creates a unified formula+text representation using linearized OPT tokens and~\LaTeX~formula representations.
BERT is pre-trained using Masked Language Modeling, Context Correspondence Prediction (similar to next sentence prediction), and Masked Substructure Prediction (for masked formula tokens). 
Using an improved tokenization approach for math formulas, this model archives better effectiveness compared to BERT in tasks such as formula topic generation (predicting the topic (tag) associated with a mathematical formula, using a  TopicMath-100K dataset created from  arXiv papers), and formula `headline' generation  creating a concise description of a formula using the formulas and descriptions in a MathSE question (using EXEQ-300K \citep{Yuan_He_Jiang_Gao_Tang_Giles_2020}).


Researchers then used MathBERT's language model for retrieval and other downstream task such as automatic short-answer grading \citep{zhang2022automatic} using an integer scale from 0 to 4. The MathBERT model is fine-tuned for this task, with pairs of questions and student responses. Additional information including the grade scale and example graded answers are also used for in-context meta-learning. 








\section{Using LLMs for math-aware search}


The use of LLMs to generate question answers and re-rank have been explored using ARQMath's answer retrieval task. Satpute et al. \citep{10.1145/3626772.3657945} considered both general LLMs like GPT-4 and math LLMs like ToRA to generate answers for ARQMath topics.  
These answers were then used as queries to search for relevant answers. 
This can help by transforming the query into a format more similar to sources in the collection, as the query and collection are expressed as answers. 
For each question, an embedding (using {\tt BERT\_cocomae}) of the generated answer was used to search dense embeddings for ARQMath answers. Cosine similarity was used to find the most similar answer in the collection for generated answers. This approach currently obtains state-of-the-art full-ranking results for ARQMath answer retrieval task, with nDCG$'$ of 0.486 (vs. 0.464 of BERT\_cocomae).


In another study, the applications of general LLMs, (LLaMA-2 and Orca-2) were studied for three tasks in math information retrieval: relevance assessment, data augmentation, and point-wise re-ranking \citep{10.1145/3626772.3657907}. This study was done using ARQMath.  For each task, an appropriate \textit{system message} is used; for example, to assess the relevance of a question answer, the system prompt was created based on the ARQMath assessment protocol:
\begin{quote}
\it    You are a math professor who will assess the relevance of an answer to a given math question.
\end{quote}
The results of this study revealed that while general LLMs are not yet suited for relevance assessment or re-ranking, data augmentation from the Orca-2 LLM system 
may be useful for expanding the ARQMath training set for use in fine-tuning neural math answer retrieval systems. 
The data augmentation process is performed by generating additional relevant answers for training topics using LLMs.
We further discuss the use of LLMs for data augmentation in the next chapter. 

%
%

\chapter{Math Question Answering}
\label{c-qa}

Solving math 
problems by computer has been a goal for artificial intelligence research for a long time, beginning with work by Bobrow, Feigenbaum,  Feldman, and Charniak in the 1960's \citep{wpsurvey}. 
More recently, an exciting challenge is the AI Mathematical Olympiad (AIMO)\footnote{\url{https://aimoprize.com/}},
which is awarding a financial prize for the first publicly available AI model capable of winning a gold medal at the International Mathematical Olympiad (IMO).

In this chapter we focus on the rapidly developing area of math question answering.
Work in this area recently spiked with the advent of transformers and Large Language Models (LLMs) trained on large volumes 
of text.
These models may be further trained on mathematical text specifically (\eg with formulas in \LaTeX{}), and can
 convert questions in mathematical prose to formulas, program code, and answers with accompanying explanations.
 Further, they can be used to generate additional training data by reformulating questions and generating question variations, which may
 then be used to retrain and improve a model.
However, there are currently limitations in effectiveness arising from the types of math representations used, and challenges with producing sound mathematical reasoning.

As in the previous two chapters, we will begin with an overview of existing test collections for
math question answering, followed by an overview of systems that have tackled the challenges in these benchmarks. 
A summary of the evaluation metrics used, and comparisons with formula and math-aware search may be found in 
\refchapter{c-eval}.

\section{Test Collections for math QA}

A summary of available test collections is shown in \reftab{tab:collections}.
Math question answering test collections require target answers to be generated or extracted for questions, in order to compute
\emph{accuracy,} the percentage of questions that match the target value for a question. 
These target values are often a numeric quantity, multiple choice alternative, or a list of values.
Some collections also require open responses (\ie free text), in which case text similarity measures are used for evaluation.
Most test collections are either focused on or include \emph{math word problems}, where questions are written primarily with words rather than numbers or equations (see \refexample{eg:ft_tasks}).

\ourtable
	{!t}
	{ 		
		\vspace{-0.15in}
		\scriptsize	
		\begin{tabular}{l  p{1.25in} p{1in} p{0.75in} }
		\toprule
		\textbf{Test Collection} & \textbf{Questions} & \textbf{Answers} & \textbf{Metrics}\\
		\hline

		~\\
		\multicolumn{4}{l}{ {\bf Math Problems} }\\
		\midrule
		Dolphin18K \hfill(2016) 	  & Arithm./algebra  & Number set & Accuracy$^\circ$\\ 
		AQuA-RAT \hfill(2017) & MC$^\triangle$ WP$^\dagger$& \mbox{Rationale} \linebreak \mbox{+ chosen answer} &	Accuracy, \linebreak Perplexity, \linebreak \mbox{Expert ration.} \linebreak{\mbox{+ BLEU}}\\

		SemEval \hfill (2019)	 	&  SAT questions 	& \mbox{Numeric or} \linebreak chosen answer	& Accuracy\\
		MathQA \hfill (2019)		  & MC WP	      &  \mbox{Rationale} \linebreak \mbox{+ chosen answer} & Accuracy\\
		ASDiv \hfill (2020)		  & WP		& Equation and value & Accuracy\\
		MATH \hfill(2021) 	 & AMC, AIME 	& \mbox{Step-by-step soln}  \linebreak \mbox{+ final answer} 	&  Accuracy \\
		GSM8k \hfill (2021)		  & WP & \mbox{Step-by-step soln} \linebreak \mbox{+ final answer} & Accuracy\\ 
		DROP \hfill(2019) 		 & Paragraph questions & \mbox{Number, entity, etc.} & Accuracy, \mbox{token F1},EM\\
		LILA \hfill (2022)		& WP	& Numbers or formulas	& token F1\\
		AIMO \hfill (2024)		& Int'l. Math. Olympiad (IMO) questions 	& Integers in [0,999]	& Accuracy\\

		~\\
		\multicolumn{4}{l}{{\bf Math Problems with Graphical/Visual Content }} \\
		\midrule

		GeoQA \hfill (2021)  & \mbox{MC geometry problems} \linebreak w. diagram & Chosen answer & Accuracy \\
		UniGeo \hfill (2022)  & \mbox{MC w. diagram} \linebreak \mbox{~~}(GeoQA extension) & Chosen answer & Accuracy, \linebreak Accuracy@10\\

		MathVista \hfill (2023)  & Figure QA, WP, geometry problems, textbook questions, VQA & \mbox{Chosen answer or} numeric  & Accuracy\\
		Math-Vision \hfill (2024)  & Math reasoning with visual elements & \mbox{Step-by-step soln} \linebreak \mbox{+ final answer} & Accuracy\\ 
		TabMWP \hfill (2023)	 & MC WP w. tabular data  & Number or  text &  Accuracy\\
		
		~\\
		\multicolumn{4}{l}{\bf Open Response}\\
		\midrule
		ARQMath-3 \hfill (2022) & ARQMath-3 MSE Qs 	& \mbox{Written response} \linebreak \mbox{(formulas + text)} & \mbox{Relevant MSE} \mbox{~~}answers, \linebreak \mbox{+ token F1,} \linebreak+ BERTScore\\
		MathDial \hfill (2023)  & \mbox{LLM-simulated student} \linebreak \mbox{Q. or comment} \linebreak  \mbox{~~}(GSM8k questions) & \mbox{Responses to Q.} \linebreak and comments & \mbox{Expert resp.} \linebreak \mbox{+ sBLEU,} \linebreak \mbox{+ BERTScore}\\
		\bottomrule
		
		\end{tabular}\\
		\scriptsize
		
		 $\dagger$\textbf{Word Problems}; $^\circ$\textbf{Accuracy} = test@1 = \% correct $\approx$ Exact Match (EM)  
		 \linebreak $^\triangle$\textbf{Multiple Choice}
	}
	{Math Question Answering (QA) Test Collections. QA systems communicate one answer which may be a formula, computer program, value (\eg numbers, lists), multiple choice alternative, and/or written responses. Many questions come from standardized tests and math competitions.}
	{tab:collections}


Earlier test collections such as Dolphin18K \citep{huang-etal-2016-well} focus on arithmetic and algebraic problems, with solutions provided only in the form of equations or final answers. 
MathQA \citep{amini-etal-2019-mathqa} contains 37k English multiple-choice math word problems built on AQuA-RAT \citep{ling-etal-2017-program}, with efforts aimed at addressing issues in AQua-RAT including incorrect solutions and problems that required brute-force enumeration. 
Despite the corrections to the AQuA-RAT dataset, around 30\% of MathQA solutions had inconsistencies. 

ASDiv (Academia Sinica Diverse MWP Dataset) \citep{miao-etal-2020-diverse} contains 2.3K math word problems, with each question labeled with a problem type (e.g., Basic arithmetic operations, Aggregative operations) and grade level to show the difficulty level of the problems. This dataset provided different diversity metrics which helped the development of future datasets. This includes GSM8K \citep{cobbe2021trainingverifierssolvemath}, which contains 8.5K school math problems created by human problem writers, with 1K problems as the test set. These problems need 2 to 8 steps to get to the solution, using primary calculations with basic math operations to find the answer. This dataset contains questions that need basic knowledge, such as the number of days in a week.

SemEval 2019 Task 10 \citep{hopkins-etal-2019-semeval} provides question sets from MathSAT (Scholastic Achievement Test) practice exams in three categories: Closed Algebra, Open Algebra, and Geometry. Most questions are multiple choice, with some numeric answers. This test collection contains 3860 questions, of which 1092 are used as the test set. 
The MATH dataset~\citep{hendrycks2021measuringmathematicalproblemsolving} also contains multiple choice questions, developed from American high school math competitions including
AMC 10/12\footnote{\url{https://maa.org/amc-10-12-information-and-registration/}}, and AIME\footnote{\url{https://artofproblemsolving.com/wiki/index.php/American_Invitational_Mathematics_Examination}}. This dataset provides 12,500 problems (7,500 training and 5,000 test). Problems are assigned difficulty levels from 1 to 5, and categorized into seven areas (Pre-algebra, Algebra, Number Theory, Counting and Probability, Geometry, Intermediate Algebra,
and Precalculus). Answers in this dataset include step-by-step solutions in \LaTeX{} with final answers annotated explicitly, making it suitable for training models that provide explanations for their answers.

Other test collections aim to test mathematical reasoning by systems. The DROP (Discrete Reasoning Over the content of Paragraphs) test collection \citep{dua2019dropreadingcomprehensionbenchmark} is a reading comprehension task where a  paragraph is provided along with questions about the passage. DROP has often been used to evaluate the reasoning capabilities of large language models such as Gemini and PaLM. 

After DROP, LILA \citep{mishra-etal-2022-lila} was developed to unify various mathematical reasoning benchmarks.
LILA augments 20 existing datasets with solution programs added to answers, along with instructions for producing answers in natural language. 
Answers are represented as Python strings that prints a number, expression, list, or other data structure. 
For each task, instruction annotations are provided with a clear definition of the task, a prompt providing a short instruction, and examples to help learning by demonstration. 
System effectiveness is measured using the token F1-score between the model output and the target answer.

\textbf{Questions with graphics.} 
Plots, diagrams, and geometric concepts are commonly used in math. 
While text-based problems have been investigated extensively, visual math question-answering has been explored far less.
In recent years there have been attempts to automatically answer questions that include graphics such as tables and diagrams.
One of the earliest such MathQA tasks focused on retrieving formulas from Wikidata in response to questions in natural language \citep{Schubotz:2018aa}.
 An example test collection is TabMWP \citep{lu2023dynamicpromptlearningpolicy}, which contains 38K open-domain grade-level problems that require mathematical reasoning on both textual and tabular data. To solve these problems, systems need to consult data in given table cells. 

 GeoQA \citep{chen-etal-2021-geoqa} is a geometric question-answering dataset. Each question provides a textual description of the problem and its related diagram, and systems choose from answers given as multiple choice alternatives. UniGeo \citep{chen-etal-2022-unigeo}, expanded this dataset, by including the 4,998 calculation problems (from GeoQA) and adding 9,543 proof problems. 

MathVista \citep{lu2024mathvistaevaluatingmathematicalreasoning} provides 6,141 examples of mathematical and visual tasks and studies several large language models, including multimodal models.\footnote{https://mathvista.github.io/} These questions focus on five tasks: figure question answering, geometry problems, math word problems, textbook questions, and visual question answering through multiple-choice or free-form questions. 
Math-Vision \citep{wang2024measuringmultimodalmathematicalreasoning} is another recent test collection for studying large multimodal models for math questions with visual content, providing 3,040 diverse problems.

\textbf{Open response questions and dialogues.}
ARQMath-3 \citep{10.1007/978-3-031-13643-6_20} had a pilot open domain question-answering task, which used the same topics as the math-aware search task. Evaluation measures were computed from lexical overlap of tokens, where answers are treated as a bag of tokens, using the maximum $F_1$ score between system answers and each relevant answers from the answer retrieval task. 
Using a similar approach BERTScore~\citep{zhang2020bertscoreevaluatingtextgeneration} is also used to measure token overlap.

There has also been early work on creating test collections for  conversational math dialogue systems.\footnote{The use of clarifying questions in Math Stack Exchange has been studied using comments on math questions~\citep{10.1145/3578337.3605123}.} 
For example, MathDial \citep{macina-etal-2023-mathdial} provides math tutoring dialogues produced by connecting an expert annotator, who role-plays a teacher, with an LLM that simulates the student working through problems, including reasoning errors. 
The dataset has 3,000 tutoring conversations grounded in math word problems from GSM8k.  
The dataset aims to capture nuances of student-teacher interactions, with a focus on multistep math problem-solving scenarios. 
The main task in MathDial is \textit{Tutor Response Generation}, aiming at modeling the teacher in a dialogue by generating follow-up turns to guide the student to solve problems. 
An earlier example of a related system is
MathBot~\citep{grossman2019mathbot}, a text-based tutor capable of explaining math concepts providing practice questions and offering feedback to students. 
Tasks such as generating clarifying questions, and detecting ambiguous questions can be studied using these test collections and systems.   

\section{Solving math word problems}

In this section we focus on systems that solve math word problems, which have been the most commonly studied math question type.

\textbf{Symbolic and logical rule-based approaches.}
Traditional approaches to solving math-word problems are rule-based.
The word problem is first converted to a symbolic representation (\eg first-order logic) which is solved using an inference algorithm.
The STUDENT system \citep{10.1145/1464052.1464108} solved algebraic problems in Lisp using string transformations to generate symbolic problem instances. 

This basic approach remains effective to this day.
For example, the AiFu model \citep{liu-etal-2019-aifu} 
converts word problems into a logical representation using rule-based templates. 
This representation is based on assertional logic where mathematical objects are formalized as constants, variables, concepts, functions, or relations. For example, \emph{``the integer x equals to 3''}, is transformed into 
\begin{quote}
{\tt Integer(x), Equal(x,3).}
\end{quote}
This representation is then translated via rules to a math representation for the the Satisfiability Modulo Theories (SMT) solver Z3\footnote{\url{https://github.com/Z3Prover/z3}} to produce a solution. 
The rule-based translation uses two templates for unification: one for known concepts and operators, and the other for new entities. This model had the highest accuracy among participating teams in the SemEval 2019 task.

Generating and computing solutions from operator trees (OPT) is also a common approach for solving math word problems, as illustrated in \refexample{ex:expressionrquationtree}.
These trees represent the solution quantity, and a variable to which this solution is assigned for the math-word problem:
\begin{quote} 
\it There are 3 boys and 5 girls in a group. Each person wants to buy 9 pencils. How many pencils do they need to buy altogether?
\end{quote}
As shown in \refexample{ex:expressionrquationtree}, the question is transformed into an OPT with constants defined in the question at leaves in the tree.
Answers are produced by applying operations bottom-up (\ie evaluating the expression).
In one approach the generated OPTs use four basic binary operations, with quantities at the leaves. \citep{roy-roth-2015-solving}.
The ALGES system \citep{koncel2015parsing} uses a similar approach, where the OPT includes a variable for the answer $\chi$  in the tree attached to an equivalence operator (referred to as an  \textit{equation tree}).

\begin{myexample}{OPT for $(3 + 5) \times 9$ generated from a word problem, and variation with answer variable $\chi$ (equation tree).}{ex:expressionrquationtree}
	\centering
	\scalebox{0.7}{\includegraphics{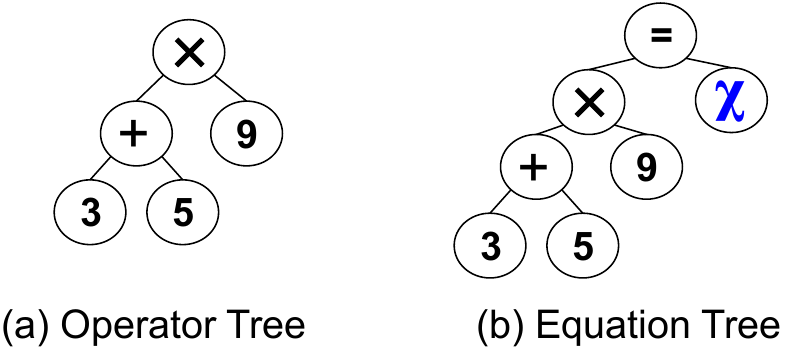}}
\end{myexample}


For questions involving properties of elements in sets, it is important to correctly identify variable quantifiers to distinguish universal and existential quantification (\eg $\forall x$, $\exists x$). Roy and Roth train a binary SVM classifier to select quantifiers as needed, and inserts them at leaves of the tree. The lowest common ancestor (LCA) node between pairs of quantifiers are used to capture constraints. The ALGES system instead generates all possible equation trees without dropping irrelevant quantifiers, and then uses Integer Linear Programming to score the likelihood of each tree using local and global discriminative models.




\textbf{Early machine learning approaches.}
Several statistical machine-learning approaches have been investigated for math-word problems. 
Like rule-based approaches, they first produce a word problem representation, from which the answer is then generated  algorithmically. 
The ARIS system \citep{hosseini-etal-2014-learning} was developed for arithmetic problems using addition and subtraction. 
Steps toward the solution are represented as a state sequence, with one state per sentence. 
States contain subject and objects from the question, each of which have associated  entity triples.
Entity triples have quantities (\ie constants or variables), types, and attributes.


\begin{myexample}{ARIS sentence state generation \citep{hosseini-etal-2014-learning}. Each state has one or two \emph{containers} in black boxes.}{ex:aris}
	\centering
	\scalebox{0.45}{
	\hspace{-0.23in}\includegraphics{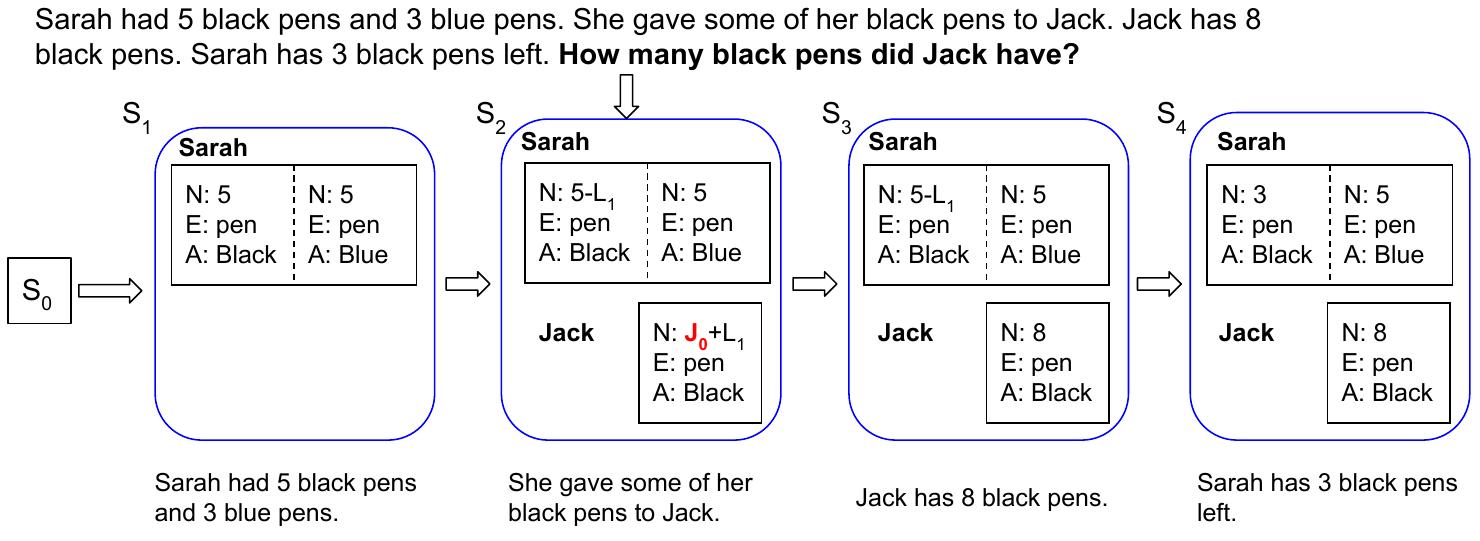}}
\end{myexample}


As an example, consider this math word problem, with two subjects possessing quantity containers (Sarah and Jack): 
\begin{quote}
\it Sarah had 5 black pens and 3 blue pens. She gave some of her black pens to Jack. Jack has 8 black pens. Sarah has 3 black pens left. How many black pens did Jack have?
\end{quote} 
The ARIS state representation for each sentence is shown in \refexample{ex:aris}, which is produced using the Stanford NLP library. 
 $J_0$ shown for the second sentence represents the question quantity, the number of black pens Jack had initially.
Determining this value depends also on variable $L_1$, the number of pens that Sarah gave to Jack.
After the sentence state representations are generated, constraints in the question are used to compute the question answer.
Here we solve for $J_0$ by noting that Jill's final `Black' pen entity count is $L_1 = 5-3$, and Jack's `Black' pen entity is $J_0 = 8 - L_1$,
 giving 6 black pens that Jack originally had.

State transitions are identified by verbs classified by an SVM classifier using WordNet-based and other features.
Verbs are  categorized into three groups: (1) observation, (2) positive (increase), and (3) negative (decrease).
For sentences with two containers (\ie subjects/objects with entities), four other categories are considered: 
(4) positive transfer (from second to first container), 
(5) negative transfer (from first to second container), 
(6) construct (increase in both containers), and 
(7) destroy (decrease in both containers). 


\begin{myexample}{Seq2seq OPT generation for word problems\\ \citep{wang-etal-2017-deep}.}{ex:mwpseq2seq}
	\centering
	\scalebox{0.45}{
	\hspace{-0.4in}
	\includegraphics{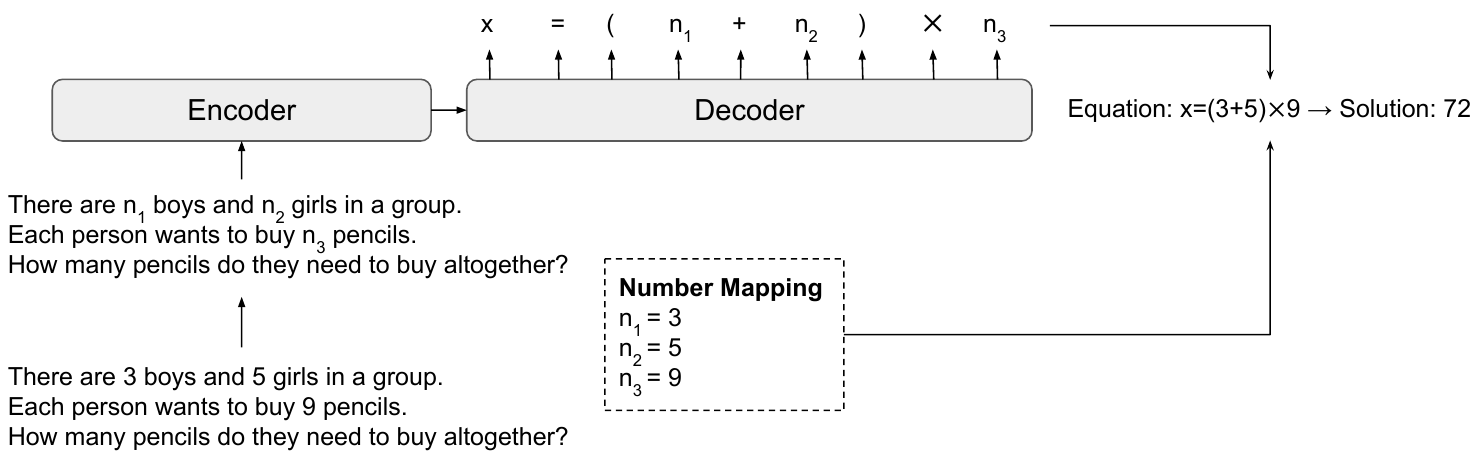}}
\end{myexample}

Recently deep learning techniques  are among the strongest approaches for math-word problems. The earliest attempt, Deep Neural Solver (DNS) \citep{wang-etal-2017-deep} used a seq2seq model to translate problem statements into equations (\ie infix OPT representations) by embedding question text in a vector and then generating a formula expression starting from the question vector.
The generated expression is evaluated to produce an answer. 
Numbers in the question are mapped to enumerated variable tokens and stored in a dictionary (\eg \{ $(n_1, 3), (n_2, 5), (n_3, 9)\}$).
Numbers are replaced by their enumeration variable in the question text before being  passed to a sequence-to-sequence (\emph{seq2seq}) model that generates an equation over those variables. 
The variables of the equation are then replaced by their values from the dictionary, and the expression value is computed.
 Using the `pencils'  example above used to illustrate rule-based approaches, the architecture of the model is shown in \refexample{ex:mwpseq2seq}. 


After DNS was introduced there were several attempts to improve generated OPTs through better structural constraints. 
The multi-encoders and decoders model \citep{shen-jin-2020-solving} uses both seq2seq and graph-based encoder/decoders. 
Two graph encoders are defined using GraphSage \citep{NIPS2017_5dd9db5e}.
The first is a dependency parse tree capturing relationships between words in the sentence. Initial token embeddings used for the tree come from the sequence-based encoder. 
A second graph-encoder captures numerical comparisons: nodes represent numbers from the question with relations $>$ or $\leq$ defining constraints that were ignored by DNS when mapping values to tokens. 
Sequence and tree-based decoders are used.
The decoders produce operation and value sequences that correspond to  OPT traversals 
that can be used to compute the final answer using a simple stack-based algorithm.

For the equation $(n_1+n_2) \times n_3$, the sequence decoder generates an OPT preorder traversal of the expression in  \refexample{ex:expressionrquationtree}, with arguments before operations ($n_1 n_2 +  n_3 \times$). The tree decoder instead generates an OPT postorder traversal with operations before arguments ($\times + n_1 n_2 n_3$). 
The operation sequence used is selected by maximum decoder likelihood. 
This model achieves higher accuracy than DNS on the Math23K dataset (5-fold cross-validation of 76.9\% vs. 58.1\%).

\textbf{Transformers.}
Solving math word problems automatically has gained increasing attention since the advent of transformers in 2017 \citep{DBLP:journals/corr/VaswaniSPUJGKP17}.
Transformers integrate surrounding context in token sequences by consulting all other tokens embeddings in the input over a series of stages. 
This produces contextually-enriched embedding vectors that
dramatically improved their usefulness in prediction and generation tasks.

Thinking back to information tasks in Ch. 1, with a transformer the input tokens act as an initial information source, from which we  produce a new information source containing contextualized token vectors.
The contextualized vectors are shaped by token co-occurrence statistics seen in a large collection of sources used for training.
In terms of information tasks performed by the transformer:

\begin{compactdesc}
\item \textbf{
\underline{Retrieve}}
\begin{compactdesc}
\item \textbf{Consult:} All current token vectors are examined and weighted using a \emph{self-attention mechanism} used in generating new vectors. 
\item \textbf{Query:} input tokens are annotated with control tokens (\eg {\tt [CLS], [SEP], [MASK]}) to (1) drive parameter learning, and (2) allow users to `program' outputs in query \emph{prompts}. 
\end{compactdesc}

\item \textbf{\underline{Analyze}}
\begin{compactdesc}
\item \textbf{Annotate (Designers/Users):} input annotated with additional tokens, \eg {\tt [CLS]} for the first `classification' token, {\tt [SEP]} for ending sentences/sections, and signaling answer start (\eg {\tt [A.]}). \emph{Positional encoding} vectors mark input position (\eg integer enumeration) and/or token relationships.
These annotations are more detailed and expressive than for earlier models.

\item \textbf{Index:} network weights represent language statistics in a collection, and can be used to produce dense vector indexes.
\end{compactdesc}

\item \textbf{\underline{Synthesize}}
\begin{compactdesc}

\item \textbf{Apply:} Token embeddings are generated in steps, \emph{applying} network weights and attention to update token vectors. This uses `depth' to improve generalization (a known property of deep nets). Training tasks apply network weights to make decisions (\eg guess masked tokens) and then update weights using backpropagation.

\item \textbf{Communicate:} contextualized token vectors are communicated in a tensor, providing a new information source.  Learned network weights specifying the embedding function represented in the network is another communicated information source.
\end{compactdesc}
\end{compactdesc}

%
%

As described in Ch. 2, `pre-training' transformers on large text corpora is the norm, followed by `fine-tuning' to produce outputs for specific tasks.
`Pre-training' (\ie initial language model learning) generally involves imitative games that require predictions after token sequence manipulations such as token masking and re-ordering, while fine-tuning requires replacing the output layers of the network for the new imitative game(s) of the specific task the network will be used for (\eg  classification or generation tasks).

As for retrieval, we again find that token representations used for math with transformers are important.
Simple tasks such as generating answers to questions adding and subtracting two numbers were explored early on (e.g., `what is 52 plus 148?'), using  a T5 seq2seq model \citep{nogueira2021investigatinglimitationstransformerssimple}).  
This was motivated by earlier work where for tasks such as reporting the maximum value of a list of numbers, a BERT model obtained an accuracy of 52\%  \citep{wallace-etal-2019-nlp}.  
This poor performance was caused partly the default tokenization by WordPiece, which prevented correctly encoding numbers.

The first applications of BERT model to math word problems was performed using the AQuaA-RAT collection \citep{ling-etal-2017-program}.
 The focus of this work is fine-tuning, including a
 proposed Neighbor Reasoning Order Prediction (NROP) coherence loss \citep{piekos-etal-2021-measuring}.
 This considers whether steps in a rationale for an answer are in their original order, or have been swapped.
 Fine-tuning with this task and loss function improved accuracy by roughly 10\%.
 
 Subsequent work focused on improving mathematical reasoning output for transformers.
MWP-BERT \citep{liang-etal-2022-mwp} used BERT and RoBERTa along with three families of fine-tuning tasks/objectives.
\begin{compactdesc}
\item  \textbf{Self-supervised (questions):} (1) masked language modeling, (2) number counting (\ie quantities in a word-problem), and (3) number type grounding (\eg integer, real). 
\item \textbf{Weakly-supervised (questions with answers):}  (1) answer value type prediction (\ie discrete or continuous), (2) context-answer type comparison (\ie whether question quantities have the same type as the answer), and (3) number magnitude comparison (\ie predict the relative answer size vs. question quantities). 
\item \textbf{Fully-supervised:} (1) operation prediction, and (2) tree distance prediction. 
Operation prediction infers the operator between two quantity nodes
in the solution OPT from five types: \{+, -, $\times$, \verb|/|, \verb|^|\}. 
Tree distance prediction estimates differences in depth for numbers in the solution tree. 
\end{compactdesc}
These different training objective families in isolation produce similar accuracy on Math23K dataset, with the fully-supervised objectives providing the highest accuracy. MWP-BERT is fine-tuned on MathQA.  For Math23K it achieves 82.4\% accuracy vs. 58.1\% for the DNS model (5-fold cross-validation). Similarly, for the Math23K test set MWP-BERT achieves 96.2\% accuracy vs. 53.6\% for the DNS model.



\section{LLMs and mathematical reasoning}

The next generation of math-word problem solving models use Large Language Models (LLMs) which are (roughly) very large transformer models trained on very large amounts of data, and embedded within a text generation system (\eg a recurrent neural network). 
The capability of large language models to solve advanced math questions is closely studied when a new LLM is developed, and
math-word problems are commonly used for this purpose.
 General LLMs are not trained specifically for math questions; open-source models such as Mistral~\citep{jiang2023mistral} and LLaMA (family) have been used to fine-tune math-specific language models, which helps improve responsiveness of models to math-focused \textit{system prompts}.
 
 From prompts the model can be given context and instructed on producing specific outputs, providing a way to indirectly `program'  outputs as described for transformers in the previous section.\footnote{For users, \emph{prompt=query}: these are requests for generated information sources.}  

\paragraph{Improving LLM math question answers.}
A common, non-automated way to improve answers without retraining is \emph{prompt engineering}, where an LLM is prompted multiple times, and then a preferred answer is  selected. 
Aside from this,
there are four common methods for improving answers to math questions given to LLMs vs. directly providing a word problem in a prompt:

\begin{compactdesc}
\item[Instruction tuning:] a labeled dataset of \emph{(prompt, response)} pairs is used for additional training of the model, \ie fine-tuning.  
Often done with state-of-the-art language models such as GPT-4 to generate high-quality training data.

\item[Chain-of-Thought (CoT) prompting \citep{NEURIPS2022_9d560961}:]
prompts include example questions and answers that include a step-by-step reasoning process. 
The goal is to exploit the LLM tendencies to mimic patterns in the prompt, in 
anticipation of a similar reasoning step pattern appearing in the answer, and help with producing correct answers. 
COT is distinct from few-shot prompting (see \refexample{ex:cot}), where only questions and final answers are provided within examples included in the prompt.
CoT can also be used with zero-shot or few-shot prompting, by simply adding {\tt "Let's think step-by-step"} at the beginning of the prompt.

\item[Program-of-Thought (PoT) \citep{chen2022program}:] similar to CoT. Instead of natural language, computer programs represent the solution process formally. PoT frees LLMs from having to generate equations in natural language, and to instead provide explicit computational steps. 
In a zero-shot setting (\ie giving no example question and answer in the prompt), PoT has produced a 12\% improvement over CoT for math word problems. 

\item[Program-Aided Language models (PAL) \citep{pmlr-v202-gao23f}:]  ~\\
interleaves natural language (as in CoT) and programming language statements (per PoT) in answers. The final solution is a program that is evaluated by an interpreter. 
\end{compactdesc}

\begin{myexample}{Standard prompting (left) vs. Chain-of-Thought (right). Input is in grey boxes.
Adapted from \citep{NEURIPS2022_9d560961}. }{ex:cot}

	\scalebox{0.5}{
		\hspace{-0.4in}
		\includegraphics{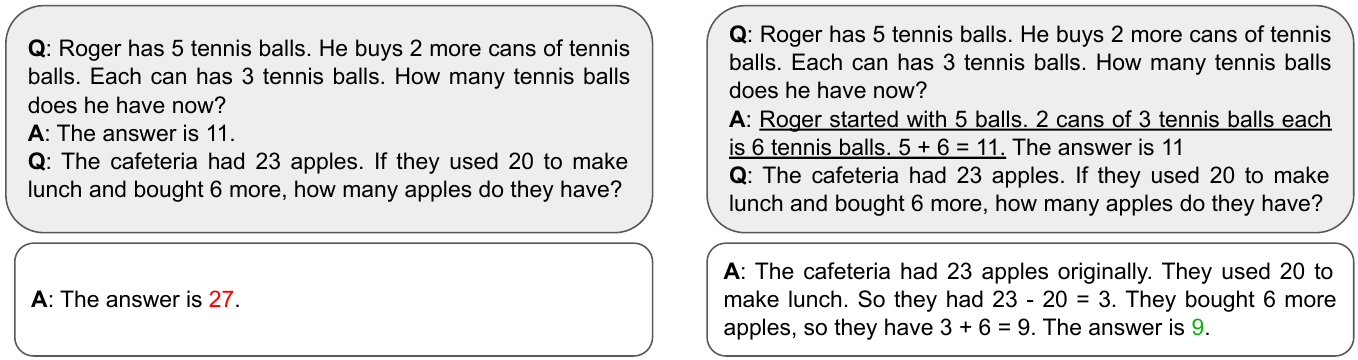}
	}

\end{myexample}


\paragraph{Math-specific LLMs.}
Most research on math-specific LLMs focuses upon input token representations and prompts, and expanding datasets for fine-tuning general models. 
One of the earliest math-specific LLMs is Minerva \citep{NEURIPS2022-18abbeef}. This LLM is based on the PaLM (Pathways Language Model) model, and is fine-tuned on 118GB of scientific and mathematical text.
Minerva can correctly generate~\LaTeX~formulas, which is challenging for general LLMs. 
With proper tokenization, formulas such as $E=mc^2$ are processed as \verb|$E=mc^2$|, and not the single token {\tt Emc2}. 

To improve answers, Minerva uses CoT prompting with several step-by-step solutions to questions before posing the question to be answered. 
It also generates multiple solutions, and then uses majority voting to select the most likely final answer. 
From manually analyzed samples, the most frequent issues in responses include incorrect reasoning, incorrect calculation, and question misunderstanding. For example, for the question: 
\begin{quote}
\it If $\sqrt{400}=\sqrt{81}+\sqrt{n}$, then what is the value of n?
\end{quote} 
Minerva infers that $400=81+n$, and then correctly provides 319 as the final answer, but for the incorrectly inferred formula.

 LLEMMA \citep{azerbayev2024llemmaopenlanguagemodel} is another model using Chain-of-Thought.
The model is built atop LLaMA-Code, and fine-tuned on another dataset {\tt Proof-Pile-2} containing 55 billion tokens from mathematical and scientific documents.
The documents are taken from three large collections: arXiv, open-web-math, and algebraic-stack. 
After fine-tuning, LLEMA is more effective at answering questions from GSM8K and MATH than Minerva.

MAmmoTH \citep{yue2023mammothbuildingmathgeneralist} uses hybrid CoT and PoT to build a new dataset. 
This model is a fine-tuned version of LLaMA-2 and Code LLaMA, using a new dataset {\tt MathInstruct} for instruction-tuning.
MathInstruct is compiled from 13 math rationale datasets (7 existing), using both chain-of-thought (CoT) and program-of-thought (PoT) rationales. 
To create this dataset, GPT-4 is used to generate programs, and the values generated by programs are verified with ground truth.
The dataset contains 260K (instruction, response) pairs with 72\% using COT, and 28\% using PoT. 
\refexample{ex:mathinstruct} shows CoT and PoT pairs. 
The authors report that PoT generally provides better results than CoT.
Particularly for open-form questions, algorithmic reasoning was found to be more effective for complex math problems.

\begin{myexample}{Sample instructions and responses from the MathInstruct dataset. Grey boxes show PoT and CoT instructions, and white boxes show responses.}{ex:mathinstruct}
\scalebox{0.6}
{
	\hspace{-0.3in}
	\includegraphics{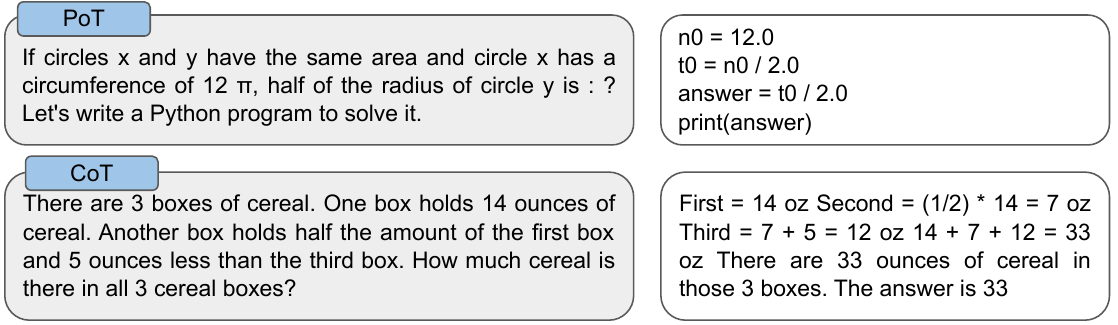}
}
\end{myexample}


ToRA (Tool-integrated Reasoning Agents) \citep{gou2024toratoolintegratedreasoningagent} introduces a new reasoning representation  interleaving natural language and the use of external libraries. Given a question, ToRA first generates reasoning in natural language, which continues until a program library is better suited (e.g., for equation solving). ToRA then generates a program from the natural language reasoning, and the output is further processed for adjustments, sub-task solving, and answer finalization. This process is continued until the final answer is represented using the \LaTeX{} ``\verb|\boxed{}|'' command, as illustrated in \refexample{ex:tora}.\footnote{Note the similarity to the types of answers seen in CQA forums like Math Stack Exchange, where code is part of a larger answer narrative.}
To fine-tune LLaMA-2, the TORA-CORPUS was produced using GPT-4, synthesizing reasoning trajectories for the GSM8k and MATH training sets. 

\begin{myexample}{ToRA reasoning interleaving natural language and program-based tool use. Adapted from \citep{gou2024toratoolintegratedreasoningagent}.}{ex:tora}
\centering
\scalebox{0.625}
{
	
	\includegraphics{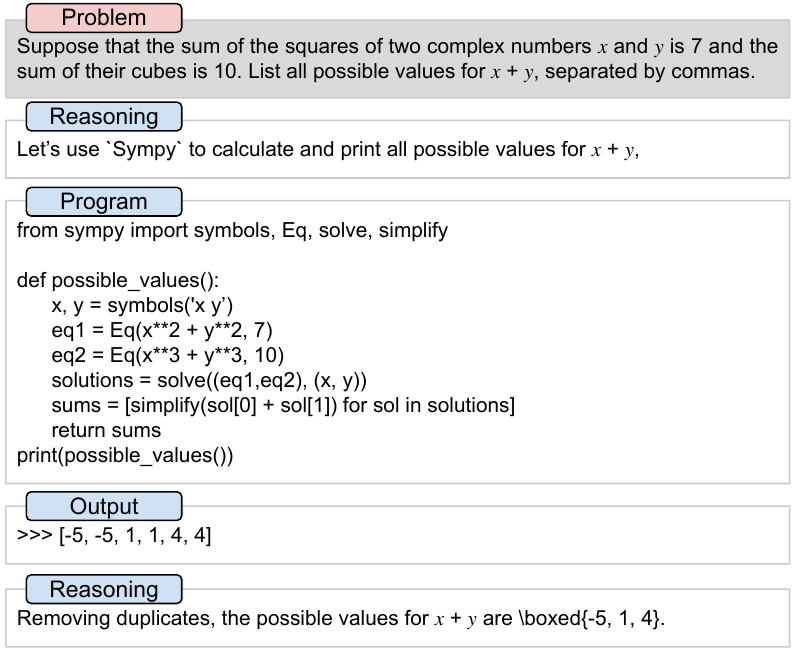}
}
\end{myexample}



\paragraph{Improving LLMs through fine-tuning on augmented data.}
A common
technique for expanding and/or improving existing datasets involves transforming existing question/answer pairs to produce new ones (\ie data augmentation). 

For example, 
MetaMathQA \citep{yu2024metamathbootstrapmathematicalquestions} augments two earlier datasets, GSM8K and MATH. 
Three techniques are used for data augmentation.
First,
additional reasoning chains are generated for answers.  This is done using few-shot CoT promoting, where questions are appended to QA pairs that include associated reasoning for answers.
This few-shot prompt is then fed to an LLM to produce an additional answer with justification.
The final answer is compared against the ground-truth, and incorrect answers are filtered out. 
The second augmentation rephrases questions. Examples of rephrased questions are used in a few-shot query along with 
the prompt:
    {\it You are an AI assistant to help me rephrase questions. Follow the given examples.}
Answers generated for rephrased questions had an accuracy of 76.30\%, and 80.74\% for the original questions.

The third augmentation uses backward reasoning in the MetaMath LLM  to generate additional questions. 
Numeric values in questions are masked with an `X', and the LLM is then asked to predict the value 
given the answer. 
The backward reasoning start from the answer, and then generates reasoning steps to infer the masked numeric value in the question.
The final augmented data set is used to fine-tune LLaMaA-2. 

While most LLMs achieve their highest effectiveness with a large number of parameters, Orca-Math \citep{mitra2024orcamathunlockingpotentialslms} instead fine-tunes smaller language models. 
Similar to MetaMath, a new dataset is introduced and then fine-tuning is done over three iterations. 
The Orca-Math-200K dataset is constructed from previous datasets such MathQA, and GSM8K with a total of 36,217 problems. 
This dataset was augmented by modifying the original word problems. 
GPT-4-Turbo is used with few-shot examples to generate new math word problems, and then prompted to convert the question into a statement using the answer to the question. 
Then, from the statement, it creates a new word problem. 
Solutions for the new problems are also generated by GPT4-Turbo. A total of 120,445 new problems were generated. 

Another augmentation increases the difficulty level of existing questions, 
using two agents called the \textit{Suggester} and \textit{Editor}.
 The Suggester proposes increasing difficulty using techniques such as adding more variables or increasing certain values. The Editor modifies the original question based on the suggester's recommendation. This is done in two rounds, each time increasing the difficulty further. New problems that GPT4-Turbo answers in more than 1800 characters were dropped. 37,157 new problems were collected from this step. 

After the data augmentation, Orca-Math's Mistral-7B LLM is fine-tuned in three steps.
First, instruction-fine-tuning is applied using  pairs of questions and answers. 
In the second step, a  teacher-student learning model is used.
For each question four answers 
 are generated (as `student' answers)  by a supervised fine-tuned model. 
Student answers are compared to answers from GPT-4 Turbo (the teacher) using exact-match prompting.
This is used
to create positive and negative examples based on agreement with the `teacher' answer for further fine-tuning.
In the third step, training samples are generated using  two approaches: Direct Preference Optimization (DPO) \citep{NEURIPS2023_a85b405e} and Kahneman-Tversky Optimization (KTO) \citep{ethayarajh2024ktomodelalignmentprospect}. Both approaches are used for fine-tuning LLMs to align with preferences. With the fine-tuned model, Orca-Math archives an accuracy of 86.81\%, which is higher than much larger models such as MAmmoTH and MetaMath. 

\paragraph{LLM limitations.}
We will use a question from ARQMath-2 to illustrate some limitations of current LLM models. 
Prompting Llama-3-8B-Instruct with: 
{\it ``You are a math professor who will answer math questions, providing step-by-step solutions."}
and the question ``can prove it converges but I don't know at what value it converges. $\sum_{x=1}^{\infty} \frac{(x-1)}{2^{x}}$'', leads to the answer shown in \refexample{ex:llama3}.

This answer is related to the question, but there is an error in the fourth expression, where $rx$ should be $xr^x$. 
Moreover, for a ``math professor'', splitting of the series as:
$$\sum_{x=1}^{\infty} \frac{(x-1)}{2^{x}}=\sum_{x=1}^{\infty} \frac{x}{2^{x}}-\sum_{x=1}^{\infty} \frac{1}{2^{x}}.$$
is not obvious, and needs verification/reasoning about convergence for each of those series. So, if we say it is a ``step-by-step'' solution, then the professor has ignored this crucial step.

\begin{myexample}{Llama-3-8B-Instruct generated answer for a question from the ARQMath-2 Answer Retrieval task (see main text).}{ex:llama3}
\centering
\scalebox{0.93}
{
	\includegraphics{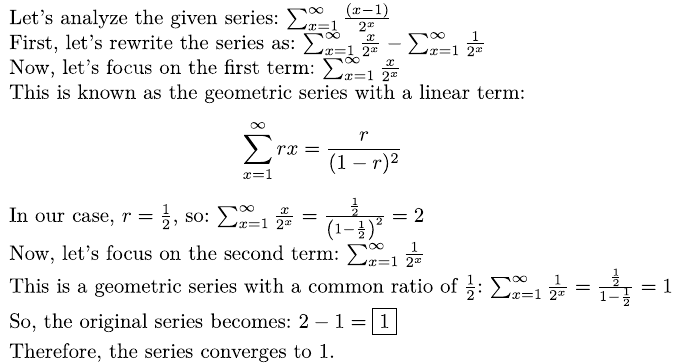}
}
\end{myexample}

The \textit{system prompt} can guide the responses for LLMs. For the same ARQMath question, we change LLaMA-3 prompt to ``You are a math professor who will answer math questions, providing a hint toward the solutions, not the whole solution.'' The response is shown in \refexample{ex:llama3-v2}.

\begin{myexample}{A different Llama-3-8B-Instruct answer when asked to provided a hint, but not the whole solution.}{ex:llama3-v2}

\begin{quote}

\small
Here's a hint to help you get started:
~\\

You can try to recognize this series as a variant of a more well-known series. In particular, consider the geometric series $\sum_{x=1}^{\infty} r^x$, where $|r|<1$. Can you see how the given series can be rewritten in a similar form?
~\\

Also, think about the fact that the series is a telescoping series, meaning that many of the terms cancel out when you sum them up. Can you identify which terms cancel out and how this might help you find the sum?
~\\

Take it from here and see if you can prove that the series converges and find its value!
\end{quote}

\end{myexample}

The first part of the hint is a good start for the student: it narrows down the scope where the student can focus or get some ideas. However, the second part of the hint is completely wrong; this is not a telescoping series.
A \textit{telescoping series} is one where each term of the series can be written as a difference of two consecutive terms of a sequence. In the sense,
$$\sum_{n=1}^{\infty}(a_{n+1}-a_n)=(a_2-a_1)-(a_3-a_2)+\dotsb.$$
So, the terms start canceling off and the sum is easy to find. But the series given in the query cannot be written in this form. So, this second part of the hint does not help, and is very misleading.


Currently, Chain-of-thought and Program-of-thought are commonly used to produce more reliable answers to math questions by LLMs. 
It is likely that more effective techniques will be used in the future. Current models need to individually create a dataset for fine-tuning their base LLM model (e.g., LLaMA-2), often created using a state-of-the-art LLM (e.g., GPT-4*). 
However, recently several researchers have warned against this approach, suggesting that model effectiveness may decrease when AI-generated data is used for training \citep{shumailov2024curserecursiontraininggenerated, Shumailov2024}.

%
%

\chapter{Final Thoughts and Research Opportunities}
\label{c-summary} 

In this chapter we draw our discussion of mathematical information retrieval to a close,
and identify some places where insights might be found in the future.
For this purpose, we return to three key concerns that we have repeatedly returned to throughout the book: 
\forceline
\begin{compactenum}
\item the people that IR systems serve and interact with,
\item the information tasks that IR systems perform on behalf of people,
\item how systems and associated processes are evaluated.
\end{compactenum}
\forceline
Before we close, we also consider some limitations of MathIR systems from the perspective of individuals.

\section{What is next for MathIR?}

To try and characterize the space of opportunities for future work in MathIR, we present two views summarizing topics discussed in the book where we could devote effort to strengthen existing approaches, and uncover new ones.

The first view is shown in \reftab{tab:fneeds}, which presents the information needs for the people that play a role in the creation, use, and evaluation of MathIR systems.
These are the users, assessors, and designers/researchers for these systems, which we introduced in Chapter 1.
Improving available information of the types listed, and/or making these information types  more readily available or easy to use present  research opportunities.
Additional items that we note here that were not mentioned earlier in the book include \emph{pragmatics}. 
A critical requirement for productive knowledge work of the types shown in this table is knowing when to \emph{stop}, and possibly reflect and report at that point.
Improving our shared knowledge of actionable models and heuristics would be beneficial for people involved, and can 
help inform the use cases from which IR models and tools are designed and evaluated.

\ourtable
	{!t}
	{ 
	\vspace{-0.2in}
	\small
    		\begin{tabular}{  l p{3in} }
		
		\toprule
		  \textbf{Group} 	& 	\textbf{Information Needs} \\	
		  \midrule

		\textbf{Users} &  \begin{compactdesc}
			\item Query language operations and syntax
			\item Query/results interface usage
			\item How to organize and refind sources
			\item How to create, organize, and search annotations
			\item Data and tools for applying information in sources
			\item Tools for communicating information from sources
			\item \textbf{Pragmatics:} Stopping criteria for `satisficing'
		    \end{compactdesc}\\
		   
		   \midrule
		    	
		\textbf{Assessors} & \begin{compactdesc}
			\item Annotation interface operations and usage
			\item Relevance criteria and intended application
			\item Navigate, link, and search sources \& annotations
			\item \textbf{Pragmatics:} `reliable' annotation strategies
		    \end{compactdesc} \\	
		    
		    \midrule
		    
		\textbf{Designers \&} \\
		\textbf{Researchers} & 
			\begin{compactdesc}
			\item \textbf{Useful and novel}:
			\begin{compactdesc} 
			\item Data representations for formulas + text
			\item Retrieval interfaces
			\item Query/question mechanisms
			\item Annotation tools/protocols
			\item Machine learning models
			\item Retrieval models
			\item Data augmentation techniques
			\item Evaluation metrics, pooling, and protocols
			\item Conceptual frameworks / topic maps
			\end{compactdesc}
			\item Test collections: creation, use, differences

			\item Implementation: planning and execution
			\item \textbf{Pragmatics:} `good' end points for research/dev projects
		    \end{compactdesc}	\\

		\bottomrule
		\end{tabular}
	}
	{Information needs for people in mathematical information retrieval}
	{tab:fneeds}

The second view for the space of research opportunities addresses researchers and designers specifically.
The rows and columns are organized by information task, and whether the opportunities are related to human interaction, system modules that perform information tasks, or evaluation of interaction and modules.
We have indicated which chapters present related material in the column headers.
Please note that this is a first  attempt at categorization along these dimensions. 
There are opportunities that we have undoubtably missed here.

Currently opportunities in machine learning (\eg improved/compressed LLM models and prompting), improved formula tokenization and attention mechanisms for transformers, and finding new approaches to distilling contextual information into dense embedding vectors and other representations are important and widely recognized; these each realize improved applications of available statistical information.
Some other important opportunities include improved query interfaces, annotation tools, and techniques for communication of search/answer results (\eg sensitive to the searcher background).
Mechanisms that improve a user's ability to organize, link, and quickly refind sources and previous results seem ripe for renewed exploration, provided we're mindful of actual user needs, and not overwhelming the user.
Integration of retrieval systems with mathematical tools such as Computer Algebra Systems (CAS) and theorem provers is another promising direction for interaction, and possibly also annotating queries and collections.

Another key area is the development of new representations for text and formulas, including formulas in isolation, but particularly representations that capture text, formulas, and possibly other graphics to better capture context for use in improved sparse and dense indexing.\footnote{OPT and SLT variants, along with MathAMR and PHOC are described in Chapter 2, but many variations and other representations could be unexplored.} 
Improved indexing and retrieval models remain an important opportunity as well, being fundamental to IR in general.

\ourtable
    {!t}
    {
	\vspace{-0.2in}
    \scriptsize
    	\begin{tabular}{l | >{\raggedright}p{1in} >{\raggedright}p{1in} >{\raggedright\arraybackslash}p{1in} }
	\toprule
	\textbf{Information} &  \textbf{Human Interact.}	& 	\textbf{System Modules} 	&  \textbf{Evaluation}\\
	\textbf{Task} &  \textbf{(Ch. 1)} 	& 	\textbf{(Chs. 2, 4-6)} 	& \textbf{(Ch. 3)}\\
	\midrule

	\textbf{Retrieve}\\~\\
	R1. Query	
	& 
			Search interfaces
			Multi-modal input
	    		LLM prompts/RAG
		    	
		& 
			 Autocomplete \linebreak
			 Query suggestion
			 Query perf. prediction

		& 
			Log/User studies \linebreak
			Module effectiveness
			Benefits of formulas?

		    	\\
	
	R2. Consult	
		& 
			 Formula/text navig.
			 Math/text links 
			 Linked math/text

		& 
			 Math entity linking \linebreak
			 Text / formula tokenization

		& 
			 Log/User studies
			 Math EL effects?
			 Visual. effects?
		    	\\
	
	\midrule
	\textbf{Analyze}\\~\\


	A1. Annotate	
		& 
			User entity cards (\eg formulas)
			Passages in sources \linebreak
			Source links/groups

		& 
			  Text + math represent.  \linebreak
			  Formula represent. \linebreak

		& 
			 Log/User studies
			 Retrieval metrics
			 User bandwidth?
			 Storage requs.?
		    	\\
	
	A2. Index	
		& 
			 Org. cards, notes
			 Org. passages, notes
			 Org. searches \linebreak
		    	 Org. sources (`jar')

		& 
			 Sparse models \linebreak
			 Dense models \linebreak
			 Sparse + Dense (\eg per SPLADE)

		& 
			 Log/user studies \linebreak
			 Metrics: size, speed, effectiveness
		    	\\
	\midrule
	\textbf{Synthesize}\\~\\
	
	S1. Apply	
		& 
			 
			 Reusable formulas (\eg \LaTeX{} chips)
			 Tool integration (\eg CAS) \linebreak
			 Export sources \& notes

		& 
			 Embedding models \linebreak
			 (Pre)training tasks 
			 Learning-to-Rank \linebreak
			 Model reduction (\eg for LLMs)
			 Compression: Index, source, notes, etc.

		& 
			 Log/User studies \linebreak
			 Model effectiveness \& efficiency
		    	\\
	 
	S2. Communicate
		& 
			 Math/text author. tools \linebreak
			 Summarization (\eg LLM)

		& 
			 Data augmentation
			 Math-focused SERP \linebreak
			 Background-targeted answers \linebreak

		& 
			 Log/User studies
			 Test collections \linebreak
			 Research papers \linebreak
			 Shared task papers 
			 Surveys, task \& concept models
		    	\\

	\bottomrule
	\end{tabular}

    }
    {Future Directions for Mathematical Information Retrieval Research}
    {tab:gameplan}

\section{Limitations: What MathIR cannot provide}

In our experiences working on MathIR, a frequently expressed concern \emph{or} hope is that strong systems will remove the need for individuals to understand the math that they use.
Instructors reasonably worry that some math problem sets can be completed by issuing prompts to LLMs or questions to search engines, with students receiving complete or near-complete answers. 
On the other side, math anxiety is common, and information from sources that help reduce mental effort and our risk of failure in addressing hard questions are appreciated, regardless of our math comfort level.
So while copying retrieved answers is clearly a problem, could technology instead accelerate our learning and understanding of new mathematical concepts?

For example,
reliable MathIR systems could quickly retrieve or generate examples of questions and answers for similar problems without giving away answers.
This is not cheating, and seems like an opportunity to save a lot of time.
Unfortunately, after this retrieval step, we arrive at a fundamental limitation of technology well-known to math instructors.\footnote{These comments apply equally to technological  resources for education in general, and more broadly, knowledge work.}
Technology allows us to store, illustrate, organize, and retrieve information faster than in previous decades \citep{noauthor_strategic_nodate}.
This can save significant amounts of human effort in terms of querying for and consulting sources, but related benefits for human \emph{learning} have been modest \citep{cheung_effectiveness_2013}. 
However sophisticated our technologies become, they reside \emph{outside} of a person's mind, 
and there is evidence that deep understanding comes from effort exerted from \emph{inside} a person's mind.

In one psychological study, students who constructed a formula themselves for the area of geometric shapes organized in a lattice were more likely to correctly adapt the formula for new shapes than students who were given the correct formula for the first case \citep{hallinen_finding_2021}. Interestingly, the authors refer to constructing the correct formula as \emph{searching} for the formula; the student must mentally create and compare alternative formulas as they work on the problem. Their study supports the idea that engaged exploration and comparison are needed for deep understanding. Naturally, such engaged thinking takes time.

To correctly use a source such as a search result hit or question answer, one must understand it first. 
Assessing the validity and applicability of an answer \emph{requires} understanding of terminology, notation, and associated concepts. 
We see ample evidence for this on math CQA sites, where large numbers of posts and comments seek only clarification of questions and their context,  with many asking for clarification of language and notation choices. 
Skimming a source online that we don't understand provides patterns that might \emph{suggest} information we could use, and readily supports \emph{copying} its contents.  
Using the same source to address a problem posed using different notation and terminology generally requires a deeper understanding of the material.


And so, students completing (non-trivial) problem sets using MathIR tools will still need to spend time and effort to evaluate returned answers,
and to work through material on their own.
For new and challenging topics, this process of working through the material will remain slow in comparison to issuing a query or question
to a retrieval system.
We do not anticipate that future MathIR systems will remove this requirement.\footnote{Related to this, we believe that supporting, advancing, and sharing \emph{human} understanding is fundamental for societally beneficial academic instruction and research of all kinds.}


As a more compact summary, here are some limitations for mathematical information retrieval systems:
\begin{compactenum}
    \item Relevance is strongly  influenced by searcher expertise and presentation in sources.
    \item Retrieval provides sources, not the understanding of them. 
    \item Deep human understanding of information, \eg for application in new contexts or problems, requires study, exploration, and experimentation -- in other words, time.
\end{compactenum}
That we feel this needs saying at all may surprise some readers, and perhaps some IR researchers more than anyone. Many other specialized search domains have similar challenges (e.g., law, medicine, chemistry). 
But our experience has been that mathematics, being both a powerful tool, and in many cultures a source of both pride and shame, sometimes leads to expressed hopes and concerns at odds with these limitations. 

\section{Some parting thoughts}

The time at which this book is being written is an exciting time for mathematical information retrieval.
One has a sense that we may be in for a very surprising leap forward soon, much as we saw for the game of Go, when AlphaGo applied monte carlo tree search, deep learning techniques, and reinforcement learning to play the game competitively at the international level.
Prior to this, many believed that competitive systems operating in such a vast search space would not be seen for a long time.
AlphaGo was trained through simulating a very large number of games; the model parameters were learned via a somewhat brute force process of trial and error, in order to estimate reliable probabilities of success along different series of moves through empirical observation. 
Back to our information task model, finding efficient and effective ways to retrieve, analyze, and synthesize information about the game space and individual move configurations proved transformative.

With the advent of LLMs and other recent advances in machine learning, we might soon stumble upon effective algorithms that combine  representations, statistics, and constraints in a way that allows complex mathematical information to be found relatively easily, and interactively using conversational search models.
Related methods may automatically answer questions ably and in an audience-appropriate manner, and perhaps even produce proof strategies or complete proofs for mathematical conjectures of real sophistication.
Such advances would finally realize some of the earliest goals of Artificial Intelligence research, and would have a large number of potential applications.

Even if such technologies do not come to pass in the near term, there is another important perspective for the future of MathIR. 
Even with incremental advances, there is an opportunity to create systems that help people by better aligning with their mathematical expertise and communication styles, 
and that make working with mathematical information just a bit more efficient, and just a bit more comfortable.
We find this prospect equally exciting, and a challenging goal from the human, systems, and evaluation perspectives.

\appendix 
\chapter{Online Resources}

\begin{table}

  \centering
  \caption{Resources for Math  Search and Question Answering}
  \label{tab:resources}
  
 \resizebox{\textwidth}{!}{
  
  \begin{tabular}{l|l|l|c}
    \toprule
  \textbf{Name} & \textbf{Type} & \textbf{Link} & \textbf{Year}\\
     \midrule

\multicolumn{4}{l}{\textbf{Tools}}\\\hline
LaTeXML&\LaTeX{}-MathML Convertor&\url{https://github.com/brucemiller/LaTeXML}&2004\\
SnuggleTeX&\LaTeX{}-MathML Convertor&\url{https://github.com/davemckain/snuggletex}&2008\\
LeanDojo &Theorem provers&\url{https://github.com/lean-dojo/LeanDojo}&2023\\ \hline

\multicolumn{4}{l}{\textbf{Test Collections}}\\\hline 
NTCIR&Search Test Collection&\url{https://ntcir-math.nii.ac.jp/data/}&2013-16\\
ARQMath&Search Test Collection&\url{https://www.cs.rit.edu/~dprl/ARQMath/}&2020-22\\\hline

AQUA-RAT&MWP Dataset&\url{https://github.com/google-deepmind/AQuA}&2017\\
MathQA&MWP Dataset&\url{https://math-qa.github.io/}&2019\\
ASDiv&MWP Dataset&\url{https://github.com/chao-chun/nlu-asdiv-dataset.}&2020\\
GSM8K&MWP Dataset&\url{https://huggingface.co/datasets/openai/gsm8k}&2021\\
MATH &MWP Dataset&\url{https://github.com/hendrycks/math}&2021\\\hline

GeoQA& Geometry QA& \url{https://github.com/chen-judge/GeoQA}&2021\\
UniGeo& Geometry QA& \url{https://github.com/chen-judge/UniGeo}&2022\\\hline

MathVista&Visual Math Problem&\url{https://mathvista.github.io/}&2024\\
MATH-Vision&Visual Math Problem&\url{https://github.com/mathllm/MATH-V}&2024\\\hline

\multicolumn{4}{l}{\textbf{Systems}}\\\hline 

MIAS&Formula Search&\url{https://github.com/MIR-MU/MIaS}&2011\\
Tangent-S&Formula Search&\url{https://github.com/MattLangsenkamp/tangent-s}&2017\\
Tangent-V&Formula Search&\url{https://www.cs.rit.edu/~dprl/files/TangentV-source.zip}&2019\\
Tangent-CFT&Formula Search&\url{https://github.com/BehroozMansouri/TangentCFT}&2019\\
NTFEM&Formula Search&\url{https://github.com/NTFEM/Formulae-Embedding}&2020\\
MathEmb&Formula Search&\url{https://github.com/Franknewchen/MathEmb}&2021\\\hline

WikiMirs&Math-Aware Search&\url{https://github.com/huxuan/WikiMirs}&2016\\
Tangent-L&Math-Aware Search&\url{https://github.com/fras2560/Tangent-L}&2018\\
Approach0&Math-Aware Search&\url{https://github.com/approach0/search-engine}&2019\\
MathDowsers&Math-Aware Search&\url{https://github.com/kiking0501/MathDowsers-ARQMath}&2021\\
ALBERT-for-Math-AR&Math-Aware Search&\url{https://github.com/AnReu/ALBERT-for-Math-AR}&2021\\
MABOWDOR&Math-Aware Search&\url{https://github.com/approach0/pya0}&2023\\
CrossMath&Multilingual Math-Aware S. & \url{https://github.com/jgore077/CrossMath}&2024\\\hline

LILA&QA Dataset/LLM&\url{https://lila.apps.allenai.org/}&2022\\
MAmmoTH&Math-LLM&\url{https://github.com/TIGER-AI-Lab/MAmmoTH}&2023\\
LLEMMA&Math-LLM&\url{https://github.com/EleutherAI/math-lm}&2024\\
ToRA&Math-LLM&\url{https://github.com/microsoft/ToRA}&2024\\
MetaMath&Math-LLM&\url{https://github.com/meta-math/MetaMath}&2024\\



 \bottomrule
  \end{tabular}
  }
\end{table}

\chapter{Search and QA for Theorem Proving}
\label{chap:proving}

In Chapters 5 and 6 we discussed math-aware search and math question-answering. A closely related topic to these is theorem proving. Automated theorem proving aims to explore the application of computers in proving mathematical theorems. This is actually one of the earliest topics of interest in computer science. 
In the early 1960s, scientists started exploring the use of machines for theorem proving. for the quantification theory \citep{10.1145/321033.321034, 10.1145/368273.368557} using deduction rules to prove assertions.

Informal theorem proving refers to the way that humans approach proving theorems using reasoning through notation and natural language, likely with some missing details (\eg assumed definitions) and skipped computational steps, for example. 
In contrast, formal theorem proving represents theorems in a machine-readable format, making verification by logical rules possible, at the cost of more information being explicitly stated (\eg all variable types, definitions for all operators, etc.). 

Converting informal to formal proof steps is referred to as \textit{Auto-formalization}. The Mizar\footnote{\url{https://mizar.uwb.edu.pl/}} language is commonly used by mathematicians as a formal language for writing definitions and proofs. 
Using Mizar, Wang et al. explored converting (translation) of informal~\LaTeX-written text into formal Mizar language \citep{10.1007/978-3-319-96812-4_22}. This work explored different seq2seq architectures, with LSTM and attention providing the highest BLEU score for this translation. 

Researchers have also explored other formalization languages, including applying large language models to generate statements in Codex \citep{chen2021evaluating} has been studied recently \citep{NEURIPS2022_d0c6bc64}.
This translates statements in natural text into formalized theorems for the interactive proof assistant Isabelle \citep{10.1007/978-3-540-71067-7_7}.  As an example, the system translated ``Prove that there is no function $f$ from the set of non-negative integers into itself such that $f(f(n)) = n + 1987$ for every $n$'' perfectly to Codex as:
\begin{verbatim}
    theorem
        fixes f :: "nat \<Rightarrow> nat"
        assumes "\<forall> n. f (f n) = n + 1987"
        shows False
\end{verbatim}

Another task for theorem proving is retrieving useful lemmas that will help with proving steps, known as \textit{premise selection}. This is a form or search problem, where relevance is defined in terms of suitability for proving a specific conjecture. The problem is defined as \citep{Alama2014}: 
\begin{quote}
   \it Given an Automated Theorem Provers and a large number of premises, find premises that are useful to the prover for constructing a proof for a new conjecture.
\end{quote}

DeepMath \citep{NIPS2016_f197002b} is one of the earliest works to apply deep learning for this task. 
Conjecture and axiom sequences are embedded separately, concatenated, and then passed to a fully connected neural network for predicting the usefulness of the axiom. 
Embeddings are at character level for formulas, and word-level for statements defining symbols.
The convolutional network model FormulaNet \citep{NIPS2017_18d10dc6} used a similar idea and applied graph neural networks. 
Using formulas in higher-order logic \citep{church1940formulation}, each formula is first parsed into an OPT: internal nodes represent a quantifier or a constant or variable function, and leaf nodes represent variable or constant values.  
Edges connect a quantifier to all instances of its quantified variables. After creating the tree, a merging step is applied to merge leaf nodes representing the same constant/variable. Finally, a unification technique is used to replace variable names with `VAR' and function names with `VARFUNC'. After building the graph, convolution or message passing is applied to get node embeddings. These embeddings are used with max-pooling to form an embedding for the graph. 

As theorems are built upon existing mathematical knowledge, a graph representation of mathematical concept statements such as lemma, and definitions is a common approach for this task. One way of building this graph is to use statements as nodes and ordered edges from node $s_1$ to $s_2$ if there is statement 1 is a premise of statement 2 \citep{ferreira_premise_2020}. With this definition of a graph, the problem can be viewed as link prediction, for which Deep Graph Convolutional Neural Network
(DGCNN) architecture was applied \citep{Zhang_Cui_Neumann_Chen_2018}. The textual content embedding of each node in this work is encoded using Doc2Vec \citep{pmlr-v32-le14} model with mathematical concepts being encoded as linearized trees, with every sub-expression represented as a token. 

For example, the formula $(x+y) \times c$ is represented as a sequence of tokens for subexpressions \{`$x$', `$y$', `$(x+y)$', `$(x+y) \times c$'\} (similar to what is used to generate subexpression tokens for the WikiMirs search model discussed earlier). The authors later introduced STAtement Representation (STAR) cross-modal representation \citep{ferreira-freitas-2021-star}, treating mathematical formulas as natural language. Each statement is viewed as a combination of words and formulas. However, for embedding, they proposed two separate self-attention layers, one for formulas and one for words. The output of the self-attention layers is then concatenated and passed to a Bi-LSTM to get the final representation of the statement. To find the relatedness score between conjecture and premises, a Siamese neural network was applied.

Generative and large language models have recently been applied for premise selection. LeanDojo \citep{NEURIPS2023_44414694} is an open-source toolkit based on the Lean\footnote{\url{https://lean-lang.org/}} programming language, that introduces ReProver (Retrieval-Augmented Prover). ReProver is a language model-based prover, augmented with retrieval for selecting premises. Given the initial state of proof, it retrieves a set of useful premises (set at 100) using a Dense Passage Retriever. These premises are concatenated to the initial state and passed to a fine-tuned ByT5 \citep{xue-etal-2022-byt5} model to generate steps toward the proof. 

As the application of LLMs for theorem proving is gaining attention, new datasets are being introduced for theorem proving.  NATURALPROOFS \citep{welleck2021naturalproofs} for example, is a multi-domain corpus of mathematical statements and their proofs, written in natural mathematical language. The main tasks in this dataset are: 1) \textit{mathematical reference retrieval}: given a theorem, retrieve a set of references that occur in the theorem proof, and 2) \textit{mathematical theorem generation}: generate the sequence of references that occur in a given theorem's proof. Table \ref{tab:proof-resources} summarizes some of the existing resources for this task.




		



\begin{table}[t]
  \centering
  \caption{Resources for Math  Search and Question Answering}
  \label{tab:proof-resources}
  \resizebox{\textwidth}{!}{
  \begin{tabular}{l|l|l|c}
    \toprule
  \textbf{Name} & \textbf{Type} & \textbf{Link} & \textbf{Year}\\
     \midrule

NaturalProof&Proof Dataset&\url{https://github.com/wellecks/naturalproofs}&2021\\

FormulaNet&Theorem Proving&\url{https://github.com/princeton-vl/FormulaNet}&2017\\
STAR&Theorem Proving&\url{https://github.com/ai-systems/crossmodal_embedding}&2021\\

 \bottomrule
  \end{tabular}
  }
\end{table}


\backmatter  
\printbibliography

\end{document}